# Atomically-Thin Tsumoite (BiTe) based All-Photonic-Isolator, Information Converter, and Logic-Gate

*Saswata Goswami[†], Caique Campos de Oliveira[†], Abhijith M.B., Varinder Pal, Vidya Kochat, Pulickel M. Ajayan, Samit K. Ray\*, Pedro A. S. Autreto\*, and Chandra Sekhar Tiwary\**

S. Goswami
School of Nano Science and Technology, Indian Institute of Technology, Kharagpur, West Bengal-721302, India

C. C. Oliveira, P. A. S. Autreto
Center for Natural and Human Sciences (CCNH)
Federal University of ABC
Rua Santa Adélia 166, Santo André 09210-170, Brazil.
Email: pedro.autreto@uafbc.edu.br

Abhijith M.B., V. Kochat
Materials Science Centre, Indian Institute of Technology, Kharagpur, West Bengal 721302, India

P. M. Ajayan
Department of Materials Science and NanoEngineering, Rice University, Houston, Texas 77005, United States

S. K. Ray
Department of Physics, Indian Institute of Technology Kharagpur, West Bengal 721302, India
Email: physkr@phy.iitkgp.ac.in

V. Pal, C. S. Tiwary
Department of Metallurgical and Materials Engineering, Indian Institute of Technology Kharagpur, West Bengal 721302, India
E-mail: chandra.tiwary@metal.iitkgp.ac.in

[†] Authors with equal contribution.

Abstract

Two-dimensional tsumoite (BiTe), a polymorph of Bi$_2$Te$_3$, has emerged as a promising candidate for nonlinear photonic devices owing to its strong spin-orbit coupling, tunable bandgap, and high carrier mobility characteristics. This work presents a thorough examination of the third-order nonlinear optical response of BiTe dispersions using spatial self-phase modulation (SSPM) spectroscopy. The nonlinear refractive index ($\boldsymbol{n_2}$) and third-order nonlinear susceptibility ($\boldsymbol{\chi^{(3)}_{monolayer}}$) are quantitatively derived from the diffraction ring patterns, demonstrating $\chi^{(3)}_{monolayer}$ values, similar to or surpassing those of advanced 2D materials. The temporal development and distortion of the SSPM rings are examined using the wind-chime model, and thermal factors influencing the SSPM pattern are analyzed. First-principles electronic band structure studies reveal that the elevated nonlinear susceptibility arises from band dispersion. Direct correlation between carrier transport and $\chi^{(3)}_{monolayer}$ is established. Utilizing these qualities, all photonic devices, including a photonic isolator based on a 2D BiTe-2D hBN heterostructure, are depicted to show asymmetric propagation. A photonic information converter and a logic gate are designed using the cross-phase modulation technique. These findings establish 2D BiTe nanostructure as a formidable nonlinear optical platform for advanced photonic signal processing and integrated photonic applications.

# 1. Introduction

All-optical devices eliminate the electronic bottleneck, ensuring ultrafast signal processing with lower latency, broader bandwidth, and reduced energy dissipation compared to traditional electro-optical systems.[1-2] To realize such high-performance all-optical functionalities, photonic materials are required with enhanced nonlinear optical properties, a high value of third-order nonlinear response $\chi^{(3)}$, ultrafast carrier dynamics, and strong quantum confinement. The layered architecture of BiTe and strong spin-orbit coupling inherent to Bi-Te systems suggest that it may also possess nontrivial optical and electronic characteristics.[3-4] In engineered bismuth telluride (Bi & Te), the combination of narrow band gaps, high carrier mobility, strong quantum confinement, and layered anisotropy gives rise to pronounced third-order nonlinear optical responses, including the Kerr effect.[5-7] Their inherently high refractive index, narrow bandgap, and topological surface states collectively enhance light-matter interaction, enabling large nonlinear refractive index modulation under continuous-wave and pulsed excitation.[8-10] When intense laser radiation propagates through a dispersion of 2D BiTe nanostructure, the spatial redistribution of the refractive index induced by electronic contribution gives rise to self-phase modulation (SSPM). Several semiconductor materials' nonlinear refractive index coefficients have been evaluated with this method, including graphene[11], $MoS_2$[12], 2D Sb[13], 2D Te NS[14], black phosphorus (BP)[15], $NiTe_2$[16], SnS[17], MoP microparticles[18], $Fe_nGeTe_2$[19], TaAs[20], black phosphorus (BP) and Violet phosphorus (VP)[21], and MXene[22]. The resulting concentric diffraction rings and intensity-dependent phase shift provide a reliable route to quantify the effective nonlinear refractive index, and third-order susceptibility. Owing to the combination of high optical nonlinearity, broadband operation, and environmental stability, BiTe-based 2D materials are a promising platform for all-optical switching, logic devices, and photonic signal processing. Photonic isolators are designed to realize nonreciprocal light propagation, playing a vital role in ensuring stable signal flow in optical communication systems and integrated photonic platforms.[23] In this work, we have synthesized 2D BiTe via an easily scalable liquid-phase exfoliation process. The total third-order nonlinear susceptibility $\chi^{(3)}_{total}$, nonlinear refractive index $n_2$, and third-order nonlinear susceptibility for a monolayer $\chi^{(3)}_{monolayer}$ are nonlinear coefficients that are computed and compared with other semiconductor materials. This work demonstrates a novel all-optical method for unidirectional light propagation in an SSPM-based photonic isolator, leveraging the strong third-order nonlinearity of 2D BiTe and the reverse saturable absorption of 2D h-BN. Our research demonstrates that 2D-BiTe has a low bandgap, enabling broadband nonlinear optical response when exposed to a laser beam. Various models were used to determine the time period required for pattern expansion by analyzing the time-dependent growth of the SSPM pattern. We investigated SSPM pattern distortion and measured changes in refractive index. The proposed scheme replaces bulky magneto-optic isolators with an SSPM-based heterostructure that exploits the unique behavior of 2D materials, with 2D BiTe and 2D hBN. A similar contrast method is applied to further estimate the value of $n_2$ in 2D BiTe. The reverse absorption property of 2D hBN was utilized to derive the proposed hybrid structure. This photonic isolator does not suffer from the typical disadvantages of a bulky magnetically based isolator. The phenomenon known as cross-phase modulation (XPM) occurs when another co-propagating pump laser beam causes the probe light to undergo additional phase accumulation or change.[11, 24] In this assignment, XPM is the optical phase shift that occurs when two beams intersect inside a nonlinear medium (2D BiTe dispersed in IPA solvent). Cross-phase modulation (XPM) was implemented to realize all photonic information converter (650 and 532 nm) and all photonic logic gate applications (λ= 650-532, 532-405, and 650-405 nm). Finally, to depict the electronic behavior with third order nonlinear susceptibility a relationship between $\chi^{(3)}_{monolayer}$ / $\chi^{(3)}(10^{-9})$ vs $m^*$ and $\chi^{(3)}_{monolayer}$ / $\chi^{(3)}(10^{-9})$ vs $\mu$. Electronic band structure of 2D BiTe was calculated through the ab-initio method, a comprehensive theory was developed to relate the high value of $\chi^{(3)}_{monolayer}$ with electronic characteristics.

## 2. Growth and Morphological Study of 2D-BiTe

High-purity bismuth (99.99%) and tellurium (99.99%) were used as precursor materials, mixed in a 1:1 molar ratio to synthesize BiTe crystals. The flame-melting method was used to synthesize the stoichiometric mixture for approximately 15 minutes in a sealed quartz ampoule under an argon atmosphere to prevent oxidation. The molten crystal was subsequently allowed to cool gradually inside the furnace to stabilize the desired crystalline phase. The solidified ingot was then annealed at 450 °C in a muffle furnace under a nitrogen atmosphere, with the sample enclosed in a sealed quartz tube to ensure phase purity and homogeneity. Using a mortar and pestle, the bulk crystal was crushed into powder. The obtained powder was subsequently dispersed in an 80 ml beaker containing isopropyl alcohol (IPA) and subjected to probe sonication using an ultrasonic probe sonicator (Rivotek, model SM250PS) for 4 hours to achieve effective exfoliation and uniform dispersion. The resulting dispersion was then allowed to stand undisturbed for 12 hours to facilitate the sedimentation of the bulk phase, thereby enabling the separation of the comparatively lighter two-dimensional (2D) nanostructure from the heavier particulates. The suspension was centrifuged at 2000 rpm for 30 minutes using a REMI PR-24 centrifuge after a 12-hour recovery period. The 2D BiTe nanostructures were synthesized efficiently using this method because the material could be removed from the solution. A Bruker D8 Advanced with Cu K$_\alpha$ radiation (K$_\alpha$ = 1.54 Å) was used for the X-ray diffraction (XRD) study. The XRD results show that the bulk and two-dimensional structures of the material share the same peaks. Figure 1a illustrates that the crystal structure is preserved throughout the exfoliation process. The reflections corresponding to the indices (005), (104), (018), (110), (0012), (0014), 0014), and (1112) demonstrate pronounced and distinct features, with associated 2θ values of 17.77°, 27.53°, 37.97°, 40.76°, 44.93°, 53.95°, and 62.41°, respectively. BiTe is classified under the P$\bar{3}$m1 space group, characterized by lattice parameters a = 4.423 Å and c = 24.002 Å. The corresponding ICDD card number is 44-667. The hexagonal cell type lattice structure of BiTe consists of 12 atomic layers, which are arranged in the sequence: - (Te$^1$-Bi-Te$^2$-Bi-Te$^1$) - (Bi-Bi) - (Te$^1$-Bi-Te$^2$-Bi-Te$^1$). To make TEM specimens, the material suspended in the IPA solution was dropped onto a carbon film-lined Ted Pella TEM sample grid, which was placed on the Cu side. STEM images of the exfoliated nanostructure of 2D BiTe were obtained with the JEOL ARM 300F2. Figure 1b presents a STEM image confirming the material's crystalline nature. Figure 1c depicts the 3D map of the selected surface, and in Figure 1b, the hexagonal lattice is visible. Image simulations were performed using an acceleration voltage of 300 kV, a spherical aberration coefficient (*Cs*) of 0.0003 mm, a chromatic aberration coefficient (*Cc*) of 1 mm, a convergence semi-angle of 26 mrad, and zero defocus (Δf = 0 nm), corresponding to a high-resolution, Cs-corrected STEM configuration.[25] The 52-200 mrad angle range was configured for the HAADF. This image simulation is shown in Figure 1d. Figure 1e illustrates the Fast Fourier Transform (FFT) of the Selected Area Electron Diffraction (SAED) pattern associated with the 2D BiTe nanostructure. EDS mapping was obtained using the EDX spectrometer operating at 300 kV. Figure 1f-g shows the elemental mapping of the 2D BiTe nanostructure and confirms the presence of Bi and Te. The EDS mapping shows a uniform distribution of the two elements in the 2D nanostructure. Figure 1h shows the overlapped elements onto the main SE (Scanning Electron) image, depicting uniform distribution of the atoms over the nanostructure.

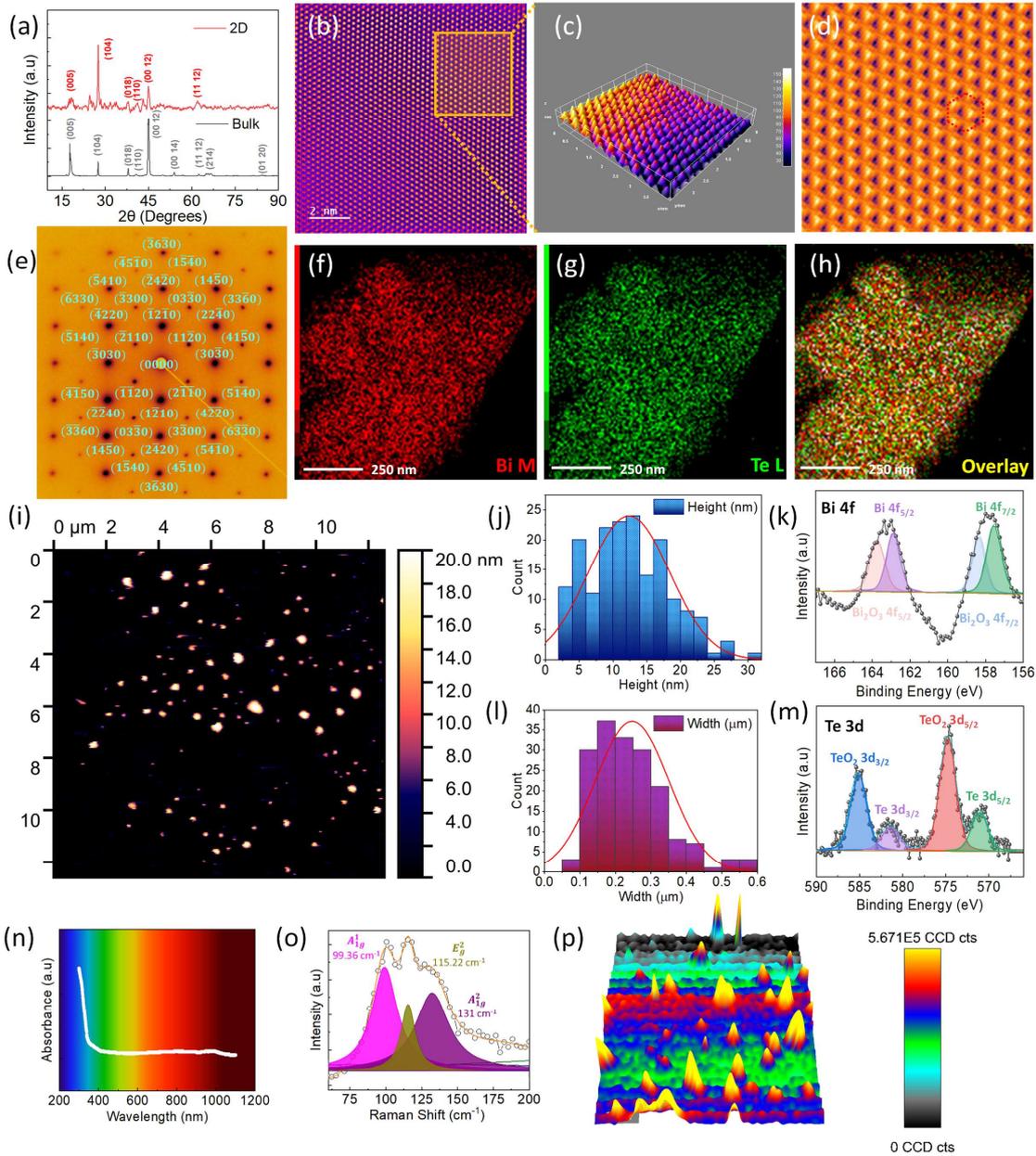

**Figure 1**. (a) X-Ray Diffraction analysis of 2D BiTe and Bulk. (b) The Scanning transmission electron microscopy image of the 2D BiTe. (c) Figure depicting a 3d map of the atomic arrangement. (d) Figure showing simulated atomic position. (e) SAED pattern of the 2D BiTe. (d) SAED pattern of the 2D BiTe. Scanning transmission electron microscopy-based elemental mapping performed on a 2D BiTe flake-type nanostructure showing the presence of (f) Bi and (g) Te. (h) Figure showing overlay of atoms on the main SEI image. (i) Figure showing AFM image of the 2D BiTe nanostructure on the Si wafer. (j) Figure showing the height distribution. (l) Figure 1 showing the width distribution. X-Ray Photoelectron Spectroscopy analysis performed on (k) Bi 4f and (m) Te 3d. (n) Absorbance spectrum derived from UV-Visible Spectroscopy. (o) Raman Spectroscopy analysis depicting $A_{1g}^1$, $E_g^2$, and $A_{1g}^2$ modes. (p) Figure showing Raman mapping performed on 2D BiTe nanostructure deposited on a Si wafer.

Figure 1i shows the atomic force microscopy (AFM) image of 2D BiTe nanostructures dropcast onto a Si wafer. A low-concentration BiTe solution was spin-coated at 2000 rpm for 25 s
6

prior to imaging. AFM (Agilent Technologies, 5500) was used to measure the dimensions of individual nanostructures, with height and width profiles presented in Figures 1j and 1l. The width of the 2D nanostructures ranges from 0.1 to 0.6 μm, with an average value of 250 nm. Figure 1j shows that the thickness distribution of the 2D BiTe ranges from approximately 1 to 30 nm, with some particles exhibiting thicknesses significantly higher than 30 nm. The mean thickness is taken to be 12 nm. 2D BiTe solution drop-casted silicon wafer, was subjected to scanning electron microscopy using a JSM-IT300HR JEOL microscope. Figure S1a shows the SEM image of the bulk BiTe after the exfoliation process. Figure S1d① displays the SEM image, while Figures S1d②-③ illustrate the EDX elemental maps of Bi and Te, thereby confirming their uniform distribution within the nanostructure. The EDX spectrum (Figure S1c) reveals a Bi:Te atomic ratio of approximately 1:1, with a minor excess of Bi indicating P-type behavior, presumably resulting from Te evaporation during the melting process. The IPA solution, which included 2D BiTe, was subsequently drop-casted onto a silicon wafer for chemical analysis. The survey spectrum, shown in Supplementary Figure S2, covered a binding energy range of 0-1100 eV and revealed the presence of Te, Bi, C, and O. Figure 1k displays the high-resolution XPS scan of the Bi 4f doublet, with $4f_{7/2}$ and Bi $4f_{5/2}$ peaks observed at ~158.5 eV and ~164 eV, respectively. Deconvolution of these peaks indicates two distinct components at 157.6 eV and 163.12 eV. Furthermore, a series of minor satellite peaks are observed at approximately 158.9 and 164.3 eV. The values closely align with those observed for $Bi_2O_3$, indicating surface oxidation. Figure S3a depicts a high-resolution XPS spectrum of the O 1s region, showing a prominent peak centered at ~530 eV. The deconvoluted components correspond to lattice oxygen and surface-adsorbed oxygen species, indicating both bulk bonding and surface-related contributions. Figure 1m displays the Te 3d core level. The photoelectron features at 572.5 eV and 582.9 eV correspond to the Te $3d_{5/2}$ and Te $3d_{3/2}$ core levels of 2D BiTe nanostructure, respectively.[26] The elevated energy peaks of the Te 3d multiplets indicate oxidized Te. Supplementary Figure S3b shows the presence of carbon in the sample. Deconvolution of the XPS signals identifies three distinct components, corresponding to C-S, aromatic C-C, and aliphatic C-C bonds.[27] Both peaks are located at 283.62 eV and 285.62 eV. The emergence of C peaks is mostly attributed to sample processing conditions.[28] The absorbance spectrum of 2D BiTe in the IPA solvent was evaluated by UV-Visible Spectroscopy, as shown in Figure 1n. 2D BiTe's direct bandgap is estimated using the Tauc plot, which is shown in Supporting Information Figure S4. The optical bandgap value is determined to be 0.9 eV. The center of inversion is drawn between the centrally situated bismuth atoms. The core unit cell consists of all 12 atoms in their positions, arranged into 6 pairs of atoms occupying equivalent positions. Consequently, the aggregate count of normal modes of BiTe is 36. The space group of the material can be deduced to be (d,c) C$_{3v}$(2).[29] At q = 0, the normal modes corresponding to the motion of each pair are as follows: $A_{1g} + A_{2u} + E_g + E_u$. By excluding the observed three acoustic modes, namely $A_{2u} + E_u$, the following exclusion principle, the 18 Raman-active modes will correspond to 12 distinct frequencies: $6A_{1g} + 6E_g$. The Raman peak locations at 59 ($A_{1g}^1$), 90 ($E_{1g}$), 101 ($E_{2g}$), and 118 ($A_{1g}$) cm$^{-1}$, respectively. These relate, respectively, to degenerate in-plane vibrations of E symmetry and non-degenerate out-of-plane vibrations of A symmetry. In ultrathin 2D nanostructures, out-of-plane vibrations are constrained than in bulk materials, unlike the in-plane vibrational modes. The Raman spectrum is represented by the convolution of three Lorentzian lines, as seen in Figure 1o pertaining to the Raman-active signals. The number of acquired peaks and their locations align with previously reported modes.[30-33] No signals are seen at around 135 cm$^{-1}$, indicating the nonexistence of the Bi$_2$Te$_3$ phase.[34-35] Figure S5 shows the Raman Spectroscopy signal from the bulk metal BiTe. Figure 1p shows the 3D map of the Raman Signals originating from the distributed particles onto the Si wafer.

## 3. Results and Discussions
### 3.1 SSPM Spectroscopy Setup and Estimation of Nonlinear Optical Coefficients

The nonlinear optical response of 2D BiTe, as measured by the SSPM phenomenon, is investigated in this experiment. A complete experimental setup is shown in Figure 2a. The process began with three separate continuous wave (CW) laser beams, a 200 mm convex lens, and a white screen. The red beam had a wavelength of 650 nm red laser beam, the green beam of 532 nm, and the violet beam of 405 nm. A 200 nm convex lens was used to focus the beam onto the quartz cuvette containing 2D BiTe-IPA solution, with its front surface placed near the focus in the laser beam path. The diffraction ring pattern was observed on the far-field diffraction screen and recorded using a charge-coupled device (CCD) camera. SSPM derives from the Kerr nonlinearity, which arises from intensity-dependent changes in the refractive index. In this case, the third-order nonlinear susceptibility $\chi_{total}^{(3)}$ controls how the Kerr medium responds to light. In the presence of a strong optical field, the induced polarization in the 2D BiTe nanostructure exhibits a large third-order nonlinear susceptibility. Supporting Information Figure S6 shows the introduction of the phase-shift due to 2D-BiTe light matter interaction, and formulation of the diffraction profile generated at the wavelength 650 nm, and intensity 10.31 W.cm$^{-2}$. This causes the refractive index to change with the intensity of the light as it traverses through the Kerr medium.

The nonlinear refractive index ($n_2$) of the 2D BiTe is defined by Equation 1,

$$n(r) = n_0 + n_2 I(r) \quad\quad\quad\quad\quad (1)$$

where $n_0$ represents the linear refractive index, and $I$ denotes the intensity of the incident laser source. The laser's operating wavelength was denoted by the symbol λ. $L_{eff}$ denotes the effective beam propagation length within the cuvette. The nonlinear refractive index $n_2$ is defined as follows:

$$n_2 = \left(\frac{\lambda}{2n_0 L_{eff}}\right) \cdot \frac{dN}{dI} \quad\quad\quad\quad\quad (2)$$

The $dN/dI$ is essential to evaluate the $n_2$ of 2D BiTe. The third-order nonlinear susceptibility $\chi^{(3)}_{total}$ is utilized to delineate the nonlinear optical properties of materials.[12, 36-37] It can be written as

$$\chi^{(3)}_{total} = \frac{cn_0^2}{12\pi^2} 10^{-7} n_2 \; (e.s.u.) \quad\quad\quad\quad\quad (3)$$

In this context, c denotes the speed of light in a vacuum, $n_0$ signifies the linear refractive index of the IPA solvent. The effective quantity of 2D materials present in the cuvette directly influences the value of $\chi^{(3)}_{total}$. Therefore, it is essential to determine the value of the third-order nonlinear susceptibility induced by a single layer of two-dimensional sheets, $\chi^{(3)}_{monolayer}$. $N_{eff}$ denotes the quantity of 2D BiTe aligned along the beam that traverses through the solution. The correlation between $\chi^{(3)}_{total}$ and $\chi^{(3)}_{monolayer}$ can be articulated as, [37-39]

$$\chi^{(3)}_{monolayer} = \frac{\chi^{(3)}_{total}}{N_{eff}^2} \quad\quad\quad\quad\quad (4)$$

A detailed discussion of the formation of the SSPM process is documented in Section 1 of the Supporting Information. Wu et al.[40] characterized the SSPM phenomena as arising from nonlocal and intraband ac electron coherence. Nearly the entire portion of the laser beam was found to diffract along the direction of propagation, with Rayleigh scattering being negligible. This observation confirms that the spatial self-phase modulation (SSPM) effect arises from a third-order nonlinear optical process. To probe the light-matter interaction, the concentration of 2D BiTe dispersions was systematically varied to determine the optimal concentration; 0.25 mg.mL$^{-1}$ yielded anticipated and consistent results. The calculation of the effective nonlinear refractive index ($N_{eff}$) is detailed in Supporting Information, Section 2.

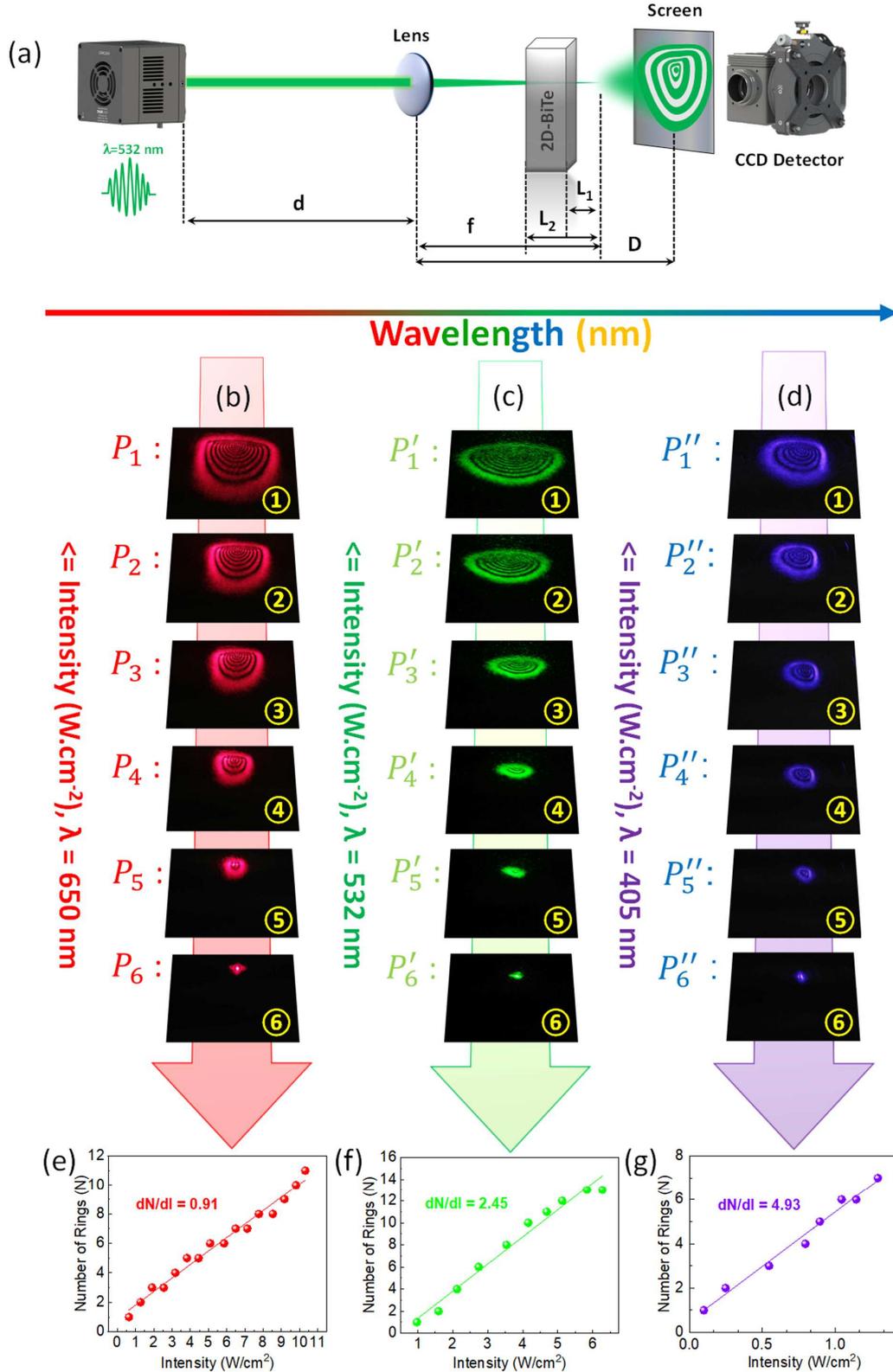

**Figure 2**. (a) SSPM Spectroscopy setup. (b-c-d) Far-field diffraction patterns obtained at various laser strengths for excitation wavelengths of 650, 532, and 405 nm. ($P_1 = 10.31\ W cm^{-2}$), ($P_2 = 9.17\ W cm^{-2}$), ($P_3 = 7.13\ W cm^{-2}$), ($P_4 = 4.45\ W cm^{-2}$), ($P_5 = 2.54\ W cm^{-2}$), ($P_6 = 0.63\ W cm^{-2}$), ($P'_1 = 6.28\ W cm^{-2}$), ($P'_2 = 4.68\ W cm^{-2}$), ($P'_3 = 3.53\ W cm^{-2}$), ($P'_4 = 2.12\ W cm^{-2}$), ($P'_5 = 1.15\ W cm^{-2}$), ($P'_6 = 0.53\ W cm^{-2}$), ($P''_1 =$

$1.29\ W cm^{-2}$), ($P_2'' = 1.04\ W cm^{-2}$), ($P_3'' = 0.89\ W cm^{-2}$), ($P_4'' = 0.79\ W cm^{-2}$), ($P_5'' = 0.54\ W cm^{-2}$), ($P_6'' = 0.14\ W cm^{-2}$). (e-f-g) Determined $dN/dI$ parameters at specified wavelengths for a fixed sample concentration of 0.25 mg·mL⁻¹ and a cuvette length of 10 mm.

**Table 1.** The measured values of the $n_2$ (nonlinear refractive index), $\chi^{(3)}_{total}$ (third-order nonlinear susceptibility), and $\chi^{(3)}_{monolayer}$ (third-order nonlinear susceptibility for one layer).

| Wavelength (nm) | Concentration (mg mL⁻¹) | L (mm) | Solvent | dN/dI (cm² W⁻¹) | $N_{eff}$ | $n_2$ (cm² W⁻¹) | $\chi^{(3)}_{total}$ (e.s.u.) | $\chi^{(3)}_{monolayer}$ (e.s.u.) |
|---|---|---|---|---|---|---|---|---|
| 650 | 0.25 | 10 | IPA | 0.91 | 577 | $2.18 \times 10^{-5}$ | $1.05 \times 10^{-3}$ | $3.16 \times 10^{-9}$ |
| 532 | 0.25 | 10 | IPA | 2.45 | 577 | $4.73 \times 10^{-5}$ | $2.3 \times 10^{-3}$ | $6.89 \times 10^{-9}$ |
| 405 | 0.25 | 10 | IPA | 4.93 | 577 | $7.18 \times 10^{-5}$ | $3.52 \times 10^{-3}$ | $10.56 \times 10^{-9}$ |

Figure 2b(①-⑥), 2c(①-⑥), and 2d(①-⑥) depict the SSPM far-field pattern for the excitation wavelengths of 650, 532, and 405 nm with varying intensity. The generation of the ring number in the diffraction pattern is directly correlated with the intensity. Raising the laser beam intensity will correspondingly enlarge the horizontal and vertical dimensions of the far-field pattern, as the modulation strength increases proportionally to the beam intensity. Figure 2e-f-g shows the relationship between intensity and ring number for 650, 532, and 405 nm laser beams. The derived values of $dN/dI$ via curve fitting are 0.91, 2.45, and 4.93 cm²W⁻¹ for excitation of λ= 650, 532, and 405 nm, respectively. Table 1 represents computed estimates of $n_2$ and $\chi^{(3)}_{total}$ derived from the aforementioned experiment using Equations 2 and 3. The experimentally determined magnitudes of $n_2$ and $\chi^{(3)}_{total}$ for 2D BiTe, evaluated using the SSPM approach, are considerably greater than those of other TMDC materials, as shown in Table 1 and in the Supporting Information (Table S1). A low-intensity laser beam is maintained to prevent thermal degradation of the 2D materials suspended in the solution. The significant size of the nonlinear optical coefficient is attributable to a specific factor. P-type BiTe has a lower Fermi level than N-type BiTe, hence facilitating a greater potential for the generation of excited carriers. This sort of hole carrier is implicated in the laser-induced hole coherence that results in elevated values of $n_2$, $\chi^{(3)}_{total}$, and $\chi^{(3)}_{monolayer}$. Section 8 illustrates the application of Miller's semi-empirical rule, where basic nonlinear optical scaling relations are employed to estimate $\chi^{(3)}_{monolayer}$ based on the corresponding linear susceptibility behavior. The electronic origin of the SSPM in the 2D BiTe nanostructure is explained in Section 4 of the Supporting Information. The gradual evolution of the diffraction pattern with time is termed "Time Evolution" and is described using the "Wind Chime Model".[40] The temporal evolution of the diffraction Pattern under different Intensities and wavelengths is discussed in Section 5 of the Supporting Information. Supporting Information Section 6 illustrates the dynamic contraction of diffraction patterns at various intensities and the corresponding variation of the nonlinear refractive index with wavelength and intensity.

## 3.2 SSPM-based Nonlinear All-Photonic Isolator Utilizing 2D BiTe - 2D hBN Heterostructure

A new all-photonic nonlinear isolator was developed with a heterostructure of 2D BiTe and 2D hBN to facilitate a more thorough investigation of photonic isolators through nonlinear optical (NLO) phenomena.[16, 41] The 2D-hBN material is synthesized via liquid-phase exfoliation and exhibits a significantly wider optical bandgap of 5.38 eV compared to 2D-BiTe. The computed bandgap of 2D-hBN from UV-Vis Spectroscopy and the Tauc method is elaborated in the Supporting Information Section 3. The all-photonic isolator has been implemented effectively utilizing continuous-wave sources (650, 532, and 405 nm) through the SSPM method. Upon the formation of the forward-biased configuration of 2D BiTe-2D hBN, as illustrated in Figure 3a, a diffraction pattern is generated. This forward-biased condition is evident in Figure 3c (①-⑦) for 650 nm, as shown by the number of rings in the generated diffraction pattern. The reverse sequence, as shown in Figure 3b, in which the 2D hBN solution precedes the 2D BiTe solution, suppresses the Gaussian beam profile ($\lambda$ = 650 nm). The observed effect can be attributed to the reverse-saturation characteristic of 2D hBN, leading to a decrease in the laser beam intensity as it passes through the cuvette containing the solution. Since the laser beam intensity is below the threshold, no diffraction pattern is observed. Owing to its reverse saturable absorption, 2D hBN can be employed to realize a photonic isolator.[42] The Gaussian beam profile is illustrated in Figure 3d (①-⑦) for the continuous wave source of 650 nm. The Supporting Information Figure S23a and S23b depict the forward-bias and reverse-bias conditions for the laser wave-source of 532 nm. The Supporting Information, Figures S23c and S23d, depict the forward-bias and reverse-bias conditions, respectively, for the 405 nm laser wave source. The determined values of $dN/dI$ for the wavelengths of 650, 532, and 405 nm are 1.10, 2.96, and 4.81 cm²W$^{-1}$, respectively, under forward bias conditions. The results observed show a substantial similarity to those derived from a single cuvette with a 2D BiTe-IPA solution. Figures 3e-f-g illustrate the linear fit graphs of $\frac{dN}{dI}$ corresponding to both forward-biased and reverse-biased states.

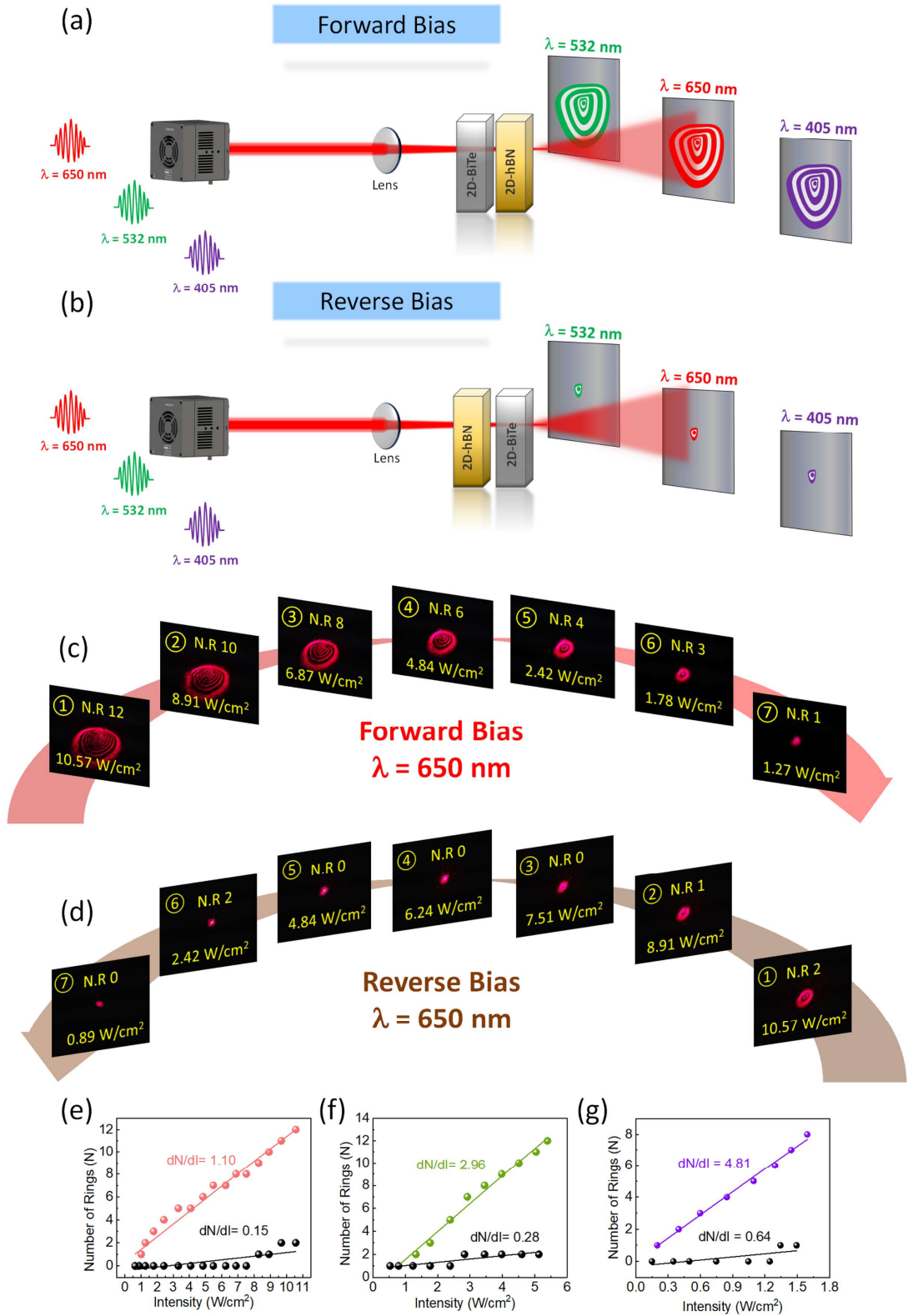

**Figure 3**. Figuration illustration of a 2D BiTe-based all-photonic isolator. Figures depicting (a) Forward-biased and (b) Reverse-biased configurations demonstrating the nonreciprocal light transmission mechanism. (c) Image illustrating diffraction pattern captured for the forward bias configuration at the far screen across multiple intensities for the wavelength λ= 650 nm. (d) Image showing the diffraction rings recorded at the far screen for varying intensity for reverse-biased arrangement for multiple intensities for the wavelength λ= 650 nm. (e-f-g) Calculated

values of $\frac{dN}{dI}$ obtained for both forward and reverse bias conditions for different excitation wavelengths (λ= 650, 532, and 405 nm).

Asymmetric light propagation is implemented using three unique laser types, each with a wavelength of λ = 650, 532, or 405 nm. The photon energy associated with each wavelength passing through the cuvette, determined from $E = \frac{hC}{\lambda}$, is greater than the 0.9 eV bandgap of 2D BiTe. Consequently, each photon of the laser beam can induce band-to-band transitions. As illustrated in Figure 4a, under forward bias, incident photons promote an interband transition. Upon relaxation to the ground state, these electrons emit photons with distinct phases, which then interfere with the incident beam, producing a diffraction pattern.[36, 39] Upon excitation to the conduction band, electrons move counter to the applied laser's electric field, giving rise to charges of opposite polarity in the suspended 2D nanostructure.[12] The incident laser beam induces alignment of the polarized 2D BiTe nanostructures along its electric-field axis, thereby minimizing interaction energy and enhancing the nonlinear optical response, thereby producing the optical Kerr effect.[39, 43] The interband transition in 2D hexagonal boron nitride (2D-hBN) is not excited by 650, 532, or 405 nm laser wavelengths, leading to energy dissipation via intraband transitions. Its reverse saturable absorption (RSA) reduces the incident beam intensity above a threshold, preventing the formation of a diffraction pattern in the 2D BiTe-IPA solution (Figure 4b). Using the Similar Comparison Method, continuous-wave source at λ = 650 nm was optimized for photonic isolator applications. The approach also indicates the potential use of wavelengths 671, 650, 532, 457, and 405 nm. This methodology can be extended to determine the nonlinear refractive index ($n_2$) of BiTe-based photonic isolator and related 2D materials under nonreciprocal light propagation, as follows:

$$n_2 = \frac{\lambda}{2n_0 L_{eff}} \cdot \frac{dN}{dI} \quad \text{..........(5)}$$

Here $\lambda/2n_0 L_{eff}$ is a constant, where λ is the excitation wavelength, $n_0$ the linear refractive index, $L_{eff}$ is the effective interaction length of the incident beam inside the cuvette. Establishing the nonlinear refractive index of a 2D BiTe-based photonic diode entails benchmarking against materials with well-characterized nonlinear refractive indices. Similar contrast ($S_C$) can be expressed as:

$$S_C = 1 - D = \frac{|n_{21} - n_{22}|}{n_{21}} \quad \text{....................(6)}$$

$$\text{Or,} \quad S_C = 1 - \frac{\left|\frac{\lambda}{2n_0 L_{eff}} \frac{N_1}{I_1} - \frac{\lambda}{2n_0 L_{eff}} \frac{N_2}{I_2}\right|}{\frac{\lambda}{2n_0 L_{eff}} \frac{N_1}{I_1}} = 1 - \frac{\left|\frac{N_1}{I_1} - \frac{N_2}{I_2}\right|}{\frac{N_1}{I_1}} \quad \text{..................(7)}$$

D represents the differential contrast, and $n_{21}$ and $n_{22}$ represent the nonlinear refractive indices of the heterostructure under forward and reverse bias conditions, respectively. A similar contrast is assessed relative to several 2D materials, with their coefficients documented in Table 3.

Table 3. Depicting Material and its corresponding $n_2$ and $S_C$.

| 2D Semiconductor materials | $n_2$ (Nonlinear Refractive Index) (cm²W⁻¹) | $S_C$ (Similar Contrast) (%) | Reference |
| --- | --- | --- | --- |
| MoS$_2$ | ≈10⁻⁷ | 74% | [43] |
| Sb | ≈10⁻⁶ | 85% | [13] |
| Bi$_2$Se$_3$ | ≈10⁻⁹ | 58% | [44] |
| SnS | ≈10⁻⁵ | 90% | [45] |
| SnS$_2$ | ≈10⁻⁹ | 54% | [36] |
| Graphene | ≈10⁻⁵ | 95% | [39] |
| Graphdiyne | ≈10⁻⁵ | 92% | [36] |
| CuPc | ≈10⁻⁶ | 75% | [39] |
| SnS$_2$ | ≈10⁻⁹ | 54% | [36] |
| NiTe$_2$ | ≈10⁻⁵ | 91.8% | [46] |
| Bi$_2$Te$_3$ | ≈10⁻⁴ | 98.3% | [47] |
| BiTe | ≈10⁻⁵ | 89% | |

Similar contrast was calculated with respect to 2D NiTe$_2$ as seen in Figure 4c. Figure 4d illustrates that the 2D BiTe exhibits a $n_2$ comparable to that of NiTe$_2$, Sb, and SnS, with a $n_2$ range of 10⁻⁵ cm² W⁻¹. This is confirmed by testing with the SSPM method. This demonstrates the reliability and reproducibility of the SSPM experiment. The 2D-BiTe exhibits a notable contrast of roughly 89%. Consequently, the heterostructure of 2D BiTe and 2D hBN may be employed in all-photonic isolator applications.[36, 38]

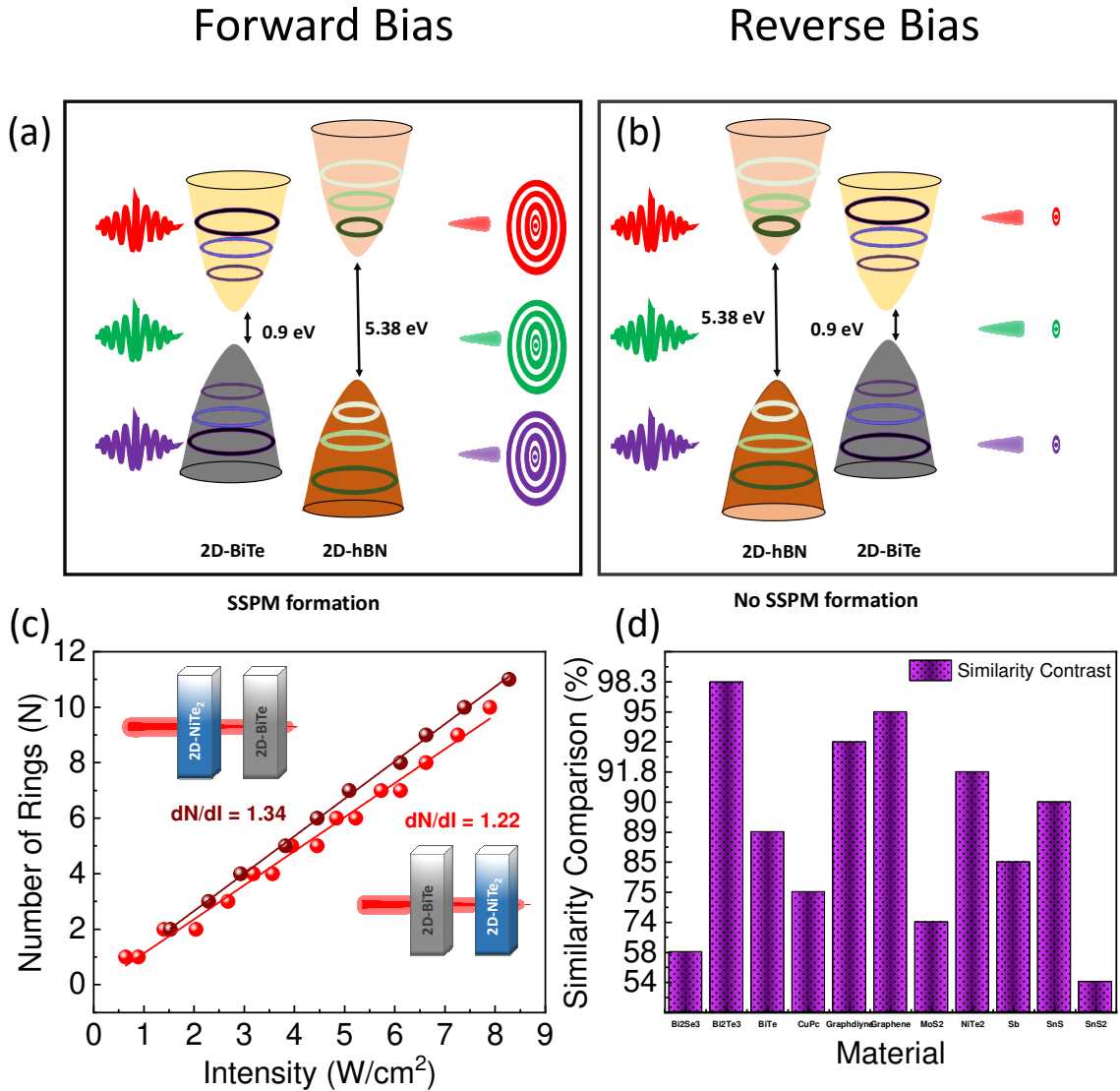

**Figure 4**. (a-b) Schematic representation of the all-photonic isolator mechanism, highlighting the distinct band structure configurations for forward-biased and reverse-biased propagation. (c) Changes in the diffraction ring numbers in both the forward and reverse directions with incident laser intensity for the 2D BiTe-2D NiTe$_2$ heterostructure. (d) The Similar contrast of 2D BiTe in relation to other semiconducting materials.

### 3.3 2D BiTe-Based All-Photonic Information Converter

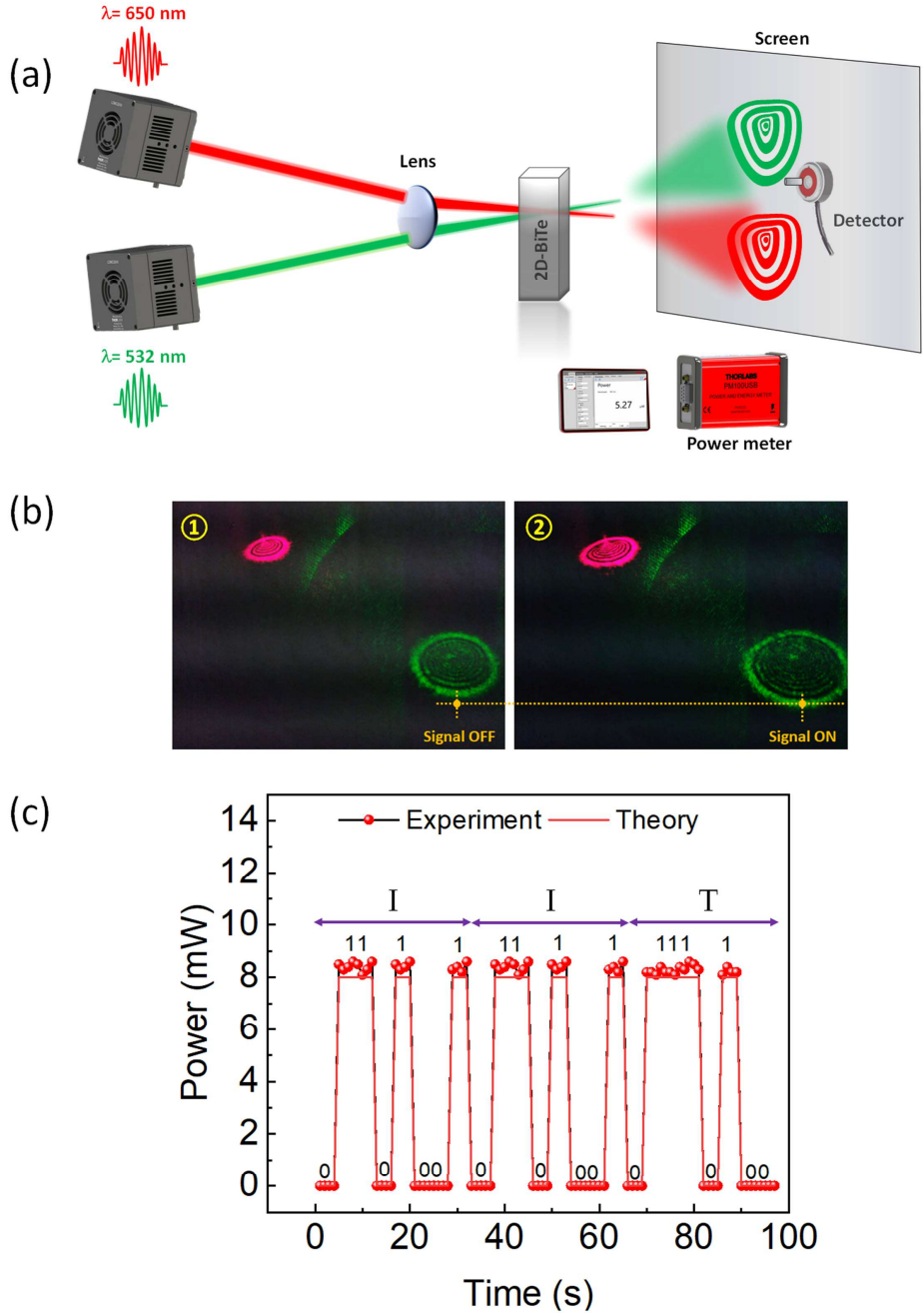

**Figure 5**. (a) The experimental setup for all-photonic information conversion. (b) The probe beam's diffraction ring patterns, which correspond to pump beam intensities between $50\ and\ 80\ mW$. (c) The power meter recordings made after the input optical signal was converted to the ASCII code for "IIT" are shown.

The design in Figure 5a has validated the viability of this laser-laser modulation system employing the Kerr effect in 2D BiTe dispersions through XPM.[48-49] As a light-modulates-light system, the laser-laser modulation configuration was created. The modified setup, shown in Figure 5a, allows the probe laser to produce diffraction rings via direct interaction between the pump laser and 2D BiTe nanosheet dispersions. To ensure the best possible overlap of the two beams within the dispersion, a 10 mm-thick quartz cuvette with a crossing angle of 15° was used. The overlap of the pump and probe beams within the BiTe dispersion generates an intensity-dependent refractive index variation via the Kerr nonlinear effect, resulting in diffraction rings (Figure 5b). The probe beam produces a strip interference signal by modulating the pump beam. Under various experimental conditions, the probe beam at 532 nm generated diffraction patterns with $N = 16$ and $N = 20$ rings at a pump power of $50\ mW$ and $80\ mW$ from the 650 nm laser.

A detector was strategically placed on the outermost ring (bright stripe) of the diffraction rings. ($\lambda = 532\ nm$), which were stimulated by probing light at $N = 16$. At the same time, we recorded the power at the detector, which was found to be $P_S = 8\ mW$. When the pump laser's intensity increases to $50\ mW$, the detector's position will correspond with the black stripe, leading to a minimum power output of $0\ mW$, as illustrated in Figure 5b①. The incident power of the pump laser steadily increases to $80\ mW$, leading to the detector aligning with the bright stripe to attain the maximum power of $P_S = 8\ mW$, as shown in Figure 5b②.

The detector's probe light intensity exhibits periodic fluctuations with increases and decreases in pump light power. This phenomenon demonstrates the promising application potential of photonic switchers/modulators by controlling the pump laser power to modulate the probe laser's phase. Upon reaching the peak value of the probe power intensity, the switcher will activate; conversely, when the power detected is at its minimum, the switcher will deactivate. By manipulating the Pump laser, we may introduce the input signal "010100101101001011101100," which corresponds to the ASCII code for "IIT" pertaining to the Probe light. Figure 5c displays the ASCII code for "IIT" as recorded by the power meter. This laser-laser modulation system work presents a concept for an all-photonic communication application with a laser-laser photonic modulation system. Simultaneously, this 2D BiTe nanostructure-based all-photonic information converter can be used for encryption-based data transmission applications.

**3.4 Cross-Phase Modulation (XPM): 2D BiTe -Based All-Photonic Switching and OR Gate Application**

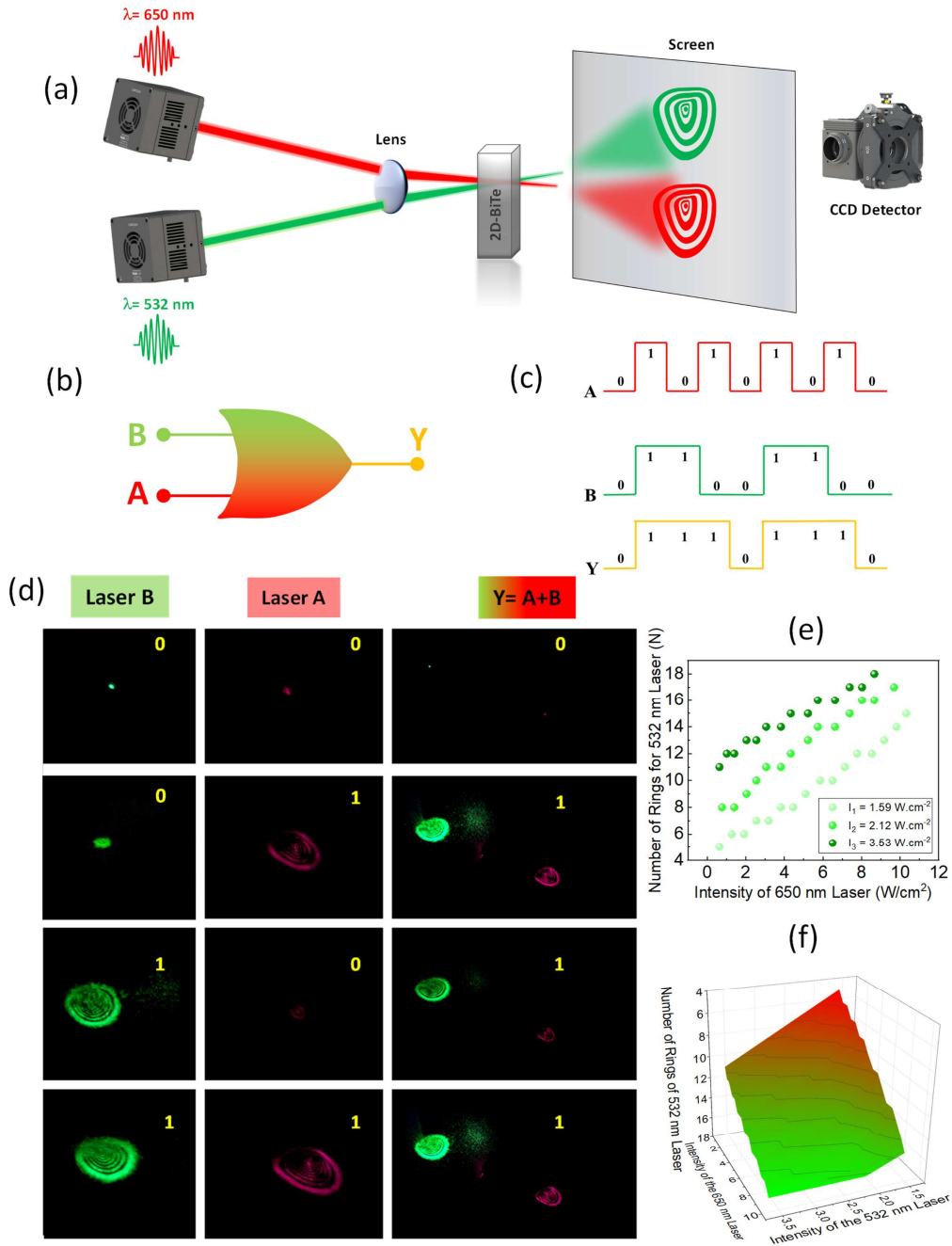

**Figure 6**. Cross-phase modulation (XPM): (a) Schematic illustration of XPM in 2D BiTe for photonic logic gate applications. (b) Symbolic and circuit representation of the logical OR gate. (c) Output waveform of the OR gate for inputs A and B, producing output Y. (d) Experimental demonstration of photonic OR gate operation using a two-color XPM scheme with 2D BiTe, tested for the laser wavelength pairs λ = 532 & 650 nm. (e) Dependence of probe laser diffraction ring count on variation in pump laser intensity. (f) Three-dimensional representation of two-color XPM for the corresponding wavelength pairs λ = 532 & 650 nm.

A purely photonic switching mechanism has been realized with a two-dimensional BiTe nanostructure, where spatial-self phase modulation (SSPM) enables the implementation of a logic function equivalent to an OR gate, as shown in Figure 6.[19, 21, 50-52] The experimental OR logic gate schematic based on the cross-phase modulation (XPM) technique is shown in Figure 6a. In this configuration, the input signals "A" and "B" for the 2D BiTe nanostructure logic device are provided by two distinct excitation wave sources (650 & 532 nm). The gate's logical outputs match the outputs caused by the XPM. The OR gate's switching circuit and symbolic representation are depicted in Figure 6b. The output "Y" likewise registers a high ("1") when either input "A" or "B" is high. Only when both inputs are low ("0") does a low output ("0") occur. The corresponding waveforms for "A," "B," and "Y" that show how an OR gate operates are shown in Figure 6c. The number of diffraction rings created by the XPM effect in this system indicates the logical states. For laser-laser modulation, the implementation uses a dual-laser setup. If the incident or probe laser intensity is below the threshold value required for nonlinear interaction with the material, the diffraction rings vanish, leaving a simple Gaussian beam profile on the screen. Diffraction rings appear for both beams when a Gaussian probe laser is cross-modulated with a high-intensity pump laser. Two different excitation wavelengths ($\lambda$ = 532 and 650 nm) are used as inputs, "A" and "B," respectively, in this arrangement. When the two lasers are cross-coupled, a combined output image known as "Y" appears on the screen. As shown in Figure 6d for the laser wavelength pairs 650 nm & 532 nm, respectively, this output can be used to realize a photonic OR logic gate. For probe beam intensities of 1.59, 2.12, and 3.53 W.cm$^{-2}$, Figure 6e illustrates the change in the number of diffraction rings of the probe beam ($\lambda$ = 532 nm) in relation to the intensity of the pump beam ($\lambda$ = 650 nm). Figure 6f depicts the change in the ring count of the diffraction profile of the probe beam ($\lambda$ = 405 nm) in relation to the intensity of the pump beam ($\lambda$ = 532 nm). As seen in Supporting Information Figure S24a and S24b, similar tests were conducted for the other two wavelength pairs, $\lambda$ = 532 and 405 nm, and $\lambda$ = 650 and 405 nm. Figures S24c and S24d show the same for the 405 nm probe and 532 nm pump beams, and 405 nm probe and 650 nm pump beams. The findings verify that the probe beam's intensity directly affects the number of diffraction profiles it produces. When the pump beam is turned on, the modulation of diffraction rings occurs according to the superposition principle. The number of diffraction rings for the probe beam clearly shows a linear relationship with the intensity of the pump beam. Three-dimensional surface maps summarizing the entire dataset are shown in Figure 6f for the wavelength pair $\lambda$ = 532 and 650 nm. The investigated maps, Figures S24e and S24f, show the correlation between the intensities of the pump and probe beams and the number of diffraction rings in the probe-beam pattern for the wavelength combinations 532 & 405 nm, and 650 & 405 nm, respectively.

## 3.5 Electronic Relationship Between $\chi^{(3)}_{monolayer}$, ($\mu$), and ($m^*$) and Band Structure Calculation

Hu et al. proposed a method combining carrier mobility, effective mass, and electronic coherence effects.[53] To uncover the mechanism behind SSPM pattern formation, an experiment was performed correlating the $\chi^{(3)}_{monolayer}$ with mobility and $\chi^{(3)}_{monolayer}$ with effective mass. The ability of charge carriers to move under an external electric field, defined as carrier mobility, is intrinsically linked to their effective mass and the band dispersion along the ab-plane of the crystal lattice. Reduced electron scattering within the 2D lattice facilitates stronger nonlinear polarization, resulting in a higher value of $\chi^{(3)}_{monolayer}$. Therefore, the optical coefficient $\chi^{(3)}_{monolayer}$ can be directly associated with the electronic parameters of the material, namely the effective mass (m*) and carrier mobility (μ). In this work, we analyze the relationship between the literature-reported mobility and effective mass values for 2D BiTe and the experimentally determined values. $\chi^{(3)}_{monolayer}$ obtained through SSPM measurements. Additionally, the value of the μ and m* is derived from pertinent literature, as are the values of the other included materials. Table S3 contains the Supporting Information. Figures 7a and 7b reveal that the reported, $\chi^{(3)}_{monolayer}$ value ($10^{-9}$ e.s.u.) exhibits a stronger correlation with electronic activity compared to graphene. These correlations are represented by the plots of $\chi^{(3)}_{monolayer}$ vs μ and $\chi^{(3)}_{monolayer}$ vs m*.

$$\chi^{(3)} = 8.00/\sqrt{m^*} \quad \ldots\ldots\ldots\ldots\ldots\ldots\ldots\ldots\ldots\ldots (8)$$

and, $$\chi^{(3)} = 0.146 \times \sqrt{\mu} \quad \ldots\ldots\ldots\ldots\ldots\ldots\ldots\ldots\ldots (9)$$

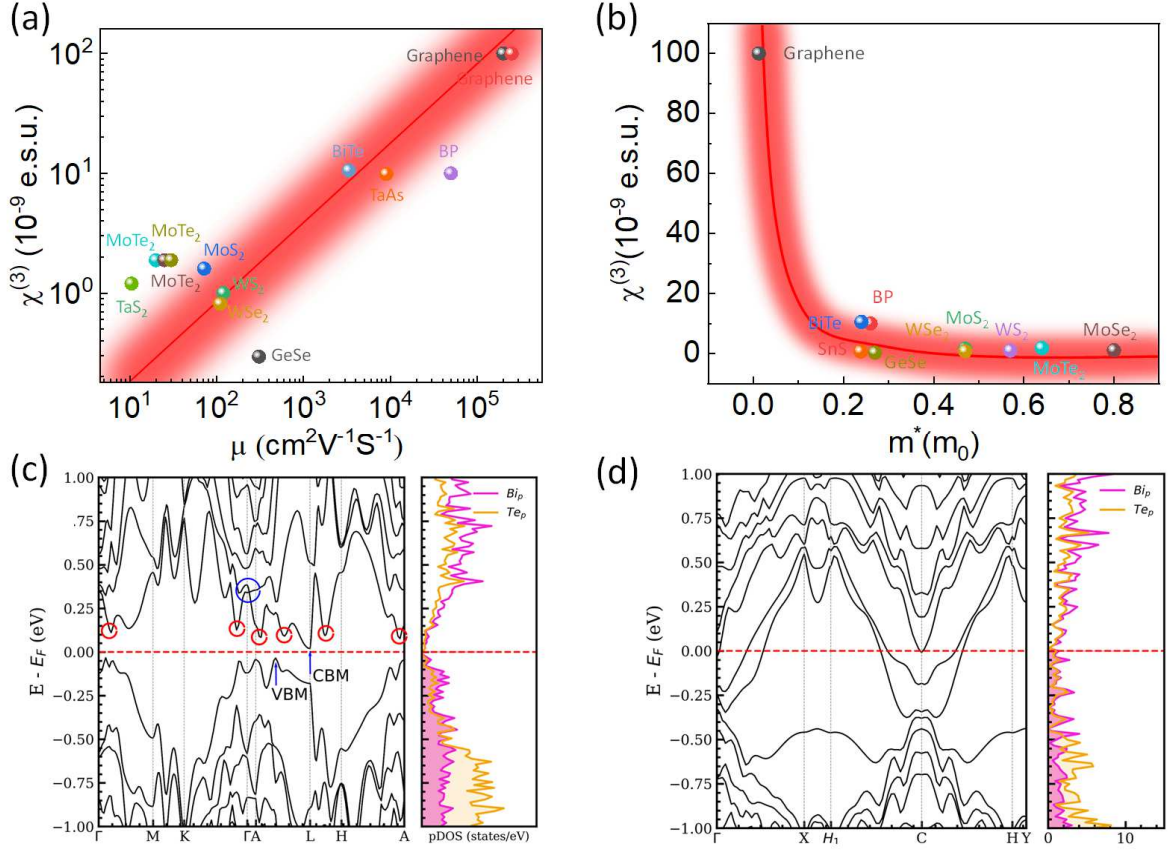

**Figure 7**. Constructive study between optical and electronic coefficients. (a) $\chi^{(3)}(10^{-9}\ e.s.u)$ versus $\mu$. (b) $\chi^{(3)}(10^{-9}\ e.s.u)$ versus effective mass ($m^*$) of several 2D materials. Electronic band structure and projected Density of States (pDOS) of (c) bulk BiTe and (d) 2D slabs, respectively. The conduction band valleys are highlighted in red, while the Dirac-like cones are highlighted in blue.

To gain insight into the electronic behavior of Tsumoite (BiTe), Density Functional Theory (DFT) calculations were carried out for both its bulk and two-dimensional (2D) phases. The optimized lattice parameters for bulk BiTe are $a = b = 4.406$ Å and $c = 23.992$ Å. These parameters are in good agreement with our experimental measurements and are consistent with previous theoretical calculations using the same parameters described in the computational details of the references. [54-55] For the 2D structures, the slabs were constructed by considering the (104) direction based on the XRD data.

BiTe is a semiconductor with a narrow indirect bandgap of 0.06 eV. The Valence band maxima (VBM) is located between points A and L on the first Brillouin zone, while the Conduction band minima (CBM) are located at the L point, as highlighted in Figure 7c. Fermi level ($E_F$) is dominated by Bi p states, as confirmed by the projected Density of States (pDOS), while Te 4p states are more pronounced at -0.5 eV below the Fermi level. Figure 7d shows that the electronic band structure of the 2D slab displays metallic character, with conduction band valleys located at the Γ and C points, and the lowest valley reaching the Fermi energy ($E_F$). The pDOS shows significant hybridization between Bi and Te states and reveals the inversion on the population of these states: deep valence states are dominated by Te, whereas conduction states located high above the $E_F$ are attributed to Bi p states. BiTe exhibits several conduction-band valleys (highlighted in red in Figure 7c). These valleys are associated with carrier pockets and strongly influence carrier mobility due to their proximity to the CBM (less than 25 meV), allowing these states to be populated by a temperature gradient [55]. Furthermore, a tilted Dirac like cone is observed at Γ → A at 30 meV above $E_F$. This exotic electronic feature has been related to enhancement in non-linear optical properties, since the linear dispersion provides resonant interband transitions [16, 56].

As seen in Figure 7a, the optoelectronic property of the 2D BiTe can be compared to that of TaAs and black phosphorus. The electron/hole pockets in the conduction band, as seen in Figure 7c, can contribute to the enhanced hole-induced mobility. From Figure 2e-f-g it was concluded that as the wavelength decreases the value of $\frac{dN}{dI}$ increases as the per photon energy increases. From Equations 2 and 3 it can be concluded that both $n_2$ and $\chi^{(3)}_{total}$ increase. The value of the $\chi^{(3)}_{monolayer}$ is calculated as seen in Equation 4, the value was found to increase with decreasing wavelength. The value of the $\chi^{(3)}_{monolayer}$ is calculated as seen in Equation 4, the value was found to increase with decreasing wavelength. The value of $\chi^{(3)}_{monolayer}$ is well comparable with the absorbance value calculated from the UV-Vis Spectroscopy, as seen in the Supporting Information Section 8. This phenomenon can be caused by the laser-induced hole coherence, as discussed by recent literature.[12, 57] The suspended 2D BiTe, having a mean atomic thickness of 12 nm, is considered bulk material. Laser irradiation induces interband transitions by promoting the majority charge carriers from the valence band to the conduction band. These photoexcited carriers traverse through the momentum space and fall into these carrier pockets, and decay through radiative recombination.[16, 57] Positioning of these 8 carrier pockets causes faster recombination, causing high mobility.[58] Hence a large amount of photoexcited carriers pass through the pockets and are involve in the nonradiative recombination process causing a larger value of $\chi^{(3)}_{monolayer}$. The interaction duration of hole-phonon scattering is considerably longer than that of hole-hole scattering, resulting in a more pronounced impact on the value of $\chi^{(3)}_{monolayer}$.[59] All free carriers, including those from the carrier pockets, contribute to the SSPM. Therefore, the "energy-matched" carriers are not the only contributors to the SSPM. All free carriers, including those near carrier pockets, can move inside the light field, therefore contributing to the SSPM. The valence band and conduction band structure is found to be strongly dispersive; hence, the presence of light electron-hole band is predicted. This type of band structure promotes fast acceleration under an electric field. This high mobility of the majority carriers contributes more to fast transport and strong optical response in the material. Thus, the carrier pockets containing lower effective mass carriers contribute to the observed SSPM patterns. The significant value of $\chi^{(3)}_{monolayer}$ translates to large optical nonlinearity we detect is associated with electronic phenomenon.

## 4. Conclusion

This study reports the synthesis of 2D BiTe via liquid-phase exfoliation. The SSPM phenomenon, including the pattern formation and its temporal evolution is analyzed and described, using the "Wind Chime model". The values of the nonlinear optical coefficients, including $n_2$ and $\chi^{(3)}_{monolayer}$, were calculated. The value of $n_2$ was determined to be $2.18 \times 10^{-5}$, $4.73 \times 10^{-5}$, and $7.18 \times 10^{-5}$ cm$^2$W$^{-1}$ at the excitation of 650, 532, and 405 nm, wavelengths respectively. The temporal evolution of the diffraction pattern is analyzed for various wavelengths and intensities, while keeping the path length 10 mm and 2D BiTe concentration constant. This work estimates the values of third-order nonlinear susceptibility for monolayer $\chi^{(3)}_{monolayer}$ as $3.16 \times 10^{-9}$ e.s.u., $6.89 \times 10^{-9}$ e.s.u., and $10.56 \times 10^{-9}$ e.s.u. at the wavelengths of 650, 532, and 405 nm, respectively. The deduced value of $\chi^{(3)}_{total}$ demonstrates a rather large magnitude in equivalence to 2D NiTe$_2$, Sb, and SnS. The temporal progression of the diffraction patterns at various wavelengths (650, 532, and 405 nm) was also computed. Vertical distortions of the diffraction patterns were measured, and the contraction period for excitation wavelengths (λ = 65, 532, and 405 nm) were measured. The 2D BiTe-2D hBN based heterostructure based photonic isolator presented in the literature exhibits $\frac{dN}{dI}$ values of 1.1, 2.96, and 4.81 Wcm$^{-2}$ for the wavelength of 650, 532, and 405 nm, respectively, under forward bias conditions. This study introduces an innovative design of a photonic isolator intended for directional isolation of propagating laser beam. The exhibited all-photonic switch utilizes high-power pump light to modulate lower-power probe light (λ = 532 & 650 nm), enabling the implementation of "ON" and "OFF" states via XPM, applicable to any photonic information conversion. Furthermore, photonic information processing has been investigated through the execution of the "OR" Gate operation utilizing cross-phase modulation (XPM) of two laser. We have established a foundation for 2D BiTe-based all-photonic switching, functioning as photonic diodes and logic devices via NLO behavior. To further differentiate this electronic coherence effect from the thermal lens effect, the dependence of $\chi^{(3)}_{monolayer}$ on m$^*$ and $\chi^{(3)}_{monolayer}$ on µ is derived, mirroring the relevant curves reported by other researchers. The dependence of the $\chi^{(3)}_{monolayer}$ on $m^*$ and µ was derived to further distinguish the electronic coherence effect from the thermal lens effect, in agreements with previously reported trends. To understand the origin of the large value of the $\chi^{(3)}_{monolayer}$, ab-initio methods were adapted to investigate the electronic band structure of the BiTe and its 2D counterpart. The elevated values of $n_2$ and $\chi^{(3)}_{total}$ are anticipated to arise from laser-induced hole coherence. The all-photonic switching characteristics have been investigated using two different lasers of different wavelength (λ = 532 & 650 nm, 532 & 405 nm, and 650 & 405 nm), causing interband coherence.

**References**

[1]     Z. Chai, X. Hu, F. Wang, et al.," Ultrafast All-Optical Switching," *Advanced Optical Materials* (2017), 5, 1600665.https://doi.org/10.1002/adom.201600665

[2]     M. Ono, M. Hata, M. Tsunekawa, et al.," Ultrafast and energy-efficient all-optical switching with graphene-loaded deep-subwavelength plasmonic waveguides," *Nature Photonics* (2020), 14, 37.10.1038/s41566-019-0547-7

[3]     Q. Wang, X. Wu, L. Wu, Y. Xiang," Broadband nonlinear optical response in $Bi_2Se_3$-$Bi_2Te_3$ heterostructure and its application in all-optical switching," *AIP Advances* (2019), 9.10.1063/1.5082725

[4]     X. Li, R. Liu, H. Xie, et al.," Tri-phase all-optical switching and broadband nonlinear optical response in Bi2Se3 nanosheets," *Optics Express* (2017), 25, 18346.10.1364/OE.25.018346

[5]     J. Cano, B. Bradlyn," Band representations and topological quantum chemistry," *Annual Review of Condensed Matter Physics* (2021), 12, 225

[6]     F. Schindler, Z. Wang, M. G. Vergniory, et al.," Higher-order topology in bismuth," *Nature physics* (2018), 14, 918

[7]     L. Wu, X. Yuan, D. Ma, et al.," Recent advances of spatial self-phase modulation in 2D materials and passive photonic device applications," *Small* (2020), 16, 2002252

[8]     Z. Zhu, Z. Zheng, F. Zhang, et al.," Causal associations between risk factors and common diseases inferred from GWAS summary data," *Nature communications* (2018), 9, 224

[9]     B. Bradlyn, L. Elcoro, J. Cano, et al.," Topological quantum chemistry," *Nature* (2017), 547, 298

[10]    M. G. Vergniory, B. J. Wieder, L. Elcoro, et al.," All topological bands of all nonmagnetic stoichiometric materials," *Science* (2022), 376, eabg9094

[11]    R. Wu, Y. Zhang, S. Yan, et al.," Purely coherent nonlinear optical response in solution dispersions of graphene sheets," *Nano letters* (2011), 11, 5159

[12]    Y. Wu, Q. Wu, F. Sun, et al.," Emergence of electron coherence and two-color all-optical switching in $MoS_2$ based on spatial self-phase modulation," *Proceedings of the National Academy of Sciences* (2015), 112, 11800

[13]    L. Lu, X. Tang, R. Cao, et al.," Broadband nonlinear optical response in few-layer antimonene and antimonene quantum dots: a promising optical kerr media with enhanced stability," *Adv. Opt. Mater.* (2017), 5, 1700301

[14]    L. Wu, W. Huang, Y. Wang, et al.," 2D tellurium based high-performance all-optical nonlinear photonic devices," *Advanced Functional Materials* (2019), 29, 1806346

[15]    J. Zhang, X. Yu, W. Han, et al.," Broadband spatial self-phase modulation of black phosphorous," *Opt. Lett.* (2016), 41, 1704

[16]    S. Goswami, C. C. de Oliveira, B. Ipaves, et al.," Exceptionally High Nonlinear Optical Response in Two-dimensional Type II Dirac Semimetal Nickel Di-Telluride ($NiTe_2$)," *Laser & Photonics Reviews* (2025), 19, 2400999. https://doi.org/10.1002/lpor.202400999

[17]     L. Wu, Z. Xie, L. Lu, et al.," Few-Layer Tin Sulfide: A Promising Black-Phosphorus-Analogue 2D Material with Exceptionally Large Nonlinear Optical Response, High Stability, and Applications in All-Optical Switching and Wavelength Conversion," *Advanced Optical Materials* (2018), 6, 1700985.https://doi.org/10.1002/adom.201700985

[18]     D. Weng, C. Ling, Y. Gao, et al.," Spatially Asymmetric Optical Propagation and All-Optical Switching Based on Spatial Self-Phase Modulation of Semimetal MoP Microparticles," *Laser Photonics Rev.*, n/a, 2401587.https://doi.org/10.1002/lpor.202401587

[19]     S. Bera, S. Kalimuddin, A. Bera, et al.," Nonlinear Optical Properties of 2D vdW Ferromagnetic Nanoflakes for Magneto-Optical Logic Applications," *Advanced Optical Materials* (2025), 13, 2402318.https://doi.org/10.1002/adom.202402318

[20]     Y. Huang, H. Zhao, Z. Li, et al.," Laser-Induced Hole Coherence and Spatial Self-Phase Modulation in the Anisotropic 3D Weyl Semimetal TaAs," *Advanced Materials* (2023), 35, 2208362

[21]     X. Xu, M. Wang, Y. Zhang, et al.," Broadband Spatial Self-Phase Modulation in Black and Violet Phosphorus and Near-Infrared All-Optical Switching," *Laser & Photonics Reviews* (2024), 18, 2300930.https://doi.org/10.1002/lpor.202300930

[22]     L. Wu, X. Jiang, J. Zhao, et al.," 2D MXene: MXene-Based Nonlinear Optical Information Converter for All-Optical Modulator and Switcher (Laser Photonics Rev. 12 (12)/2018)," *Laser & Photonics Reviews* (2018), 12, 1870055

[23]     L. Del Bino, J. M. Silver, M. T. Woodley, et al.," Microresonator isolators and circulators based on the intrinsic nonreciprocity of the Kerr effect," *Optica* (2018), 5, 279

[24]     A. Ferreira, C. Sobrinho, G. Guimarães, et al.," All-optical logic gates based on XPM effect under the PAM-ASK modulation in a symmetric dual NLDC," *Microsystem Technologies* (2019), 25, 447

[25]     J. Barthel," Dr. Probe: A software for high-resolution STEM image simulation," *Ultramicroscopy* (2018), 193, 1

[26]   K. Morgan, I. Zeimpekis, Z. Feng, D. Hewak," Enhancing thermoelectric properties of bismuth telluride and germanium telluride thin films for wearable energy harvesting," *Thin Solid Films* (2022), 741, 139015

[27]   A. Ghosh, S. Shukla, M. Monisha, et al.," Sulfur Copolymer: A New Cathode Structure for Room-Temperature Sodium–Sulfur Batteries," *ACS Energy Lett.* (2017), 2, 2478.10.1021/acsenergylett.7b00714

[28]   E. A. Hoffmann, T. Körtvélyesi, E. Wilusz, et al.," Relation between C1s XPS binding energy and calculated partial charge of carbon atoms in polymers," *J. Mol. Struct.:THEOCHEM* (2005), 725, 5

[29]   D. L. Rousseau, R. P. Bauman, S. Porto," Normal mode determination in crystals," *Journal of Raman Spectroscopy* (1981), 10, 253

[30]   O. Concepcion, M. Galván-Arellano, V. Torres-Costa, et al.," Controlling the epitaxial growth of Bi2Te3, BiTe, and Bi4Te3 pure phases by physical vapor transport," *Inorganic Chemistry* (2018), 57, 10090

[31]   V. Russo, A. Bailini, M. Zamboni, et al.," Raman spectroscopy of Bi-Te thin films," *Journal of Raman Spectroscopy: An International Journal for Original Work in all Aspects of Raman Spectroscopy, Including Higher Order Processes, and also Brillouin and Rayleigh Scattering* (2008), 39, 205

[32]   P. Kuznetsov, V. Yapaskurt, B. Shchamkhalova, et al.," Growth of $Bi_2Te_3$ films and other phases of Bi-Te system by MOVPE," *Journal of crystal growth* (2016), 455, 122

[33]   D. Mallick, S. Mandal, Y. Bitla, R. Ganesan, P. A. Kumar," Emergence of electron-phonon coupling in a dual topological insulator BiTe," *Materials Research Express* (2019), 6, 126321

[34]   O. Caha, A. Dubroka, J. Humlicek, et al.," Growth, structure, and electronic properties of epitaxial bismuth telluride topological insulator films on $BaF_2$ (111) substrates," *Crystal growth & design* (2013), 13, 3365

[35]     J. Yao, Z. Zheng, G. Yang," All-layered 2D optoelectronics: a high-performance UV–vis–NIR broadband SnSe photodetector with $Bi_2Te_3$ topological insulator electrodes," *Advanced Functional Materials* (2017), 27, 1701823

[36]     L. Wu, Y. Dong, J. Zhao, et al.," Kerr nonlinearity in 2D graphdiyne for passive photonic diodes," *Adv. Mater.* (2019), 31, 1807981

[37]     Y. Shan, L. Wu, Y. Liao, et al.," A promising nonlinear optical material and its applications for all-optical switching and information converters based on the spatial self-phase modulation (SSPM) effect of $TaSe_2$ nanosheets," *Journal of Materials Chemistry C* (2019), 7, 3811

[38]     Y. Liao, Y. Shan, L. Wu, Y. Xiang, X. Dai," Liquid-Exfoliated Few-Layer InSe Nanosheets for Broadband Nonlinear All-Optical Applications," *Advanced Optical Materials* (2020), 8, 1901862

[39]     K. Sk, B. Das, N. Chakraborty, et al.," Nonlinear Coherent Light–Matter Interaction in 2D $MoSe_2$ Nanoflakes for All-Optical Switching and Logic Applications," *Advanced Optical Materials* (2022), 10, 2200791

[40]     Y. Wu, Q. Wu, F. Sun, et al.," Emergence of electron coherence and two-color all-optical switching in $MoS_2$ based on spatial self-phase modulation," *Proceedings of the National Academy of Sciences* (2015), 112, 11800.doi:10.1073/pnas.1504920112

[41]     Y. Gao, C. Ling, D. Weng, et al.," Demonstration of Spatial Asymmetric Light Propagation Performance Using Violet Phosphorus Quantum Dots with Tunable Bandgap," *Laser & Photonics Reviews* (2024), 18, 2301062.https://doi.org/10.1002/lpor.202301062

[42]     P. Kumbhakar, A. K. Kole, C. S. Tiwary, et al.," Nonlinear optical properties and temperature-dependent uv–vis absorption and photoluminescence emission in 2d hexagonal boron nitride nanosheets," *Advanced Optical Materials* (2015), 3, 828

[43] G. Wang, S. Zhang, X. Zhang, et al.," Tunable nonlinear refractive index of two-dimensional $MoS_2$, $WS_2$, and $MoSe_2$ nanosheet dispersions," *Photonics Research* (2015), 3, A51

[44] X. Li, R. Liu, H. Xie, et al.," Tri-phase all-optical switching and broadband nonlinear optical response in $Bi_2Se_3$ nanosheets," *Opt. Express* (2017), 25, 18346

[45] L. Wu, Z. Xie, L. Lu, et al.," Few-layer tin sulfide: a promising black-phosphorus-analogue 2D material with exceptionally large nonlinear optical response, high stability, and applications in all-optical switching and wavelength conversion," *Advanced Optical Materials* (2018), 6, 1700985

[46] S. Goswami, C. C. de Oliveira, B. Ipaves, et al.," Exceptionally High Nonlinear Optical Response in Two-dimensional Type II Dirac Semimetal Nickel Di-Telluride ($NiTe_2$)," *Laser & Photonics Reviews* (2025), 2400999

[47] B. Shi, L. Miao, Q. Wang, et al.," Broadband ultrafast spatial self-phase modulation for topological insulator $Bi_2Te_3$ dispersions," *Applied Physics Letters* (2015), 107.10.1063/1.4932590

[48] Y. Liao, Y. Shan, L. Wu, Y. Xiang, X. Dai," Liquid-Exfoliated Few-Layer InSe Nanosheets for Broadband Nonlinear All-Optical Applications," *Advanced Optical Materials* (2020), 8, 1901862.https://doi.org/10.1002/adom.201901862

[49] L. Wu, W. Huang, Y. Wang, et al.," 2D Tellurium Based High-Performance All-Optical Nonlinear Photonic Devices," *Advanced Functional Materials* (2019), 29, 1806346.https://doi.org/10.1002/adfm.201806346

[50] K. Sk, B. Das, N. Chakraborty, et al.," Nonlinear Coherent Light–Matter Interaction in 2D $MoSe_2$ Nanoflakes for All-Optical Switching and Logic Applications," *Advanced Optical Materials* (2022), 10, 2200791.https://doi.org/10.1002/adom.202200791

[51] X. Xu, Z. Cui, Y. Yang, et al.," Large Optical Nonlinearity Enhancement and All-Optical Logic Gate Implementation in Silver-Modified Violet Phosphorus," *Laser & Photonics Reviews* (2025), 19, 2401521

[52] D. Weng, C. Ling, Y. Gao, et al.," Spatially Asymmetric Optical Propagation and All-Optical Switching Based on Spatial Self-Phase Modulation of Semimetal MoP Microparticles," *Laser & Photonics Reviews* (2025), 19, 2401587

[53] L. Hu, F. Sun, H. Zhao, J. Zhao," Nonlinear optical response spatial self-phase modulation in MoTe$_2$: correlations between χ (3) and mobility or effective mass," *Optics Letters* (2019), 44, 5214

[54] M. Samanta, K. Pal, U. V. Waghmare, K. Biswas," Intrinsically Low Thermal Conductivity and High Carrier Mobility in Dual Topological Quantum Material, n-Type BiTe," *Angewandte Chemie International Edition* (2020), 59, 4822.https://doi.org/10.1002/anie.202000343

[55] R. Pathak, A. Joseph, P. Dutta, et al.," Impact of Pressure on Metavalent Bonding in BiTe Influencing Electronic Topological Transitions," *Angewandte Chemie* (2025), 137, e202422652

[56] M. Baudisch, A. Marini, J. D. Cox, et al.," Ultrafast nonlinear optical response of Dirac fermions in graphene," *Nature Communications* (2018), 9, 1018

[57] Y. Huang, H. Zhao, Z. Li, et al.," Laser-Induced Hole Coherence and Spatial Self-Phase Modulation in the Anisotropic 3D Weyl Semimetal TaAs," *Advanced Materials* (2023), 35, 2208362.https://doi.org/10.1002/adma.202208362

[58] M. Samanta, K. Pal, U. V. Waghmare, K. Biswas," Intrinsically low thermal conductivity and high carrier mobility in dual topological quantum material, n-type BiTe," *Angewandte Chemie International Edition* (2020), 59, 4822

[59] Z. Tian, Q. Zhang, Y. Xiao, et al.," Ultraweak electron-phonon coupling strength in cubic boron arsenide unveiled by ultrafast dynamics," *Physical Review B* (2022), 105, 174306

## Acknowledgements

((Acknowledgements, general annotations))

## Data Availability Statement

The data that support the findings of this study are available in the Supporting Information of this article.

2D BiTe exhibits strong Kerr nonlinearity and a broadband optical response, enabling spatially self-phase-modulation-driven photonic applications. This work demonstrates all photonic isolators with nonreciprocal transmission and cross-phase modulation, enabling all-photonic information-conversion and logic-gate applications. These showcase the 2D BiTe as a versatile material for integrated nonlinear photonics.

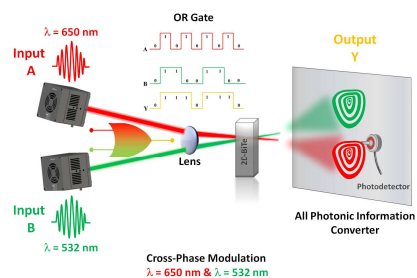

**Atomically-Thin Tsumoite (BiTe) based All-Photonic-Isolator, Information Converter, and Logic-Gate**

*Saswata Goswami[†], Caique Campos de Oliveira[†], Abhijith M.B., Varinder Pal, Vidya Kochat, Pulickel M. Ajayan, Samit K. Ray\*, Pedro A. S. Autreto\*, and Chandra Sekhar Tiwary\**

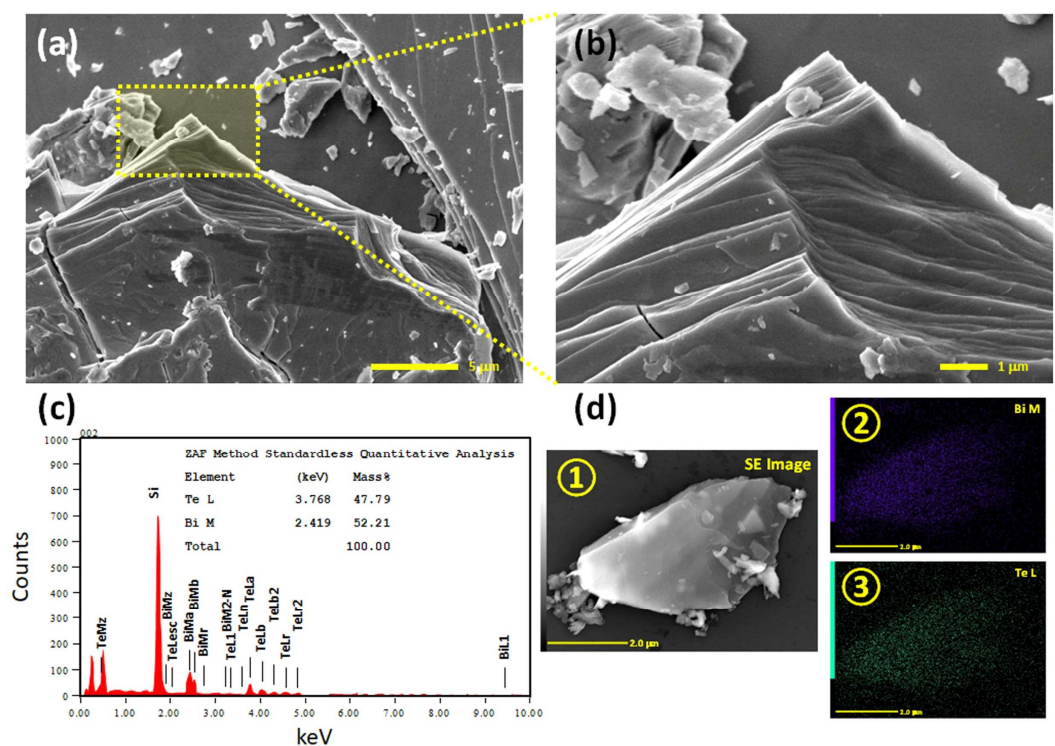

**Figure S1.** (a) Scanning Electron Microscope picture of Bulk powder after a brief exfoliation period. (b) Depiction of a zoomed-in view showcasing layered morphology. (c) Energy-dispersive X-ray spectroscopy provides quantitative analysis of elemental mass percentages. (d) ① Figure depicts the SE image of the 2D BiTe nanostructure. ②-③ Figure depicting elemental mapping of Bi, Te. Showing uniform distribution of both elements throughout the structure.

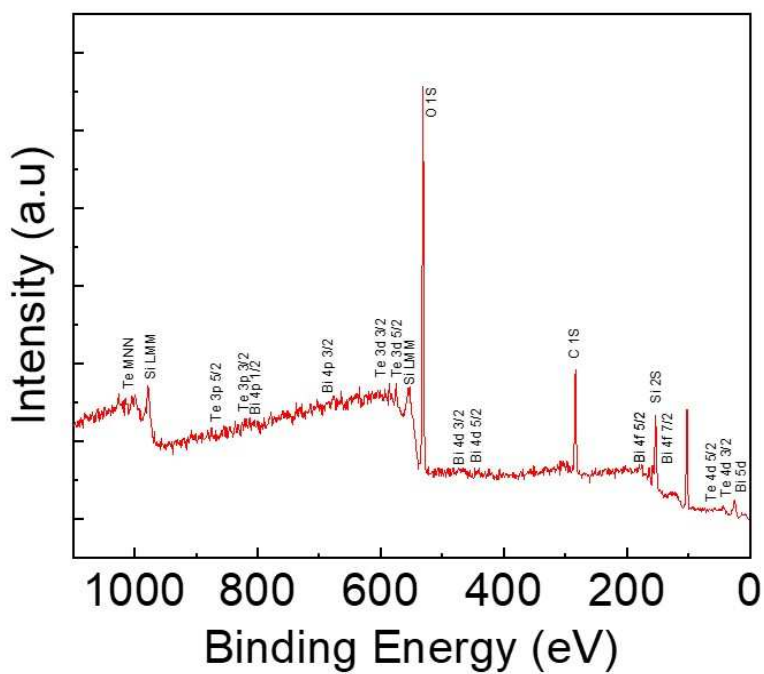

**Figure S2.** X-Ray Photoelectron Spectroscopy of 2D BiTe.

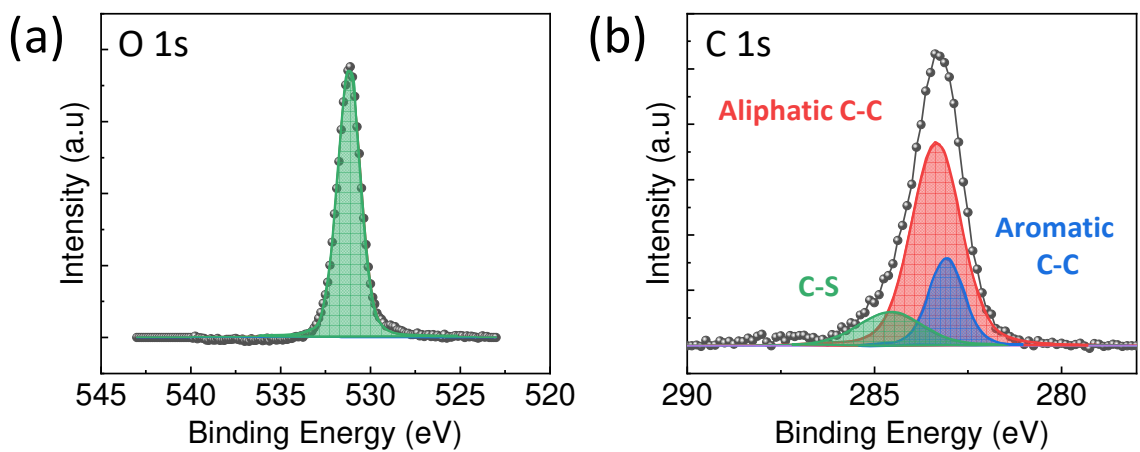

**Figure S3**. The recorded XPS spectra display well-defined signals at the (a) O 1s and (b) C 1s binding energies, indicating the incorporation of oxygen- and carbon-containing groups on the sample surface.

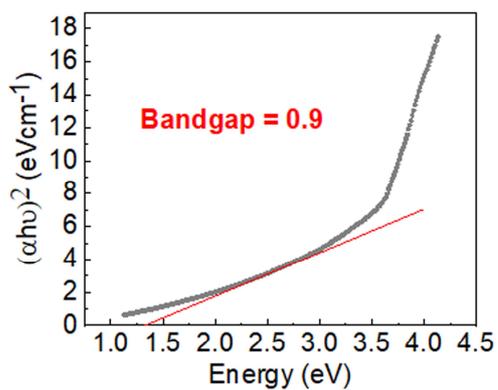

**Figure S4.** Bandgap estimation from bandgap using Tauc plot

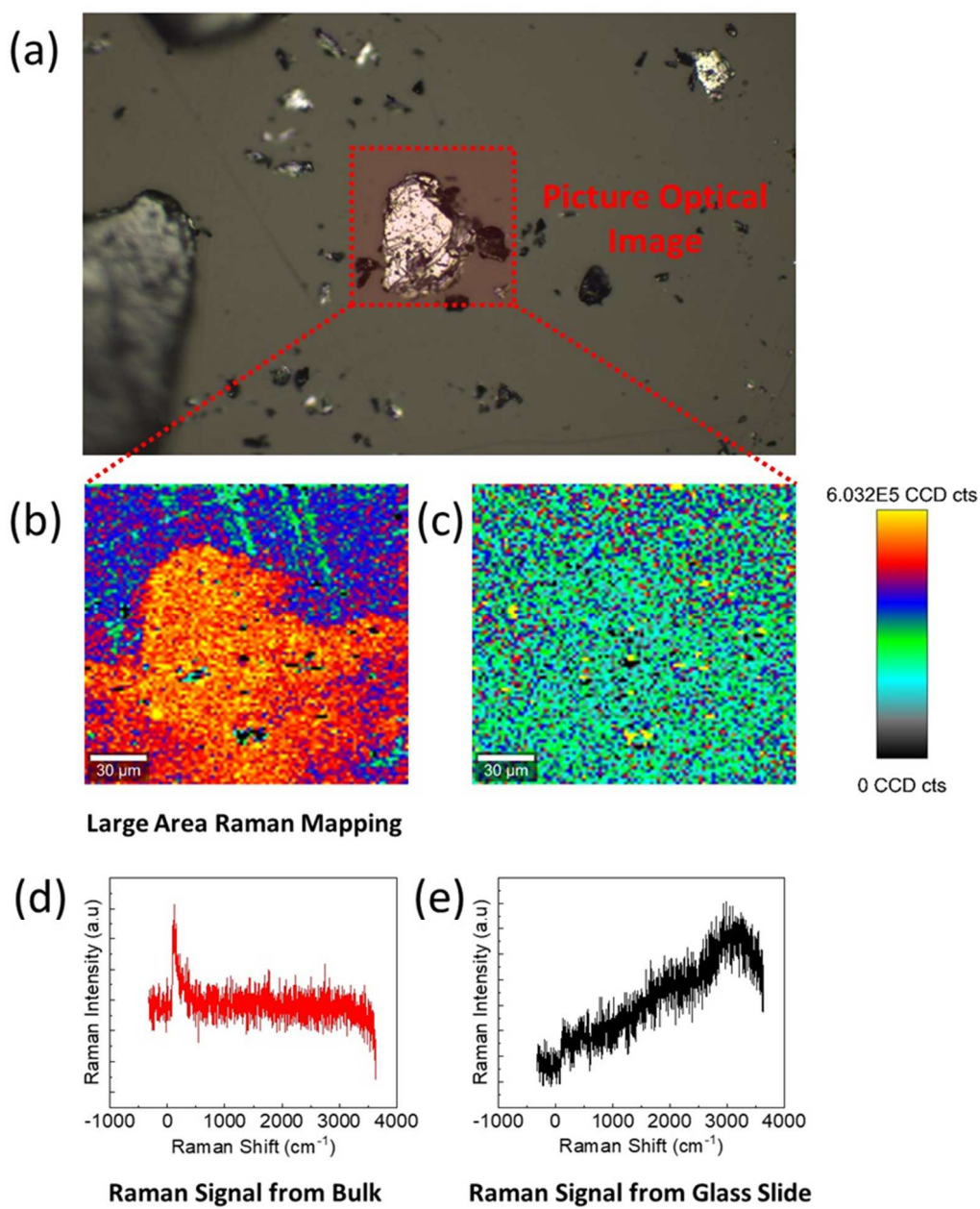

**Figure S5**. (a) Optical image depicting Bulk BiTe particle. (b) Raman signal acquired from the bulk BiTe particle while performing large area mapping. (c) Raman Signal acquired from the Si substrate. (d) Corresponding element-wise Raman signal collected by the detector.

**Table S1**. Recent investigations employing SSPM spectroscopy have calculated $n_2$ and $\chi^{(3)}_{total}$ for various materials

| 2D Material | Type of laser | $n_2$ | $\chi^{(3)}_{total}$ | $\chi^{(3)}_{monolayer}$ | References |
|---|---|---|---|---|---|
| Graphene | CW 532 nm | $2.5 \times 10^{-9} m^2 \cdot W^{-1}$ | $1 \times 10^{-4}$ | $1 \times 10^{-7}$ | [1] |
| $MoS_2$ | CW 478 nm | $10^{-7}\ m^2 \cdot W^{-1}$ | $1.44 \times 10^{-4}$ | $1.6 \times 10^{-9}$ | [2] |
| $Ti_3C_2Ti_x$ | CW 457 / 532 /671 nm | $11 \times 10^{-4}/$ $4.75 \times 10^{-4}/$ $4.72 \times 10^{-4}\ cm^2 W^{-1}$ | | $4.34 \times 10^{-7}/1.68 \times 10^{-7}$ $/0.15 \times 10^{-7}$ | [3] |
| 2D Te NS | CW 457 / 532 /671 nm | $6.14 \times 10^{-5}/$ $6.202 \times 10^{-5}/$ $7.37 \times 10^{-5}\ cm^2 W^{-1}$ | ------ | ------- | [4] |
| Sb FS/QD | CW 532 / 633 nm | $FS - 2.88 \times 10^{-5}$ $/0.979 \times 10^{-4}$ $QD - 1.91 \times 10^{-5}$ $/0.719 \times 10^{-5}$ | $FS$ $- 3.98 \times 10^{-9}$ $/ 1.74 \times 10^{-9}$ $QD$ $- 2.87 \times 10^{-5}$ $/ 1.29 \times 10^{-5}$ | ----- | [5] |
| $Bi_2Te_3$ | CW 1070 nm | $2.91 \times 10^{-9}\ m^2 \cdot W^{-1}$ | $10^{-3}$ | $10^{-8}$ | [6] |
| $Nb_2C$ | 400 nm | $0.62 \times 10^{-5}\ cm^2 W^{-1}$ | | $2.43 \times 10^{-9}\ (e.s.u.)$ | [7] |
| Graphene Oxide | CW 532 / CW 671 nm | $3.57 \times 10^{-6}/$ $1.1 \times 10^{-6}\ cm^2 W^{-1}$ | $1.7 \times 10^{-6}$ $5.32 \times 10^{-6}$ | ------ | [8] |
| $MoTe_2$ | CW 473 /532 / 750 / 801 nm | ------ | ------- | $1.88 \times 10^{-9}$ $1.3 \times 10^{-9}$ $1.14 \times 10^{-9}$ $0.98 \times 10^{-9}\ (e.s.u.)$ | [9] |
| $NbSe_2$ | 532 / 671 / 405 nm | $1.352 \times 10^{-5}\ cm^2 W^{-1}$ $/ 2.0 \times 10^{-5}\ cm^2 W^{-1}/$ $1.07 \times 10^{-5}\ cm^2 W^{-1}$ | $1.352 \times 10^{-5}$ $/9.354 \times 10^{-6}$ $/5.03 \times 10^{-6}$ | $3.34 \times 10^{-9}/2.59 \times 10^{-9}$ $/3.39 \times 10^{-9}$ | [10] |
| $Bi_2Se_3$ | 600/ 700 / 1160 nm | $1.16 \times 10^{-8}/ 3.53 \times 10^{-9}/$ $2.5 \times 10^{-9}/$ $1.65 \times 10^{-9}\ (m^2 W^{-1})$ | $5.76 \times 10^{-3}$ $/1.82 \times 10^{-3}$ $/ 1.29 \times 10^{-3}$ $/8.53 \times 10^{-3}$ $(e.s.u.)$ | $10^{-8}/10^{-8}/10^{-9}/10^{-9}$ | [11] |
| Black Phosphorus | Pulsed Laser 35 − 1160 nm | $10^{-5}\ cm^2 W^{-1}$ | $10^{-8}\ e.s.u$ | ----- | [12] |
| SbSI nanorods | 400 nm | $0.278 \times 10^{-5}\ cm^2 W^{-1}$ | ----- | ----- | [13] |

| Material | Wavelength | $\beta$ | $n_2$ | $\chi^{(3)}$ | Ref |
|---|---|---|---|---|---|
| $WSe_2$ | CW 532 /671 /457 nm/ | $2.94 \times 10^{-6}$ /$8.66 \times 10^{-6}$ /$6.402 \times 10^{-6}$ | $1.371 \times 10^{-6}$/ $4.04 \times 10^{-6}$/$2.98 \times 10^{-6}$ (e.s.u.) | $8.14 \times 10^{-10}$/ $8.44 \times 10^{-11}$ /$3.69 \times 10^{-9}$ | [14] |
| $TaS_2$ | CW 532 /671 /457 nm | $1.14 \times 10^{-5}$, $0.88 \times 10^{-5}$, $0.69 \times 10^{-5} cm^2W^{-1}$ | ------ | $1.2 \times 10^{-6}$/ $0.9 \times 10^{-6}$/ $0.7 \times 10^{-6}$ | [15] |
| $TaSe_2$ | 532 / 671 nm | $8.0 \times 10^{-7}$/$3.3 \times 10^{-7}$ $(cm^2.$ | $1.37 \times 10^{-7}$ /$1.58 \times 10^{-7}$ | $3.1 \times 10^{-10}$/$1.64 \times 10^{-10}$ | [16] |
| $GeSe$ | 532 nm | $4.841 \times 10^{-6}$ $(cm^2.W^{-1})$ | $2.258 \times 10^{-6}$ | $2.945 \times 10^{-10}$ | [17] |
| Boron NS | CW 457 /532 /671 nm | $1.25 \times 10^{-5}$ / $3.43 \times 10^{-6}$ / $9.45 \times 10^{-6}$ $(cm^2W^{-1})$ | $1.75 \times 10^{-7}$ / $0.64 \times 10^{-6}$ / $0.48 \times 10^{-6}$ (e.s.u.) | $4 \times 10^{-9}$/ $1.8 \times 10^{-9}$ / $1.8 \times 10^{-9}$ (e.s.u) | [18] |
| SnS NS | CW 532 / 633 nm | $4.531 \times 10^{-5}$ / $0.323 \times 10^{-5} (cm^2.W^1)$ | $2.317 \times 10^{-5}$ /$0.165 \times 10^{-5}$ (e.s.u. | $6.995 \times 10^{-10}$ / $2.037 \times 10^{-10}$ (e.s.u) | [19] |
| $Bi_2S_3$ | CW 457/532 / 671 nm | $3.34 \times 10^{-5}$/$1.26 \times 10^{-6}$/ $1.62 \times 10^{-7} (cm^2.W^1)$ | ------ | ------- | [20] |
| $MoSe_2$ | CW 532 nm | $3.24 \times 10^{-10} \, m^2W^{-1}$ | -------- | $1.1 \times 10^{-9}$ (e.s.u.) | [21] |
| TaAs | 405 / 532 / 671 / 841 nm | ------- | $6.06 \times 10^{-4}$ / $5.68 \times 10^{-4}$ / $5.30 \times 10^{-4}$/ $4.65 \times 10^{-4}$ (e.s.u.) | $10.50 \times 10^{-9}$ / $9.86 \times 10^{-9}$ / $9.19 \times 10^{-9}$ / $8.07 \times 10^{-9}$ (e.s.u.) | [22] |
| 2D $NiTe_2$ | CW 650 / 532 / 405 nm | $3.22 \times 10^{-5}$/ $6.15 \times 10^{-5}$ / $68 \times 10^{-5}$ $(cm^2W^{-1})$ | $1.56 \times 10^{-3}$ / $2.99 \times 10^{-3}$ / $3.76 \times 10^{-3}$ | $4.06 \times 10^{-9}$ / $7.79 \times 10^{-9}$ / $9.89 \times 10^{-9}$ (e.s.u.) | [23] |
| CuPc Nanotubes | CW 671 / 532 / 405 nm | $2.62 \times 10^{-5}$/ $2.65 \times 10^{-5}$ / $3.66 \times 10^{-5}$ $(cm^2W^{-1})$ | $1.44 \times 10^{-3}$ / $1.45 \times 10^{-3}$ / $2.01 \times 10^{-3}$ | $5.12 \times 10^{-9}$/ $5.14 \times 10^{-9}$/ $7.13 \times 10^{-9}$ (e.s.u.) | [24] |
| $Sb_2Se_3$ Nanorods | 671/ 532 / 405 nm | $6.67 \times 10^{-5}$/ $7.12 \times 10^{-5}$/ $9.12 \times 10^{-5}$/ | $4.23 \times 10^{-3}$ / $3.43 \times 10^{-3}$ / $2.28 \times 10^{-3}$ | $1.05 \times 10^{-9}$/ $0.85 \times 10^{-9}$/ $0.57 \times 10^{-9}$ (e.s.u.) | [25] |
| Violet Phosphorus NS | CW 405 / 473 / 532 / 671 / 721 nm | -- | -- | $3.54 \times 10^{-8}$/ $1.65 \times 10^{-8}$/ $9.64 \times 10^{-9}$/ $4.63 \times 10^{-9}$/ $2.31 \times 10^{-9}$ (e.s.u.) | [26] |
| Hybrid Bismuth Halide $(PPA)_3BiI_6$ | CW 405 / 473 / 532 nm | -- | -- | $1.77 \times 10^{-8}$/ $1.26 \times 10^{-8}$/ $7.74 \times 10^{-8}$ | [27] |

| Material | Wavelength | $\beta_{eff}$ | $n_2$ | $\chi^{(3)}$ (e.s.u.) | Ref. |
|---|---|---|---|---|---|
| MoP Microparticles | CW 532 nm | $1.91 \times 10^{-5}$ ($cm^2W^{-1}$) | -- | -- | [28] |
| $Fe_{3-x}GeTe_2$, $Fe_{4-x}GeTe_2$, $Fe_{5-x}GeTe_2$ | 671 / 532 nm | $2.49 \times 10^{-5}$ / $2.16 \times 10^{-5}$, $1.23 \times 10^{-5}$ / $1.09 \times 10^{-5}$, $0.93 \times 10^{-5}$ / $0.92 \times 10^{-5}$ | -- | $2.75 \times 10^{-8}$/ $2.37 \times 10^{-8}$/ $1.57 \times 10^{-8}$/ $1.39 \times 10^{-8}$/ $1.3 \times 10^{-8}$ / $1.29 \times 10^{-8}$ | [29] |
| $Bi_2Te_3$ | CW 650 / 532 / 405 nm | $2.7 \times 10^{-4}$/ $8.14 \times 10^{-5}$ / $10.1 \times 10^{-5}$ ($cm^2W^{-1}$) | $1.56 \times 10^{-3}$ / $4.741 \times 10^{-3}$ / $5.88 \times 10^{-3}$ | $1.2 \times 10^{-7}$/ $3.63 \times 10^{-7}$/ $4.51 \times 10^{-9}$ (e.s.u.) | [30] |
| Galistan $Ga_{0.63}In_{0.24}Sn_{0.13}$ | 405/ 532/ 766/ 1064/ 2000/ 3500 nm | $1.33 \times 10^{-5}$ / $1.68 \times 10^{-5}$ / $4.95 \times 10^{-5}$ / $6.95 \times 10^{-5}$ / $6.5 \times 10^{-5}$ / $3.48 \times 10^{-5}$ / $5.95 \times 10^{-5}$ ($cm^2W^{-1}$) | -- | -- | [31] |
| $NiPS_3$ | 405/ 473/ 532 / 671/ 721 nm | $3.77 \times 10^{-6}$ / $2.87 \times 10^{-6}$ / $2.77 \times 10^{-6}$ / $2.10 \times 10^{-6}$ / $1.20 \times 10^{-6}$ / ($cm^2W^{-1}$) | $2.05 \times 10^{-4}$ / $1.56 \times 10^{-4}$ / $1.51 \times 10^{-4}$ / $1.15 \times 10^{-4}$ / $0.65 \times 10^{-4}$ / (e.s.u.) | $3.59 \times 10^{-9}$ / $2.73 \times 10^{-9}$ / $2.64 \times 10^{-9}$ / $1.66 \times 10^{-9}$ / $1.14 \times 10^{-9}$ / (e.s.u.) | [32] |
| $FePS_3$ | 405/ 473/ 532 / 671/ 721 nm | $3.09 \times 10^{-6}$ / $2.17 \times 10^{-6}$ / $1.72 \times 10^{-6}$ / $1.49 \times 10^{-6}$ / $1.18 \times 10^{-6}$ / ($cm^2W^{-1}$) | $1.68 \times 10^{-4}$ / $1.18 \times 10^{-4}$ / $9.40 \times 10^{-5}$ / $8.10 \times 10^{-5}$ / $6.40 \times 10^{-5}$ / (e.s.u.) | $2.67 \times 10^{-9}$ / $1.87 \times 10^{-9}$ / $1.49 \times 10^{-9}$ / $1.29 \times 10^{-9}$ / $1.02 \times 10^{-9}$ / (e.s.u.) | [32] |
| $MnPS_3$ | 405 nm | $8.40 \times 10^{-7}$ ($cm^2W^{-1}$) | $4.60 \times 10^{-5}$ | $7.30 \times 10^{-10}$ (e.s.u.) | [32] |
| 2D BiTe (Our work) | CW 650 / 532 / 405 nm | $2.18 \times 10^{-5}$/ $4.73 \times 10^{-5}$/ $7.13 \times 10^{-5}$ ($cm^2W^{-1}$) | $1.05 \times 10^{-3}$ / $2.3 \times 10^{-3}$ / $3.52 \times 10^{-3}$ | $3.16 \times 10^{-3}$ / $6.89 \times 10^{-3}$ / $10.56 \times 10^{-3}$ | |

**Table S2.** Recent literature discusses the characteristic time associated with the emergence of the SSPM-induced diffraction pattern.

| Material | Laser Specification | Solvent | Intensity | Formation time ($\mathcal{T}$) | Ref. |
|---|---|---|---|---|---|
| $MoTe_2$ | 473/532 /750 nm (CW) | NMP | 252 $W.cm^{-2}$ | 0.45 / 0.6 / 0.62 s | [9] |
| $MoSe_2$ | 671 nm (CW) | NMP/ Acetone | 12 $W.cm^{-2}$ | 0.41 / 0.22 s | [10] |
| TaAs | 589 / 532 / 473 nm (CW) | NMP | 90 $W.cm^{-2}$ | 2.5 / 2.5 / 2.3 s | [22] |
| Graphene Oxide | 532 nm (CW) | IPA | ---- | 0.43 s | [33] |
| Black Phosphorus | 700 nm (CW) | NMP | 18.9 $W.cm^{-2}$ | 0.7 s | [12] |
| $Sb_2Se_3$ nanorods | 671 / 532 / 405 nm | NMP | 90 / 90/ 44.5 mW | 0.4 / 0.26/ 0.3 s | [25] |
| Violet Phosphorus | CW 405 / 473 / 532 / 671 / 721 nm | NMP | 1.55 / 3.55 / 9.80 / 19.65 / 18.64 $W.cm^{-2}$ | 0.3 / 0.47 / 0.7/ 0.83 / 0.77 s | [34] |
| $Fe_{3-x}GeTe_2$, $Fe_{4-x}GeTe_2$, $Fe_{5-x}GeTe_2$ | 671 / 532 nm | NMP | 15 $W.cm^{-2}$ | 0.28/ 0.31, 0.16 / 0.23, 0.15 /0.18 s | [29] |
| MoP Microparticles | CW 532 nm | NMP | 219.8 $W.cm^{-2}$ | 0.4 s | [28] |
| $Bi_2Te_3$ | 650 / 532 / 405 nm | NMP | 10.318, 5.39, and 1.244 $W.cm^{-2}$ | 0.266, 0.5 ,0.3999 s | [30] |

**Section 1:**

The nonlinear Kerr effect, crucial for interpreting the optical nonlinearity of 2D-BiTe, is analyzed in this study. Equation 1 expresses the dependability of the refractive index ($n$) on the incident laser intensity ($I$).

$$n = n_0 + n_2 I \ \ldots\ldots\ldots\ldots\ldots.. (1)$$

Here are the terms, "$n_0$" and "$n_2$" signifies the linear and nonlinear refractive indexes of the IPA solution and 2D BiTe suspended in the IPA, respectively.[35] The "$n_0$" of the 2D BiTe is found to be minimal compared to the IPA solution discussed earlier.

Furthermore, a variation in the refractive index of the IPA-2D BiTe solution will necessitate a phase shift into the incoming laser beam. The incoming continuous monochromatic wave traversing the medium undergoes intensity modulation, referred to as self-phase modulation (SPM). The effect of this modulation is visible through the generated diffraction pattern. The propagation of the Gaussian beam through a nonlinear medium of length L along the z-axis can be represented by the following electric field distribution:

$$E(r,z) = E(0,z)\frac{\omega_0}{\omega(z)}\exp\left(-\frac{r^2}{\omega(z)^2}\right) \times \exp\left(-i\left[kz - \arctan\frac{z}{z_0} + \frac{kn_0 r^2}{2R(z)}\right]\right) \quad \ldots\ldots (2)$$

Here, $\lambda$ denotes the wavelength of the excitation source, and the corresponding wave vector is given by $k = \frac{2\pi}{\lambda}$. The Rayleigh length, which specifies the axial distance over which the laser beam retains its minimum spot size, is defined as $z_0 = \pi\omega_0^2/\lambda$. The radius of curvature of the beam's wavefront at any axial position $z$ is described by $R(z) = z\left[1 + \left(\frac{z}{z_0}\right)^2\right]$, while $\omega_0$ represents the beam waist radius evaluated at $z = 0$. Using the beam waist as the reference point for propagation, the incident electric field $E(r, z_0)$ interacting with the 2D BiTe medium can be written as,

$$E(r, z_0) = E(0, z_0)\exp\left(-\frac{r^2}{\omega(z_0)^2}\right) \times \exp\left(-i\frac{kn_0 r^2}{2R(z)}\right) \quad \ldots\ldots (3)$$

The continuous wavelength laser source intensity distribution can be defined as,

$$I(r,z) = I_0(1 + z^2/z_0^2)^{-1}\exp\left(-2r^2/\omega_0^2\right) \quad \ldots\ldots (4)$$

Where, $I_0 = 2P_{ave}/[\pi\,\omega(z)^2]$ is the strength of the laser light at the central region. The average power of the continuous-wave excitation source is denoted by $P_{ave}$. Hence, the refractive index is defined as:

$$n(r) = n_0 + n_2 I(r) \quad \ldots\ldots (5)$$

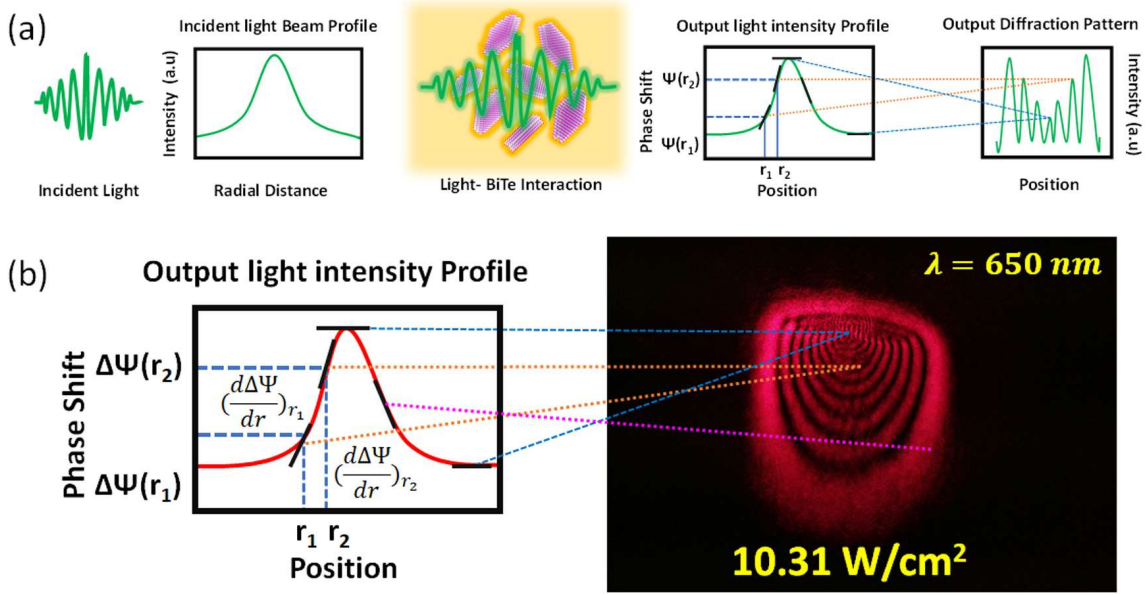

**Figure S6**. (a) Introduction of phase shift due to light-matter interaction and diffraction pattern formation at the far screen. (b) inclusion of a real image for a 650 nm wavelength.

It is assumed that the sample is thinner in dimension, we can conclude that, Total Phase shift

$$\Delta\phi = \Delta\psi_L(r) + \Delta\psi_{NL}(r),$$

Here, $\Delta\psi_L(r)$ is denoted as the linear phase shift. $\Delta\psi_L(r) = kn_0 r^2/[2R(z)]$. The term $\Delta\psi_{NL}(r)$ is considered the nonlinear phase shift. This term, depicted as a nonlinear phase shift, can be correlated with the light intensity in the following manner,

$$\Delta\psi_{NL}(r) = \frac{2\pi n_0}{\lambda} \int_0^L n_2 I(r,z) dz \quad \ldots\ldots\ldots\ldots\ldots (6)$$

" $\Delta\psi_{NL}(r)$ ", depicted in Equation 6, is induced by self-inducing Nonlinear Kerr Effect, which is indirectly a result of the high value of $\chi_{total}^{(3)}$. This high value of $\chi_{total}^{(3)}$ can be due to the intense laser light irradiation, or the crystal symmetry. The lateral component of the wave vector associated with beam propagation is formulated as:

$$\delta k(r) = \frac{d\Delta\psi_{NL}}{dr} = \frac{-8krn_2 I(r,z)\exp(-2r^2/\omega_0^2)}{n_0} \quad \ldots\ldots\ldots\ldots\ldots (7)$$

Here, $\lambda$ denotes the wavelength of the incident light beam, $L_{eff}$ quantifies the effective propagation distance traversed by the beam through the cuvette, $r \in [0, +\infty)$ specifies the radial position, and $I(r,z)$ represents the excitation signal's radial strength profile.[14]

The laser beam is absorbed by the 2D BiTe suspended in the IPA solution, and a photon is emitted when the electron-hole pair recombines. These photons exhibit a phase shift relative to the incoming photons in the laser beam; hence, they can both interact and produce a diffraction pattern in the far field.

For a Gaussian beam emerging from the medium, two or more radial positions, $r_1$ and $r_2$, exist where $(d\Delta\psi/dr)_{r=r_1} = (d\Delta\psi/dr)_{r=r_2}$ and the corresponding phases coincide. This follows from the Gaussian nature of the nonlinear phase shift, as depicted in Figure S6 of the Supporting Information. Hence, the resulting intensity profile demonstrates a fixed phase difference while preserving equivalent slope regions.

If both of these points in the far-field satisfy the condition of the interference, a diffraction ring appears. This condition can be stated as $\Delta\psi_0 \geq 2\pi$, hence fulfilling the condition for a bright ring to appear. In the far-field region, diffraction from the self-induced phase modulation gives rise to visual bright and dark ring patterns, whose positions are defined by Equation 8.

$$\Delta\psi_{r_1} - \Delta\psi_{r_2} = 2M\pi \quad \ldots\ldots\ldots\ldots\ldots (8)$$

Within this formulation, $M$ is an integer variable. The dark-field pattern arises for odd values of $M$, while the bright-field pattern appears for even values.[1] Figure S6 of the Supporting

Information shows that the phase variation underlying the SSPM effect results in a far-field diffraction pattern. The effective transmission length ($L_{eff}$), evaluated using Equation 9, plays a vital role in determining ($n_2$) as depicted in Equation 13.[12]

$$L_{eff} = \int_{L_1}^{L_2}(1 + \frac{z^2}{z_0^2})^{-1} \, dz = z_0 \arctan(Z/z_0)\Big|_{L_1}^{L_2}, \; z_0 = \frac{\pi\omega_0^2}{\lambda} \quad \ldots\ldots\ldots\ldots (9)$$

Here, $L_1$ and $L_2$ specify the distances from the beam focus to the front and rear walls of the quartz cuvette, respectively, giving a cuvette thickness of $L_2 - L_1$. The transmitted beam intensity follows a Gaussian profile, peaking at the center, where $I(0, z) = 2I$. The parameters $I$, $z_0$, and $\omega_0$ denote the average input intensity, diffraction length, and beam waist radius at $\frac{1}{e^2}$, respectively.

The depiction of $\Delta\psi_{NL}(r)$,

$$\Delta\psi_{NL}(r) = \frac{2\pi n_0 n_2}{\lambda} I_0 L_{eff} e^{-2r^2/\omega_0^2} \quad \ldots\ldots\ldots\ldots (10)$$

Also, $\Delta\psi(r_1) - \Delta\psi(r_2) = M\pi$, The observed ring brightness is determined by the phase correlation between two radial points, $r_1$ and $r_2$. The Gaussian beam satisfies the center and infinity limits as:

$$\Delta\psi(0) - \Delta\psi(\infty) = 2N\pi \quad \ldots\ldots\ldots\ldots (11)$$

Here, the ring count due to SSPM pattern formation is considered to be $N$,

$$\Delta\psi(\infty) = 0, \; \Delta\psi(0) = \frac{2\pi n_0 n_2}{\lambda} I_0 L_{eff} \quad \ldots\ldots\ldots\ldots (12)$$

The nonlinear refractive index is dependent on the term $\frac{dN}{dI}$. Hence, the estimation of $\frac{dN}{dI}$ is considered to be an important factor, the $n_2$ is depicted as,

$$n_2 = \left(\frac{\lambda}{2n_0 L_{eff}}\right) \cdot \frac{dN}{dI} \quad \ldots\ldots\ldots\ldots (13)$$

To assess the nonlinear optical properties of the material, the third-order nonlinear susceptibility ($\chi_{total}^{(3)}$) is employed as a key parameter.[2, 16, 36] Hence the $\chi_{total}^{(3)}$ is depicted as,

$$\chi_{total}^{(3)} = \frac{cn_0^2}{12\pi^2} 10^{-7} n_2 \; (e.s.u) \quad \ldots\ldots\ldots\ldots (14)$$

Here, $c$ denotes the speed of light in free space, $n_0$, and $n_2$ is defined in earlier paragraphs. The suspension of 2D material layers dispersed into the solution cuvette directly influences the total third-order nonlinear susceptibility, $\chi_{total}^{(3)}$, as the 2D material directly causes the light-matter interaction. Therefore, it becomes essential to evaluate the third-order nonlinear susceptibility monolayer of the two-dimensional material, denoted as $\chi_{monolayer}^{(3)}$. The dependence of the overall electric field ($E_{total}$) on the electric field within a single BiTe monolayer ($E_{monolayer}$) can be formulated as:[1, 37]

$$E_{total} = \sum_{j=1}^{N_{Eff}} E_j \cong N_{eff} E_{monolayer} \quad \dots\dots\dots\dots\dots (15)$$

In this context, $N_{Eff}$ depicts the 2D-BiTe layer count in the cuvette through which the beam passes. The link between $\chi^{(3)}_{total}$ and $\chi^{(3)}_{monolayer}$ is defined as, [16, 37-38]

$$\chi^{(3)}_{total} = N^2_{eff}\chi^{(3)}_{monolayer} \quad \dots\dots\dots\dots\dots (16)$$

**Section 2: The number of 2D BiTe along the laser optical path in the solution.**

BiTe conjures a molecular weight of 800.7608 g.mol$^{-1}$. The concentration of the 2D material-IPA solution inside the cuvette is $3.122 \times 10^{-4}$ mol. L$^{-1}$. The cuvette volume is taken as $4.5 \times 10^{-3}$ L. The Total number of molecules of BiTe in the solution is $M = \rho \times V \times N_A$. The total molecular count of BiTe in the prepared solution was obtained from the expression $M = \rho \times V \times N_A$, where $\rho$, $V$, and $N_A$ represent the density, volume, and Avogadro constant, respectively. Here the term $N_A$ stands Avoagdro's Constant. The space group of the BiTe ($P\bar{3}m1$) is different from Bi$_2$Te$_3$ ($R\bar{3}m$) which suggests different electronic and optical responses. The structure is identified as hexagonal, with lattice parameters $a$ = 4.423 Å and $c$ = 24.002 Å. Hence, the number of 2D BiTe nanostructures present in a single cuvette system with effective layer estimation can be expressed as:[39]

$m = 1 \times 4.5 \, cm^2 / (Sin \, 90°) \times (4.423)^2 = 2.079 \times 10^{15}$ molecules. The layer count of the 2D BiTe suspended in the direction of the effective path length can be estimated as, $n = M/m = 361$.

**Table S3:** The values of $\chi^{(3)}$, Mobility (µ) & Effective Mass ($m^*$)

| Material & Corresponding Wavelength | $\chi^{(3)}_{monolayer}$ (Third-order nonlinear susceptibility) | Mobility (µ) & Effective Mass ($m^*$) | Reference |
|---|---|---|---|
| Graphene | $1 \times 10^{-3}$ (e.s.u.) | [40-41] | [1, 42] |
| BP | $10^{-5} \, cm^2 W^{-1}$ (e.s.u.) | [43-44] | [12] |
| MoS$_2$ | $1.44 \times 10^{-4}$ (e.s.u.) | [45] | [2] |
| WSe$_2$ | $1.371 \times 10^{-6} / 4.04 \times 10^{-6} / 2.98 \times 10^{-6}$ (e.s.u.) | [46] | [14] |
| MoSe$_2$ (532 nm) | $1.76 \times 10^{-4}$ (e.s.u.) | [47] | [38] |
| WS$_2$ | $8.14 \times 10^{-10} / 8.44 \times 10^{-11} / 3.69 \times 10^{-9}$ (e.s.u.) | [48] | [49] |
| MoTe$_2$ | $1.88 \times 10^{-9} / 1.3 \times 10^{-9} / 1.14 \times 10^{-9} / 0.98 \times 10^{-9}$ (e.s.u.) (CW 473 /532 / 750 / 801 nm) | [50] | [9] |
| TaS$_2$ | $1.2 \times 10^{-6} / 0.9 \times 10^{-6} / 0.7 \times 10^{-6}$ (e.s.u.) | [51] | [15] |
| GeSe | $2.945 \times 10^{-10}$ (e.s.u.) | [52] | [17] |
| SnS | $6.995 \times 10^{-10}$ / $2.037 \times 10^{-10}$ (e.s.u.) | [53] | [19] |
| TaAs (405 nm/ 532 nm/ 671 nm/ 841 nm) | $6.06 \times 10^{-4}$ / $5.68 \times 10^{-4}$ / $5.30 \times 10^{-4}$ / $4.65 \times 10^{-4}$ / $6.06 \times 10^{-4}$ / $5.68 \times 10^{-4}$ / $5.30 \times 10^{-4}$ / $4.65 \times 10^{-4}$ (e.s.u.) | [22] | [22] |

**Section 3: Estimation of the optical bandgap of 2D-hBN utilizing Tauc method**

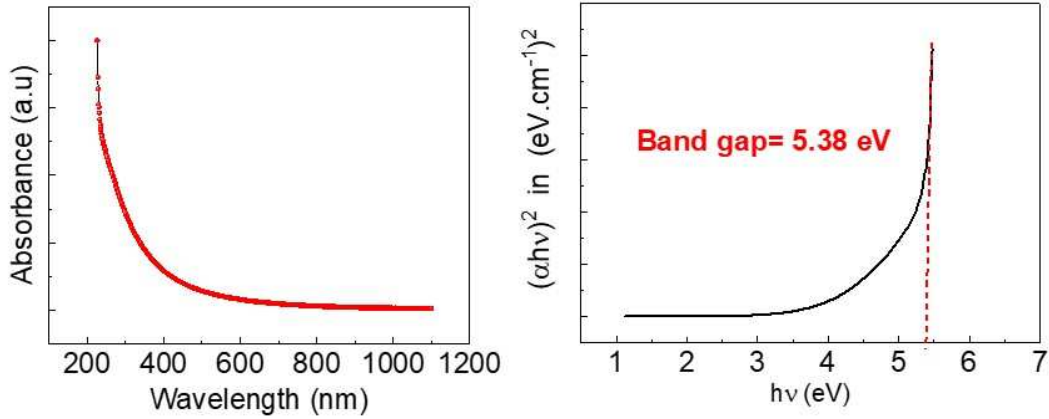

**Figure S7.** (a) UV-Vis absorbance data of exfoliated 2D-hBN. (b) Tauc plot showing the estimation of the direct optical bandgap, which is approximately 5.38 eV.

**Section 4:**

This work examines the nonlinear optical properties of 2D BiTe via the SSPM diffraction method. A continuous-wave laser source is used in the SSPM Spectroscopy method. Recently, it was concluded that SSPM does not arise from a thermally induced reaction between the laser and the 2D material.[30, 54] The subsequent conclusion points and experiments will elucidate the arguments. Hu et al.[12] predicted that $\chi^{(3)}_{monolayer}$ exhibits a proportional relationship with carrier mobility. These two parameters signify the capability of charge accumulation, illustrating the energy-storing mechanism, but $\chi^{(3)}_{monolayer}$ exhibits an inverse relationship with the effective mass of the majority charge carriers ($m^*$). The effective mass signifies an attenuative property, emphasizing that energy is a dissipative process. The electric field associated with the incoming laser beam enhances carrier velocity, resulting in less scattering during transit, which increases the value of $\chi^{(3)}_{monolayer}$. The increased value of $\chi^{(3)}_{monolayer}$ ascends as the wavelength decreases, correlating with a rise in photon energy. The reported $\chi^{(3)}_{monolayer}$ values from SSPM measurements are analyzed concerning carrier mobility (μ) and effective mass ($m^*$). $\chi^{(3)}_{monolayer}$, μ, and $m^*$ respectively describe the nonlinear optical, transport, and excited-state electronic properties. Their strong interdependence leads to pronounced laser-induced coherence.

Beyond Kerr effects, laser-induced thermal variations in the medium can alter the material's refractive index, leading to self-phase modulation. Dabby et al.[55] referred to the phenomenon as "thermal lens effect." The study presented here investigated the stimulation of electronic behavior in contrast to thermal behavior, "Wind Chime Model''. This model emphasizes the polarization characteristics of the dispersed nanostructures and their reorientation upon laser exposure. In contrast, the thermal lens effect corresponds to a linear optical response.[23] To validate the aforementioned coherence phenomenon, a mechanical chopper was used to make the laser source excitation discreet, before the laser source was focused through a lens. The corresponding optical configuration is depicted in Figure S8a. The chopper was allowed to rotate at 1-3kHz (with a 500 Hz interval) at a constant intensity of 11.45 W.cm$^{-2}$ to investigate any changes in the diffraction pattern. Variation in chopping frequency (1-3 kHz) produced no change in the diffraction ring count, indicating stable electronic excitation. Hence, it can be concluded that the nonlinear optical response is primarily governed by electronic coherence rather than the thermal lens effect. An Optris PI 640i thermal camera was used to monitor the temperature distribution at various mechanical chopper modulation frequencies under constant laser intensity, as illustrated in Figure S10a. The laser–solvent thermal response was evaluated for frequencies between 0 - 3 kHz (0.5 kHz step) to identify any manifestation of the thermal lens effect. Figures S10b①-③-⑤-⑦-⑨-⑪show the open end of the cuvette prior to laser exposure at 0-3 kHz (0, 1, 1.5, 2, 2.5, and 3 kHz), while Figures S10b ②-④-⑥-⑧-⑩-⑫ present the corresponding images after irradiation. Figure S10c shows that higher chopping frequencies reduce laser-solvent interaction, resulting in lower temperature rise. Each 0.5 kHz increase in frequency corresponded to a decrease in the temperature difference.

As shown in Figures S8b (①-⑥), the ring count remains constant with increasing chopping frequency. The slight reduction observed at lower intensities results from weakened electronic coherence. Since temperature changes (Figure S8c) do not affect ring formation, SSPM is attributed to electronic coherence rather than to thermal effects such as the thermal lens effect. To confirm the generality of the previous findings, an analogous experiment was performed using a 532 nm laser source (if the finding follows the change in wavelength) (shown in Figure S9a). The laser-solvent thermal interaction was analyzed at various chopper frequencies; the light and solution were allowed to interact over different time scales. The discreet excitation induces periodic interaction between the laser light and a 2D material contained in a cuvette. temperature variation was recorded before and after laser exposure at 0 kHz and at frequencies between 1 and 3 kHz, with increments of 0.5 kHz. Figure S11b (①-③-⑤-⑦-⑨-⑪) presents the top view of the cuvette prior to laser irradiation for each corresponding frequency. Figures S11b (②-④-⑥-⑧-⑩-⑫) display the thermal images of the cuvette's open end following laser exposure at 0-3 kHz (0.5 kHz step). Figure S11c presents the corresponding temperature differentials recorded before and after laser irradiation for each chopping frequency. Increasing chopping frequency decreases the laser-solvent interaction duration, thereby reducing localized heating. Decreasing temperature differences do not affect the number of diffraction rings, which remains constant, as observed in Figure S9b. The dependence of ring count on chopping frequency is summarized in Figure S9c, spanning 0-3 kHz at 0.5 kHz intervals.

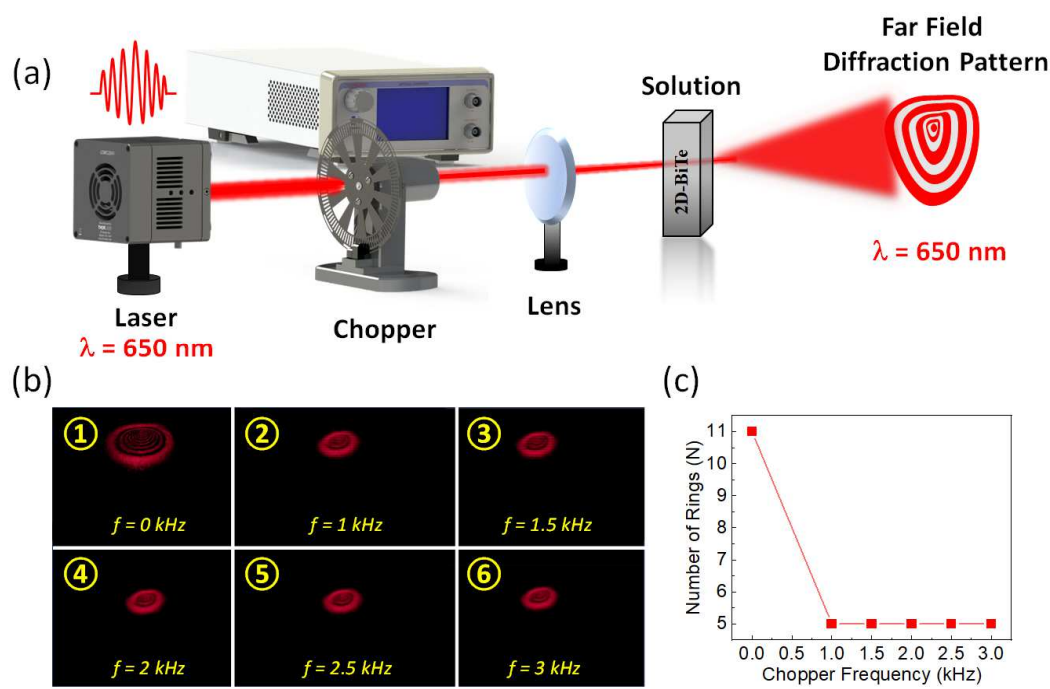

**Figure S8.** (a) Figure depicting the modified SSPM setup with a mechanical chopper for the wavelength 650 nm. (b) The diffraction pattern obtained for an excitation of the laser wavelength 650 nm was recorded while the mechanical chopper was operated at elevated modulation frequencies. (0,1-3 kHz with 0.5 kHz of interval frequency). (c) The dependence of the ring count of the diffraction profile on the chopper frequency reveals the nonlinear optical characteristics of the material for 650 nm.

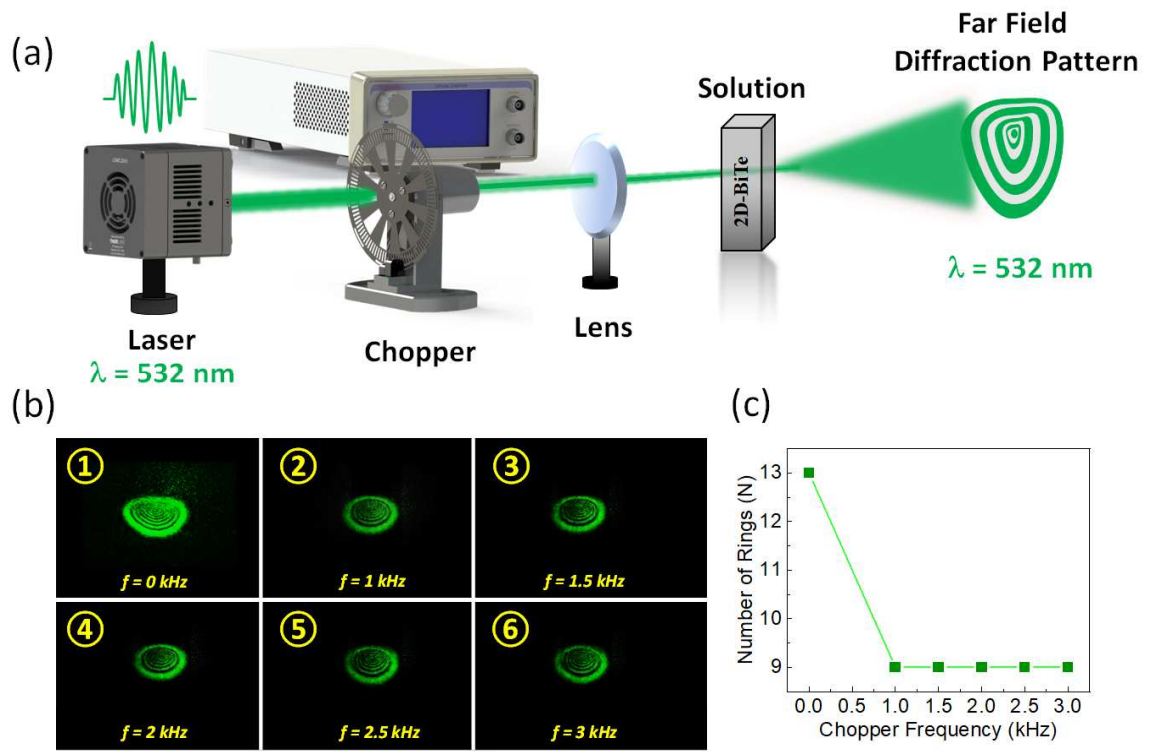

**Figure S9.** (a) Figure depicting the modified SSPM setup with a mechanical chopper for the wavelength 532 nm. (b) The diffraction pattern obtained with an excitation at 532 nm was recorded while the mechanical chopper was operated at elevated modulation frequencies. (0,1-3 kHz with 0.5 kHz of interval frequency). (c) The dependence of the ring count of the diffraction profile on the chopper frequency reveals the nonlinear optical characteristics of the material for 532 nm.

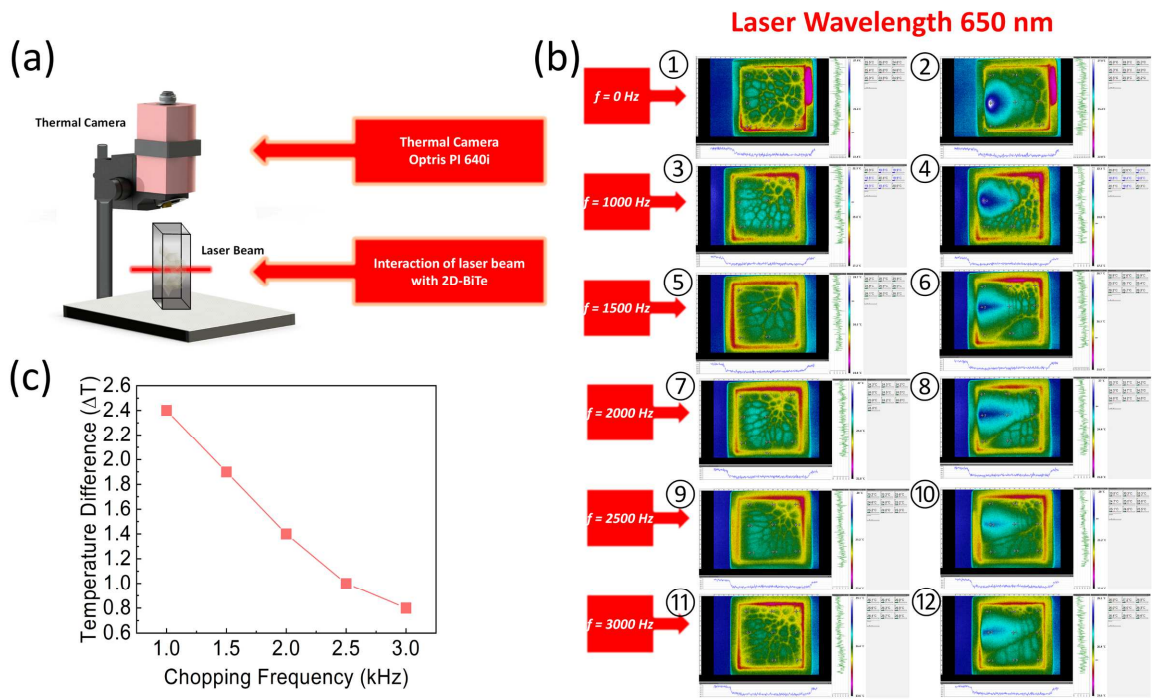

**Figure S10.** (a) Experimental arrangement showing the thermal imaging setup used to record the top-view temperature distribution of the cuvette. (b)①, (b)③, (b)⑤, (b)⑦, (b)⑨, and (b)⑪ display the thermal maps of the open-end section of the cuvette prior to laser exposure at modulation frequencies of 0, 1-3 kHz (step size 0.5 kHz) for the 650 nm excitation. (b)②, (b)④, (b)⑥, (b)⑧, (b)⑩, and (b)⑫ show the corresponding thermal maps after laser exposure under identical frequency conditions. (c) Variation in temperature of the cuvette before and after laser irradiation plotted as a function of chopper frequency for the 650 nm wavelength.

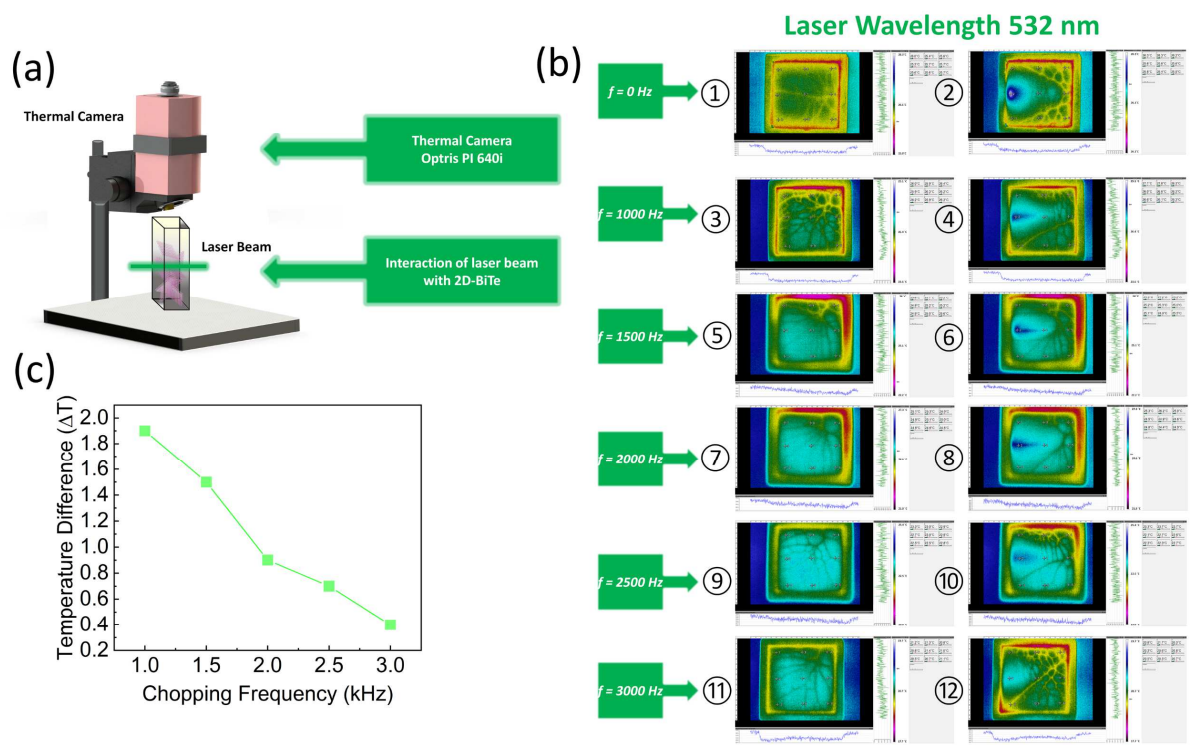

**Figure S11.** (a) Experimental arrangement showing the thermal imaging setup used to record the top-view temperature distribution of the cuvette. (b)①, (b)③, (b)⑤, (b)⑦, (b)⑨, and (b)⑪ display the thermal maps of the open-end section of the cuvette prior to laser exposure at modulation frequencies of 0, 1-3 kHz (step size 0.5 kHz) for the 532 nm excitation. (b)②, (b)④, (b)⑥, (b)⑧, (b)⑩, and (b)⑫ show the corresponding thermal maps after laser exposure under identical frequency conditions. (c) Variation in temperature of the cuvette before and after laser irradiation plotted as a function of chopper frequency for the 532 nm wavelength.

**Section 5:**

**Wind Chime Model: Generation of SSPM-Based Diffraction Profile under Different Intensities and Wavelength:**

This section explains pattern generation and the SSPM mechanism. Depicted in Figure S12a, the 2D BiTe flakes are randomly oriented in the solvent. Initially, the interaction between the incoming laser beam and the 2D BiTe flakes is negligible. Over time, the diffraction pattern begins to emerge on the far screen. Wu et al. defined the temporal SSPM diffraction profile generation, incorporating the characteristic response time necessary to establish a steady-state intensity distribution within the nonlinear medium.[39] According to this theorem, the initial interaction between the incident laser and the 2D BiTe layer occurs at random orientations relative to the laser's electric field. The solution's suspended 2D BiTe is polarized by this laser beam.[56] Energy relaxation induces the alignment of the polarized 2D BiTe axis with respect to the electric field of the incoming laser light.[39] The macroscopic angle between 2D

BiTe decreases with time, aligning it more precisely with the electric field and increasing the ring count in the diffraction pattern. When all 2D-BiTe in the laser beam's path length are fully aligned, the ring count reaches its maximum number. The electron coherence is further investigated to examine the formation of the diffraction pattern. Expressing the electronic wave function at $r_A$ is defined as $\psi_A(\mathbf{r}_A) = \sqrt{\rho_A(\mathbf{r}_A)}e^{i\phi(\mathbf{r}_A)}$. Denoting the local electron density $\psi_A^*(\mathbf{r}_A) \cdot \psi_A(\mathbf{r}_A) = \rho_A(\mathbf{r}_A)$ and the phase is defined as $\phi(\mathbf{r}_A) = \mathbf{k} \cdot \mathbf{r}_A - \omega t + \phi_0(\mathbf{r}_A)$. The phase $\phi(\mathbf{r}_A)$ is entirely dictated by the extrinsic light field (Figure S12b). This denotes an enforced oscillation, in which electrons are required to conform to the local phase established by the extrinsic field, even in the presence of scattering and interactions. In this context, $\phi_0(\mathbf{r}_A)$ depicts the lag in the phase magnitude as a response to the extrinsic field. The initial arbitrary phase remains. The same concept is valid for the electronic wave functions at $(r_B)$, regardless of whether they belong to the same nanostructure or different ones. Because both wave functions remain synchronized with the incident laser field, they also exhibit mutual coherence. This phenomenon represents dynamic (AC) electron coherence, which differs from the conventional steady-state (DC) electron coherence commonly identified in electronic transport studies. The electrons stay near their equilibrium positions. In accordance with the wind-chime configuration, electron coherence increases as more electrons become correlated, while the collision frequency decreases due to the perfectly oriented nanostructures. The intensity on the screen is $I(\mathbf{r}_C) = E^*(\mathbf{r}_C) \cdot E(\mathbf{r}_C)$, here $E(\mathbf{r}_C)$ is denoted as the electric field of the laser beam at $r_C$. The term $E(\mathbf{r}_C)$ denotes the electric field component of the laser beam evaluated at the spatial coordinate $r_C$. The magnitude of the intrinsic electric field constituent in the light wave becomes $E(\mathbf{r}_C) = \varsigma(\chi^{(3)})\Psi(\mathbf{r}_C)$, with the phase being $\Psi(\mathbf{r}_C) = e^{i[\mathbf{k}\cdot\mathbf{r}_C - \omega t + \phi_0]}\left[\sqrt{\rho_A}(\mathbf{r}_A)e^{-i\mathbf{k}_1\cdot\mathbf{r}_A} + \sqrt{\rho_B}(\mathbf{r}_B)e^{-i\mathbf{k}_1\cdot\mathbf{r}_B}\right]$, the wave vector $k_\perp = d\Delta\phi/dr$ is included due to Kerr nonlinearity. The precise coefficient is dictated by the photonic nonlinearity $\chi^{(3)}_{monolayer}$. The behavior is dominated by the charge-carrier characteristics of the material.

(Figure S12b) depicts the influence of the laser beam electric field on the electrical wave function. We examined the polarization of intense laser light on 2D BiTe to validate the proposed model.

$$N = N_{max}\left(1 - e^{t/\tau_{rise}}\right) \quad \ldots\ldots\ldots\ldots (17)$$

Figures S12c, S12d, and S12e present the time-dependent evolution of the diffraction patterns corresponding to excitation sources with excitation of wave sources of 650, 532, and 405 nm, respectively. The respective laser beam intensities are 10.318, 6.28, and 1.29 W.cm$^{-2}$. Table S2

shows the temporal evolution of the 2D BiTe nanostructure under the influence of different excitation wave sources of 650, 532, and 405 nm.

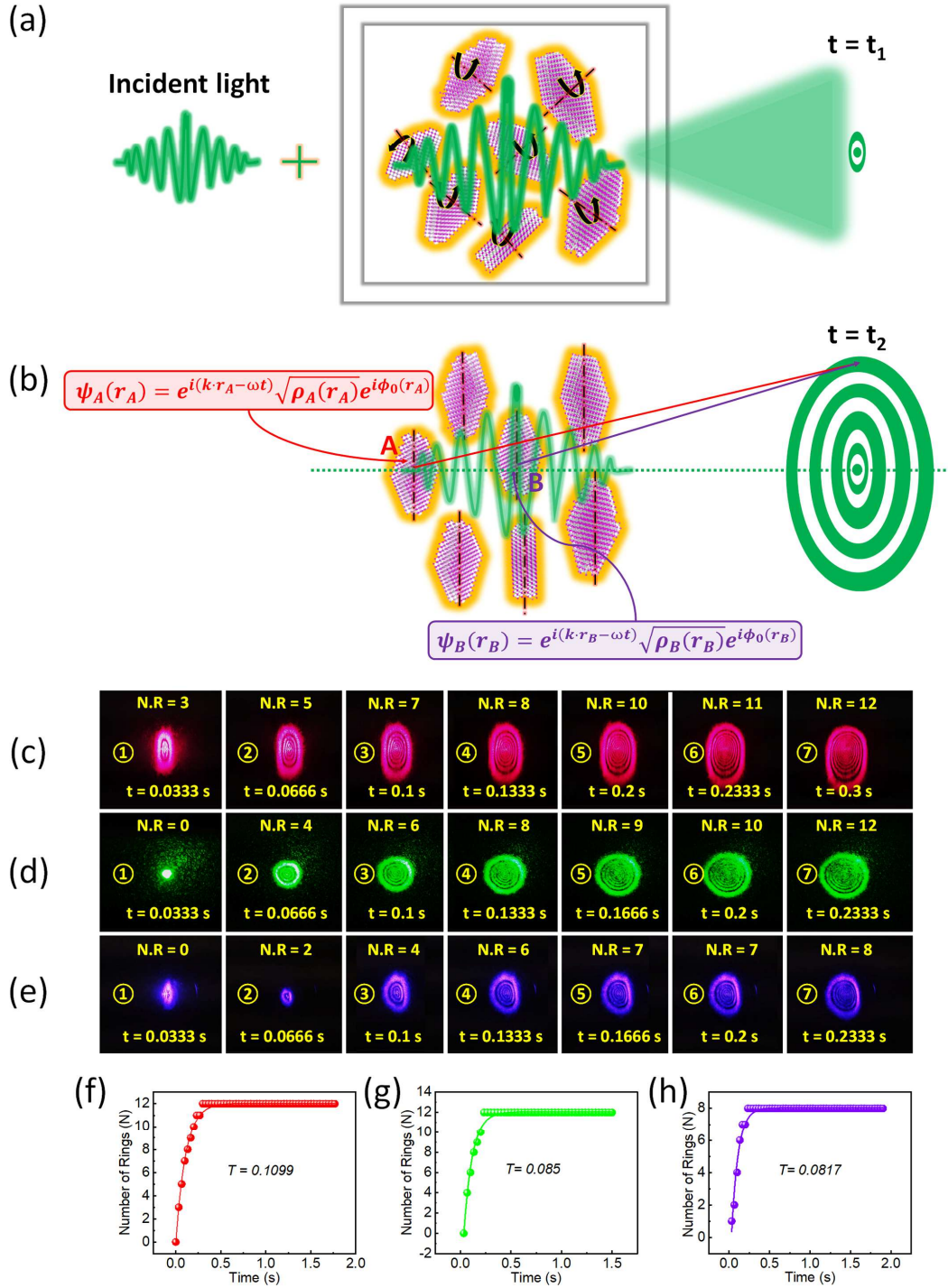

**Figure S12.** (a) Initial period of SSPM pattern on the far screen at $t = t_1$. (b) Depiction of windchime model and full diameter diffraction pattern formation on the far screen due to enhanced electronic coherence at time t = $t_2$. (c-d-e) Temporal progression of SSPM patterns across various excitation laser sources. ($\lambda$ = 650, 532, and 405 nm). (f-g-h) Time evolution under different wavelengths. ($\lambda$= 650, 532, and 405 nm).

$N$ denotes the visible ring count visible seen on the diffraction profile, while $N_{max}$ represents the maximum ring count produced under a steady laser intensity. The term $\tau_{rise}$ corresponds to the time required for the diffraction rings to fully develop. The Wind Chime Model effectively describes the characteristic time ($\mathcal{T}$) required to form the complete diffraction pattern, including the full vertical and horizontal diameters.

$$\mathcal{T} = \frac{\epsilon_r \pi \eta \xi R_C}{1.72(\epsilon_r - 1) I h} \quad \ldots\ldots\ldots\ldots (18)$$

In this context, $\varepsilon_r$ denotes the relative dielectric constant of 2D BiTe, numerically estimated as 43.42, 20.09, and 6.03 for the wavelength of 650, 532, and 405 nm, respectively. Detailed computational details for the calculation of the value $\varepsilon_r$ are given in the Supporting Information Section 9. $\eta$ represents the viscosity coefficient of the solvent, which for IPA is approximately $2.4 \times 10^{-3}\ Pa.s$ for 20 °C. The parameter $\xi$ corresponds to the fraction of the fluid sphere in direct contact with the 2D nanostructure, estimated to be 0.00735. $R_C$ denotes the effective radius of the 2D BiTe nanostructure. While (h) refers to the vertical dimension of the 2D BiTe, and $I$ represent the intensity of the excitation source. The effective radius ($R_C$) determined via AFM is 125 nm, and the height (h) of the nanostructure is deduced to be around 12 nm. Theoretical estimates ($\mathcal{T}$) for the NMP solvent are 0.31, 0.55, and 0.287 s at excitation wave sources of 650, 532, and 405 nm, respectively. The corresponding curve-fitted results closely match the experimental observations, yielding 0.333, 0.5333, and 0.3 s for the same wavelengths. These results, illustrated in Figures S12f-g-h, correspond to laser intensities of 10.318, 6.28, and 1.29 Wcm$^{-2}$, respectively.

The ring count at 532 nm exceeds that at other wavelengths, indicating enhanced light-matter interaction and an extended duration of diffraction pattern production. The diffraction profile requires a duration to attain the maximum ring count, which is expressed in Table S4 for various excitation wave sources.

Table S4. The experimental and theoretical estimates of the time evolution corresponding to the diffraction pattern formation.

| Sample | Solvent | Wavelength (nm) | Intensity (W.cm$^{-2}$) | C [mg/mL] | L [mm] | $\tau_c$ [s] | Observational $\mathcal{T}$ [s] | Theoretical $\mathcal{T}$ [s] |
|---|---|---|---|---|---|---|---|---|
| 2D-BiTe | IPA | 650 | 10.318 | 0.25 | 10 | 0.1099 s | 0.333 s | 0.31 s |
| 2D-BiTe | IPA | 532 | 6.28 | 0.25 | 10 | 0.085 s | 0.533 s | 0.55 s |
| 2D-BiTe | IPA | 405 | 1.29 | 0.25 | 10 | 0.0817 s | 0.3 s | 0.287 s |

**Section 6:**

**Dynamic Reduction of the Diffraction Profile Influenced by Incident Intensity, along with the variation of the nonlinear refractive index under different excitation wavelengths and Intensities.**

A detailed analysis of the SSPM behavior reveals that the self-diffraction pattern deforms progressively over time. At the stage where the pattern reaches its maximum vertical extent, the upper portion begins to contract inward toward the central core. As illustrated in Figure S13a for the 532 nm wavelength, vertical aberration appears more significant than horizontal distortion. Such deformation, first described by Wang et al.[42], arises from non-axisymmetric thermal convection, revealing the fundamental mechanism that drives the contraction of the diffraction pattern. Here, parameters $R_H$ and $\theta_H$ represent the maximum radius and the associated half-cone angle, respectively, before contraction begins.

The relationship can be illustrated as, [19]

$$\theta_H = \frac{R_H}{D} \quad \ldots\ldots\ldots\ldots\ldots\ldots (19)$$

Equation 14 is applicable exclusively under the condition that D significantly exceeds $R_H$. The value D specifies the distance from the cuvette to the screen. The thermal convection effect induces distortion in the diffraction profile, leading to alterations in both the vertical diffraction radius and the half-cone angle, depicted as $R_h$ and $\theta_h$ after the contraction process. If D is significantly more than $R_h$. Then the relationship can be expressed as follows:

$$\theta_h = \frac{R_h}{D} \quad \ldots\ldots\ldots\ldots\ldots\ldots\ldots\ldots (20)$$

Accordingly, the distortion in the diffraction profile can be depicted as the distortion radius ($R_D$) and angle ($\theta_D$), depicted as:

$$\theta_H - \theta_h = \frac{R_D}{D} \quad \ldots\ldots\ldots\ldots\ldots\ldots\ldots (21)$$

Moreover, the half-cone angle can be depicted as,

$$\theta_H = \frac{\lambda}{2\pi}\left(\frac{d\psi}{dr}\right)_{max}, r \in [0, +\infty] \quad \ldots\ldots\ldots\ldots\ldots (22)$$

Here, $r$ denotes the transverse beam position. For a Gaussian beam profile, the preceding relation can be rewritten as a function of the half-cone angle $\theta_H$.[49]

$$\theta_H = n_2 I C \quad \ldots\ldots\ldots\ldots\ldots (23)$$

Where $C = [-\frac{8rL_{eff}}{\omega_0^2}\exp(-\frac{2r^2}{\omega_0^2})]$ is constant when $r \in [0, +\infty)$. Figure S13b depicts the progression of the diffraction pattern with a wavelength of 532 nm. The distortion variance is quantified from the angle between the 2D nanostructure cuvette and the far-field, with the deformation amount increasing with beam intensity.[57] Therefore, the contraction angle can be depicted as,

$$\theta_D = \theta_H - \theta_h = (n_2 - n_2')IC = \Delta n_2 IC \ldots\ldots\ldots\ldots (24)$$

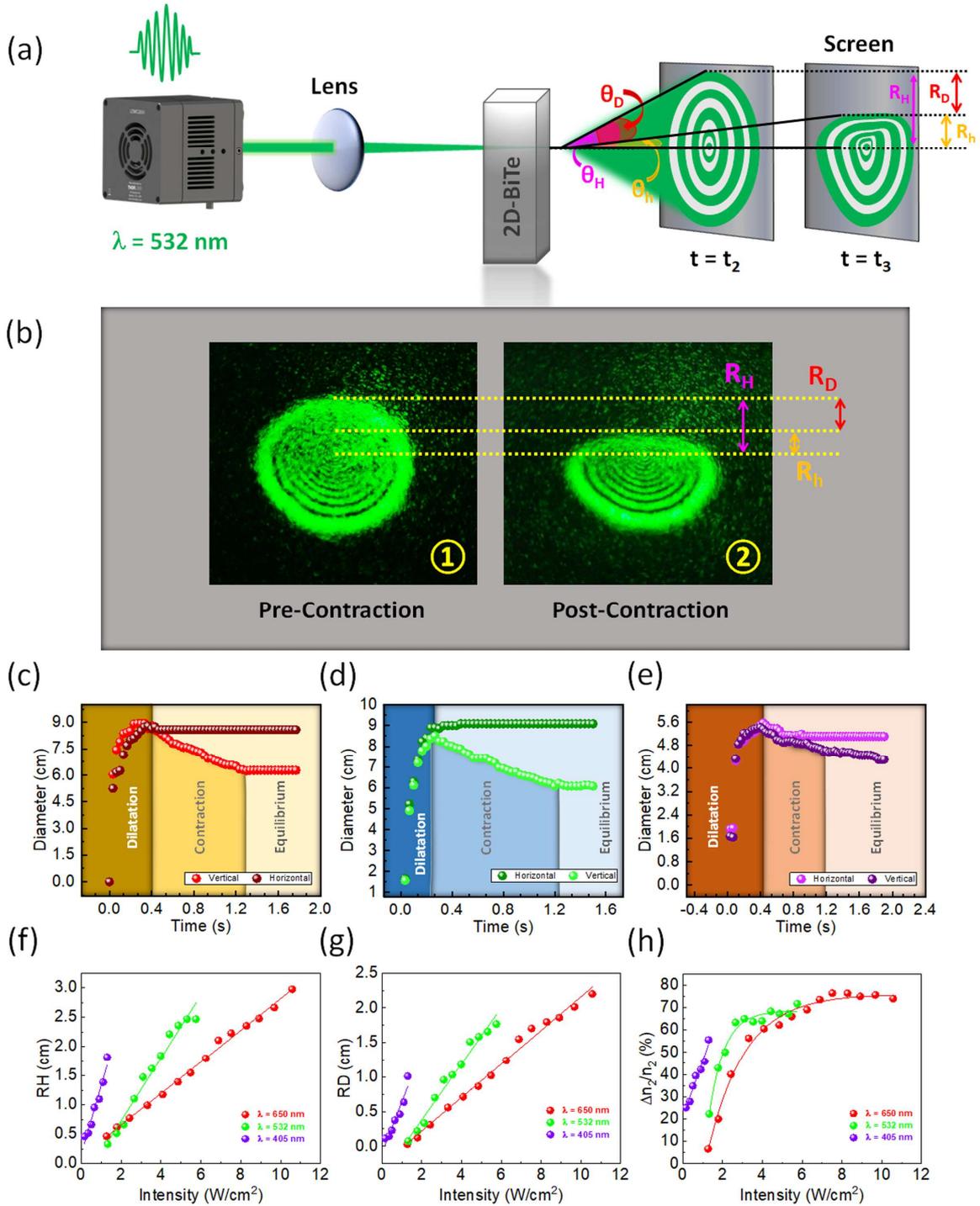

**Figure S13.** (a) The figure depicts the contraction phenomenon in the far-field SSPM diffraction profile, exhibiting a deformation angle. (b) Figurative depiction of the dynamic contraction process, ① pre-contraction and ② post-contraction. (c-d-e) The progression of the vertical and horizontal diameters varies with the progress of time at wavelengths (λ = 650, 532, and 405 nm). (f) The variation in the maximum radius ($R_H$) as a function of laser intensity is examined for excitation at 650, 532, and 405 nm. (g) The change in the distorted radius ($R_D$) with changing intensity for different wavelengths (λ = 650, 532, and 405 nm). (h) The illustration depicts ($\Delta n_2$) changes with intensity for different excitation wavelengths (λ = 650, 532, and 405 nm).

Heat causes the fluid to move in an asymmetric pattern, distorting the diffraction pattern and triggering contraction.[42] Because the absorption coefficient remains constant, the solvent is heated by thermal convection as the laser beam travels through the solution.[57-58]

A temperature gradient orthogonal to the laser axis is created by the indirect effect of laser heating, which amplifies thermal convection. From the data presented in Equations 18 and 19, it can be deduced that,

$$\frac{\Delta n_2}{n_2} = \frac{\theta_D}{\theta_H} = \frac{R_D}{R_H} \qquad\qquad\qquad\qquad\qquad ……………………..(25)$$

The relative variation of the nonlinear refractive index can be determined by using Equation 25 and keeping track of the dynamic change ($\frac{\theta_D}{\theta_H}$) or ($\frac{R_D}{R_H}$) in the diffraction ring distortion. The ratio $\Delta n_2/n_2$ is affected by several factors, including laser wavelength, temperature, and duration.[59] The relative variation $\Delta n_2/n_2$ is predominantly governed by the intensity of the incident laser beam.[59] This parameter $\Delta n_2/n_2$ can be quantitatively evaluated through an appropriate formula-based calculation. [17, 49] Figure S13b (①–②) presents the diffraction pattern recorded by the detector. After the diffraction profile starts exhibiting the maximum ring count and the largest vertical diameter under stable conditions. Following the expansion phase, a contraction star happens when irradiated with a 650 nm laser. Figures S13c-d-e illustrate the temporal evolution of the horizontal and vertical diameters of the diffraction rings corresponding to laser intensities of 10.57, 6.27, and 1.29 Wcm$^{-2}$, associated with wavelengths of 650, 532, and 405 nm, respectively.

All three lasers demonstrated an increase in horizontal diameter over time until achieving their maximum diameter; after that, the diameters remained constant. The vertical diameters reach the maximum magnitude and subsequently decrease over duration to an equilibrium value. Distortions in the diffraction pattern appear when the thermal convection process becomes considerable, which occurs at a certain ring number. The vertical deformation becomes increasingly pronounced over time. Figure S13f-g illustrates the determined maximum vertical radius through pre-contraction and the deformed radius value, respectively. The change observed in the diffraction profile was attributed to heat convection. The $n_2$ variable is used to measure distortion, as outlined in Equation 25. The distortion index, represented as $\Delta n_2/n_2$, increases with an increase in intensity value. The distortion index $\Delta n_2/n_2$ is expressed as a numerical percentage (%) form, corresponding to three light wave sources of 650, 532, and 405 nm, illustrated in Image S12h. The $\Delta n_2/n_2$ is measured at 73.95% for 650 nm, 71.64% for 532 nm, and 55.55% for 405 nm, corresponding to intensities of 10.57, 6.27, and 1.29 Wcm$^{-2}$, respectively. At the same intensity levels, the 532 nm laser causes greater deformation than the 650 nm laser because the 532 nm beam contains more energetic photons per discrete number than the 650 nm laser beam. The deformation induced by the excitation wave source at 405 nm is attributed to being maximum because the 405 nm photons have higher energy than those at 532/650 nm. The distortion increment follows a linear pattern and reaches its maximum when the excitation wave source exceeds a threshold. Beyond this threshold, saturation dominated, as shown in Figure S13h. The overall pattern formation time is approximately equal to the time required for the diffraction rings to reach their maximum counts, plus the time needed for the system to stabilize. The basic Supporting Information Section 6 emphasizes upon the origin of the distortion process.

**Section: 7**

To analyze the generation of the diffraction profile, a continuous-wave laser source with a Gaussian intensity profile is considered to propagate in the horizontal direction through a stationary, isothermal fluid. The optical field's complex amplitude at the point where it enters the sample medium can be formulated as,

$$U(x, y, t, z = 0) = U_0 \exp\left(-\frac{x^2+y^2}{\omega^2}\right) \exp\left(-ik\frac{x^2+y^2}{2R}\right) \quad \text{.........................(26)}$$

The laser source is considered to have an amplitude along the optical axis, denoted by $U_0 = \sqrt{\frac{2P}{\pi\omega^2}}$, P represents the net power, k is defined as $2\pi/\lambda$, where $\lambda$ is the wavelength of the light beam in space. The radius of the beam of light depending on the cross section is given by $\omega$, which is defined as the half-width at ($\frac{1}{e^2}M$). $R$ denotes the radius of curvature of the wavefront at a specific point, with the coordinate origin defined at the beam waist, while r represents the radial distance from the optical axis. At the initial stage, radiation absorption and thermal conduction are the predominant mechanisms, producing radially symmetric distributions of temperature and irradiance in the fluid surrounding the beam axis. The evaluation of the temperature field within the liquid requires consideration of the relative motion between successive fluid layers. Hence, the laser's thermal self-action leads to temperature-dependent density variations in the liquid. As a result of the density gradient, variations in buoyant force arise, causing fluid movement that helps transport heat away from the illuminated zone. This convective motion leads to an asymmetric temperature distribution within the fluid. Fluid motion generated by spatial changes in buoyancy is represented through the Navier–Stokes and continuity equations. When combined with the heat-transport relation, they form a coupled set of equations that describe the dynamics of convective heat transfer and thermal instability. To quantify the influence of thermal convection, a steady vertical convective velocity within the beam region can be considered, which induces optical distortion through thermally driven effects. For ease of calculation, convection velocity is considered to be the same across the beam as its maximum value at the beam center, where it is set by the balance between the fluid's viscosity and the buoyant force generated by heating.[60] The center of the excitation source profile exhibits the highest amplitude of convection velocity at the specified location by Equation 27.

$$v_x = \frac{\beta g [\Delta T]_{max} \pi h^2}{16\mu} \quad \ldots\ldots\ldots\ldots (27)$$

Here, g is depicted as the acceleration caused by gravity itself (g = 9.8 m/s²), $\beta$ represents the thermal expansion coefficient of the solvent, $\mu$ denotes the viscosity coefficient of the liquid, and $h$ It represents the minimum radial distance from the beam center to the location where convection remains active. $[\Delta T]_{max}$ indicates the maximum temperature rise. Estimating a uniform upward convective velocity is beneficial, as it enables a concise and effective investigation of the 2D-temperature distribution (thermal map). Assuming a liquid flow velocity of constant magnitude, along the x-axis ($v_x$), (the vertical direction) and perpendicular to the laser source, the temperature profile depicted by the equation is as follows:

$$c_P = \rho \left( \frac{\partial}{\partial t} [\Delta T(x,y,t)] + v_x \frac{\partial [\Delta T(x,y,t)]}{\partial x} \right) \quad \ldots\ldots\ldots (28)$$

$$-K\nabla^2 [\Delta T(x,y,t)] = Q(x,y)$$

$$\Delta T(x,y,0) = 0 \quad x,y < \infty$$

$$\Delta T(\infty, y, 0) = 0 \quad t > \infty$$

$$\Delta T(x, \infty, t) = 0 \quad t > 0$$

Here, $c_p$ corresponds to the specific heat capacity, $\rho$ to the material's density, and $K$ to its thermal conductivity. The thermal power generated per unit volume and per unit time by the Gaussian beam at a point $r$ inside the sample is described by $Q(x,y)$.[61]
where α is the absorption coefficient and,

$$I(x,y) = \frac{2P}{\pi \omega^2} \exp\left(-\frac{2(x^2+y^2)}{\omega^2}\right) \quad \ldots\ldots\ldots\ldots (29)$$

The Green's function is employed as a fundamental solution to the heat conduction–convection equation. It represents the system's response to a point heat source, allowing the temperature distribution within the medium to be expressed as an integral over the heat sources. By using the Green's function approach, one can analytically model how heat propagates and dissipates through both conduction and convective flow within the sample. This method significantly simplifies the solution of complex thermal problems, especially when the boundary conditions or heat sources vary spatially.[61] The calculated alteration in the temperature profile resulting from absorbing the laser excitation source. Equation 30 depicts the temperature distribution inside the cuvette,

$$\Delta T(x,y,t) = \frac{\alpha P}{\pi \rho c_p} [\int_0^t \frac{dt'}{8Dt'+\omega^2} \\ \times \exp(-2[(x-v_x t')^2 + y^2]/[8Dt'+\omega^2])], \quad \ldots\ldots (30)$$

Here $D = \frac{K}{\rho c_p}$ is the thermal diffusivity coefficient of the solution, considered IPA in this work. The assessment of the temperature profile estimation in Equation 30 serves as the foundation for determining the spatial variation of the $(\Delta n)$, which is expressed as,

$$\Delta n(x,y,t) = n_0 + \frac{dn}{dT}\Delta T(x,y,t) \quad \ldots\ldots (31)$$

Here $dn/dT$ is the thermo-optic coefficient of the medium. Assuming that the change in thermal refractive index remains almost constant along the beam's propagation path, an assumption valid for thin samples induced phase shift can be calculated as,

$$\Delta \phi(x,y,t) = \frac{2\pi}{\lambda} L[n(x,y,t) - n(0,0,t)] \quad \ldots\ldots (32)$$

where λ represents the wavelength of the laser source, and $L$ denotes the beam propagation length through the sample. The phase shift can be determined using equations (5)-(7).

$$\Delta \phi(x,y,t) = \frac{\theta}{2t_c} \times \int_0^t \frac{dt'}{1+\frac{2t'}{t_c}} \\ \times [\exp\left(-\frac{2[(x-v_x t')^2+y^2]}{\left[\left(1+\frac{2t'}{t_c}\right)\omega^2\right]}\right) \quad \ldots\ldots (33) \\ - \exp\left(-\frac{2(v_x t')^2}{\left[\left(1+\frac{2t'}{t_c}\right)\omega^2\right]}\right)]$$

Where $t_c = \frac{\omega^2}{4D}$ is the diffusion time and $\theta = \left(\frac{dn}{dT}\right)\alpha \frac{PL}{\lambda K}$ is the on-axis phase shift. Consequently, following transmission through the 2D BiTe-IPA nonlinear medium, the laser source field's complex amplitude at the output interface of the sample is mathematically described as,

$$U(x,y,t,z=0) = U_0 \exp\left(-\frac{\alpha L}{2}\right) \exp\left(-\frac{x^2+y^2}{\omega^2}\right) \\ \times \exp\left(-ik\frac{x^2+y^2}{2R}\right) \exp(i\Delta\phi(x,y,t)). \quad \ldots\ldots (34)$$

By employing the Fraunhofer approximation of the Fresnel-Kirchhoff diffraction formulation, the laser beam's complex amplitude and its far-field intensity pattern at the detector plane located in the far-field region can be described as,

$$U(x',y',t) = U_0 \frac{i\pi\omega^2}{\lambda d} \exp(ikd) \exp\left(-\frac{\alpha L}{2}\right) \times \int_{-\infty}^{\infty} dx \int_{-\infty}^{\infty} dy \exp\left(-\frac{x^2+y^2}{\omega^2}\right) \times$$
$$\exp\left[i\left(-k\frac{x^2+y^2}{2R} + \Delta\phi(x,y,t)\right)\right] \times \exp(-ik(xx'+yy')/d)\ldots\ldots\ldots (35)$$

And,

$$I(x',y',t) = \left|U_0 \frac{i\pi\omega^2}{\lambda d} \exp(ikd)\exp\left(-\frac{\alpha L}{2}\right)\right.$$
$$\times \int_{-\infty}^{\infty} dx \int_{-\infty}^{\infty} dy \exp\left(-\frac{x^2+y^2}{\omega^2}\right)$$
$$\times \exp\left[i\left(-k\frac{x^2+y^2}{2R} + \Delta\phi(x,y,t)\right)\right]$$
$$\left. \times \exp(-ik(xx'+yy')/d)\right|^2 \ldots\ldots\ldots (36)$$

$x_0$ and $y_0$ represent the spatial variables in the detector plane, while $d$ denotes the displacement along the detection plane and the sample's departure plane. Applying equation (35) allows the numerical determination of the far-field intensity pattern produced by a Gaussian laser beam as it propagates through the liquid medium. In the subsequent calculations, the parameters for the sample and laser beam are established as $\lambda = 532\ nm$ (laser wavelength). D = 0.35 mm (laser source beam radius $HW1 = e^2M$), $R = 3500\ mm$ (radius of curvature), $D = 40\ m^{-1}, \rho = 789\ kg\ m^{-3}, c_p = 2440\ J\ kg^{-1}K^{-1}, \alpha = 750 \times 10^{-6}\ K^{-1}, \nu = 1.36 \times 10^{-6} m^2 s^{-1}, dn/dT = 4 \times 10^{-4}\ K^{-1}, K = 0.17\ W\ m^{-1}\ K^{-1}, d = 2100\ mm$, and h is measured as 1.3 mm. The convection velocity must initially be estimated utilizing equations (27) and (30). Figure S20 illustrates the temporal evolution of the convection velocity in the liquid sample at an applied laser power of 40 mW. It is evident that the ring count and the degree of vertical distortion in the diffraction profile increase with increasing laser light power. The simulation outcomes exhibit good agreement with the experimental observations.

To further validate the numerical model, the temporal evolution of the outermost ring diameters along the horizontal and vertical directions for a laser power of 40 mW is shown in Figure S20a. In accordance with the experimental results (see Figure S11c), the simulations reveal that the diameter of the outermost ring increases over time. The horizontal diameter eventually reaches a steady maximum, while the vertical diameter initially grows to a peak before stabilizing at a smaller size than its horizontal counterpart. Figures S13c and S20b present a quantitative comparison between the experimental observations and the simulated outcomes, illustrating the horizontal and vertical diameters of the outermost diffraction rings obtained from both analyses. The numerical predictions derived from the theoretical model show minor deviations from the experimental measurements (see Figures S13c-d-e for laser wave sources 650, 532, and 405 nm). These discrepancies are likely due to the use of an approximate convection velocity in the simulation. In principle, the buoyancy-driven convective velocity should be determined directly from the Navier-Stokes and continuity equations. Moreover, the simulation accounted for IPA's relevant physical properties, including its thermo-optic coefficient and thermal conductivity. The SSPM brief setup was depicted in Figure S14a. The laser beam was focused at the center of the face, the laser beam passing through the cuvette in the midsection, as seen in Figure S14b. The Navier-Stokes and continuity equations were used to simulate fluid flow inside the cuvette. The fluid dynamics caused by the laser heating can be understood through analyzing the diffraction profile. Initially, the laser is not capable of heating the solution. As time progresses, the maximum fluid velocity increases (as seen in Figure S14c), and the temperature also increases. The pattern starts to emerge from the wind chime model depicted in Figure S12c; the ring count and the vertical and horizontal diameters reach their maximum values. After some time, as indicated in S13c, the equilibrium is reached, although the temperature at the center starts to increase, indicating greater fluid motion. Hence, with greater fluid motion, the vertical diameter starts to shrink, as described in Figure S13c. The vertical diameter starts to decrease with passing time, indicating distortion due to time-dependent thermal fluid motion. Figure S14c①-②-③-④-⑤-⑥ refers to the fluidic motion caused by laser heating in the solution for the laser wave source of 650 nm. Figure S15c①-②-③-④-⑤-⑥ and Figure S16c①-②-③-④-⑤-⑥ refers to the fluidic motion caused by the laser wave source of 532 and 405 nm. Figures S14d, S15d, and S16d depict the fluidic motion at the center of the cuvette over time. Figure S14e-S15e-S16e show the temperature increase caused by the laser wave source of $\lambda$ = 650, 532, and 405 nm. The top surface of the cuvette was considered to be open boundary during the simulation. Figure S12f illustrates the relationship between temperature variation and laser beam intensity, with the estimated findings closely resembling those of the original Figure S15f. The Optris PI 640i is used to collect thermal pictures of the top surface profile of the fluid inside the cuvette. Figure S17 a①-b①-c①-d①-e① illustrates the thermal imaging data before impact of the laser and the Figure S17 a②-b②-c②-d②-e② illustrates the thermal image profile after the laser impact. In the Figure S17 the change in temperature with intensity is plotted and is found to be very similar to that of Figure S14f. The same numerical simulation in MATLAB environment experiment is done for the case of 532, and 405 nm, it is depicted in the Figure S15, and Figure S16. Corresponding original experiment was also performed for the wavelength 532 nm and 405 nm. These experiments involving thermal camera is depicted in the Figure S18 and S19 for the wavelengths 532, 405 nm respectively.

Figure S15(f)-16(f)-17(f) depicts the temperature profile for different excitation source which further show similar characteristics to the Figure S 17(f)- S 18(f)- S 19(f).

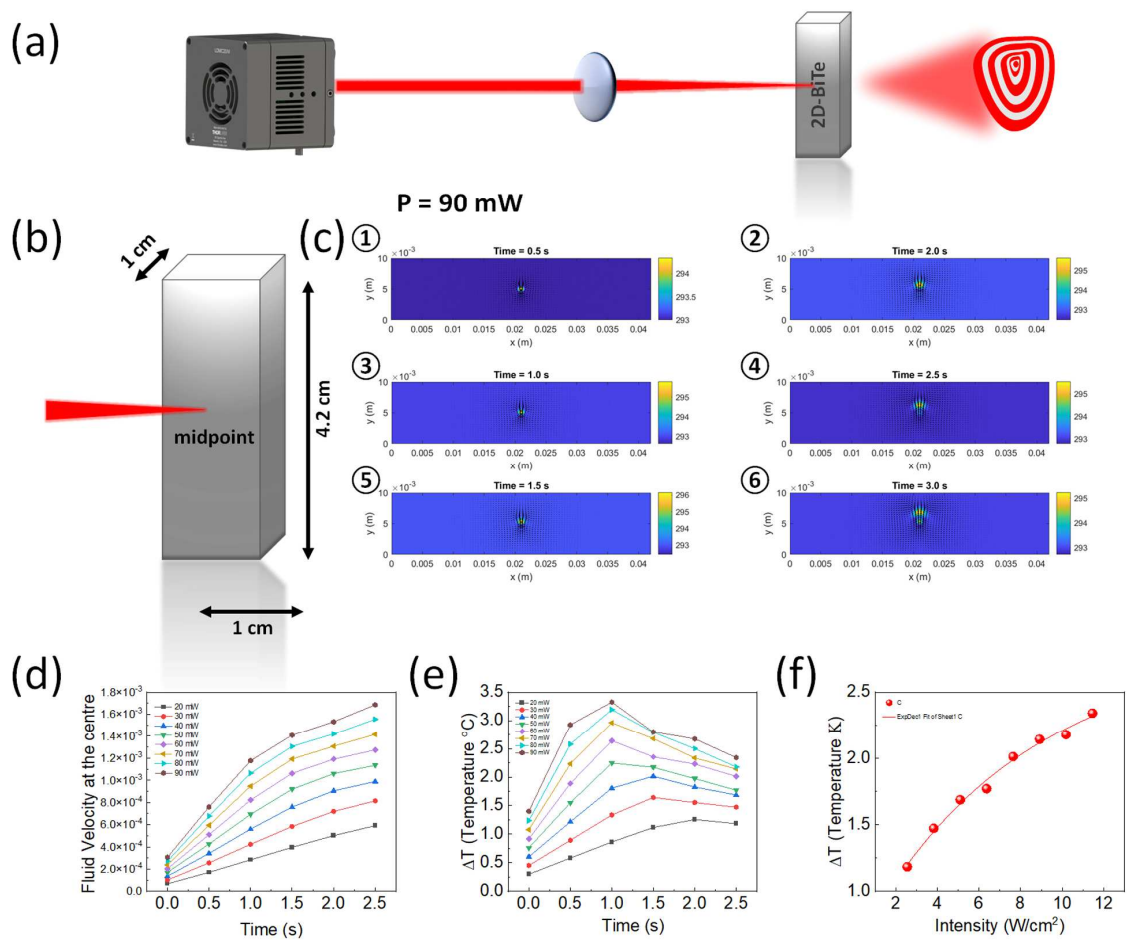

**Figure S14.** (a) SSPM setup used to generate the diffraction pattern. (b) Dimension of the cuvette to visualize the incoming beam and the part of the face that is exposed to the beam. (c) ①-⑥ Figure showing the fluid flow inside the cuvette with respect to time. (d) Maximum Fluid velocity inside the cuvette with respect to time under different laser intensities at the wavelength 650 nm. (e) Maximum Fluid temperature inside the cuvette with respect to time under different laser intensities at the wavelength 650 nm. (f) The temperature difference between before and after the laser impact.

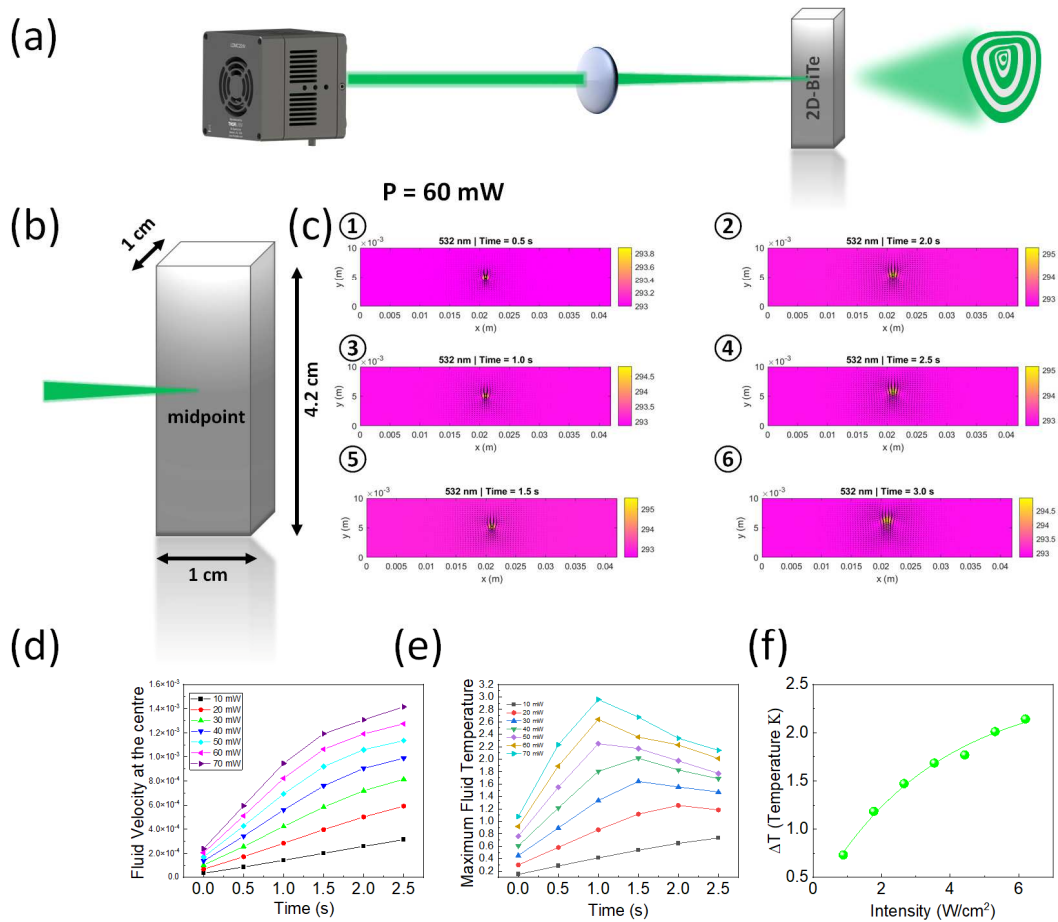

**Figure S15.** (a) SSPM setup used to generate a diffraction pattern at 532 nm. (b) Dimension of the cuvette to visualize the incoming beam and the part of the face that is exposed to the beam. (c) ①-⑥ Figure showing the fluid flow inside the cuvette with respect to time. (d) Maximum Fluid velocity inside the cuvette with respect to time under different laser intensities at the wavelength 532 nm. (e) Maximum Fluid temperature inside the cuvette with respect to time under different laser intensities at the wavelength 532 nm. (f) The temperature difference between before and after the laser impact.

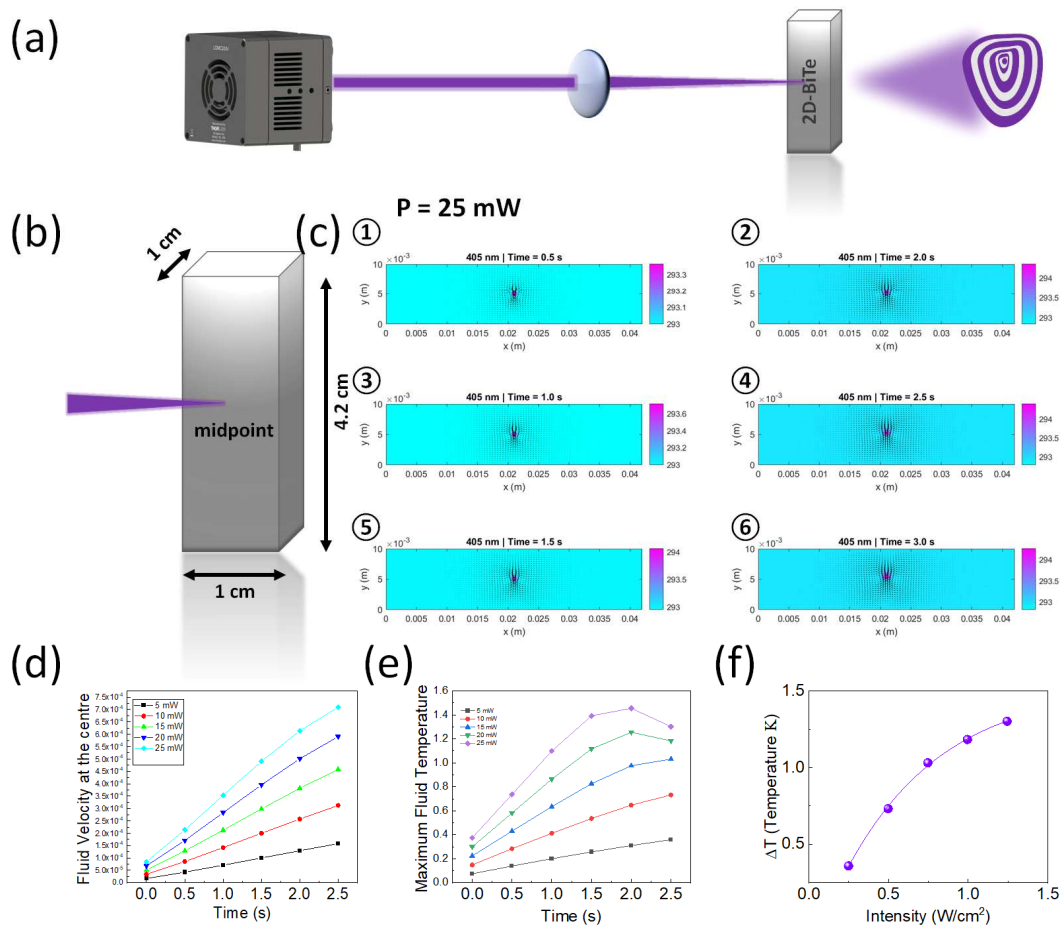

**Figure S16.** (a) SSPM setup used to generate a diffraction pattern at 405 nm. (b) Dimension of the cuvette to visualize the incoming beam and the part of the face that is exposed to the beam. (c) ①-⑥ Figure showing the fluid flow inside the cuvette with respect to time. (d) Maximum Fluid velocity inside the cuvette with respect to time under different laser intensities at the wavelength 405 nm. (e) Maximum Fluid temperature inside the cuvette with respect to time under different laser intensities at the wavelength 405 nm. (f) The temperature difference between before and after the laser impact.

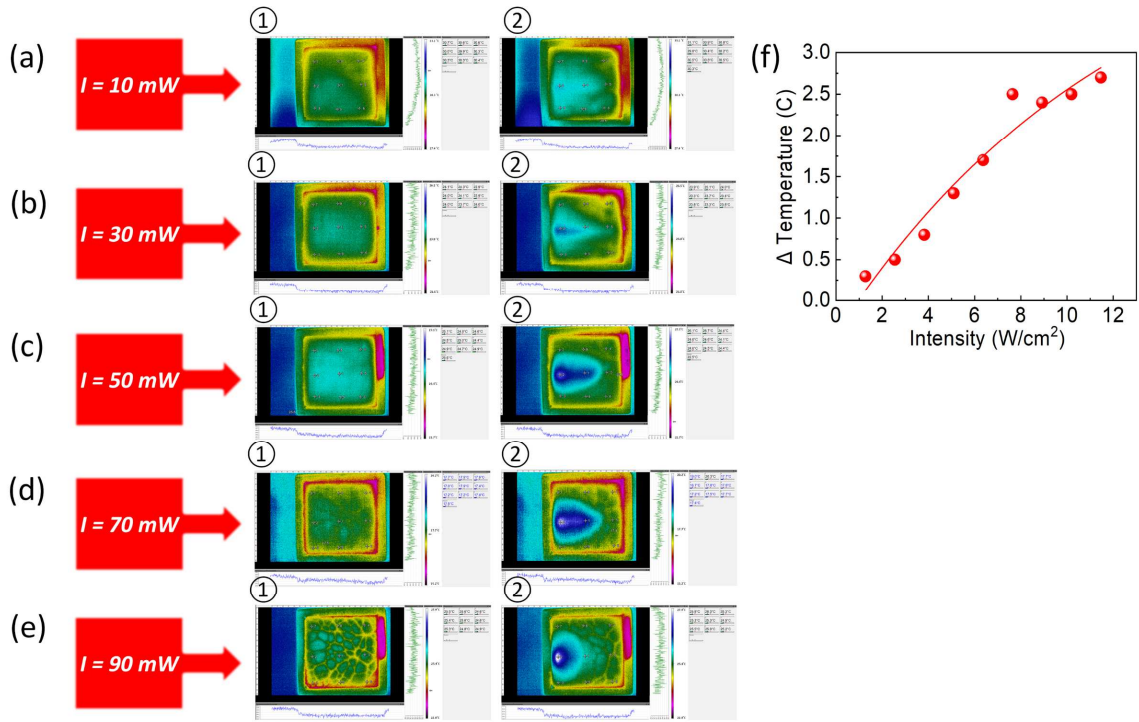

**Figure S17.** (a-c-d-e) Thermal camera image taken for different laser beam intensities (10 − 90 $mW$ with 20 $mW$ interval at the wavelength 650 nm. Adjacent images showing results for before (a)①-(b)①-(c)①-(d)①-(e)① and after laser impact (a)②-(b)②-(c)②-(d)②-(e)②. Compared to data obtained with a thermal camera, the observed temperature increase is quite comparable.

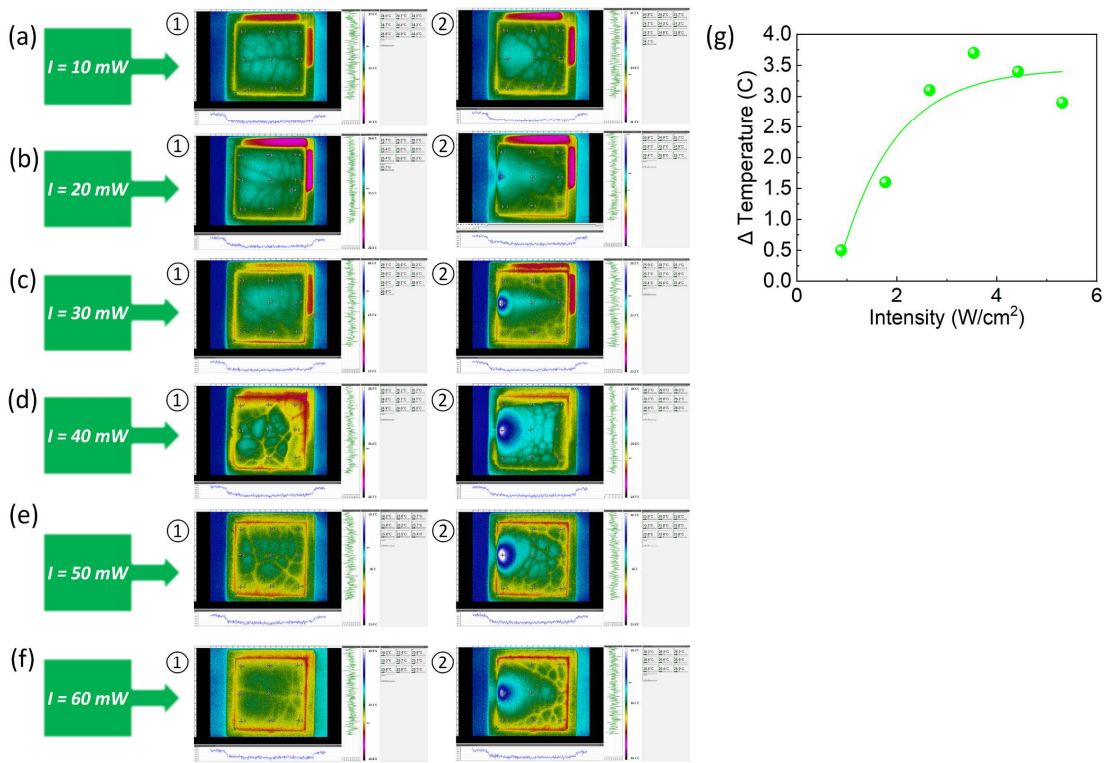

**Figure S18.** (a-c-d-e-f) Thermal camera images captured at varying laser beam intensities ($10 - 60\ mW$ with $10\ mW$ increments) at a wavelength of 532 nm. Adjacent photos depicting outcomes before laser impact (a)①-(b)①-(c)①-(d)①-(e)①-(f)① and subsequent to laser impact (a)②-(b)②-(c)②-(d)②-(e)②-(f)②. The recorded temperature rise is very similar to data collected with a thermal camera.

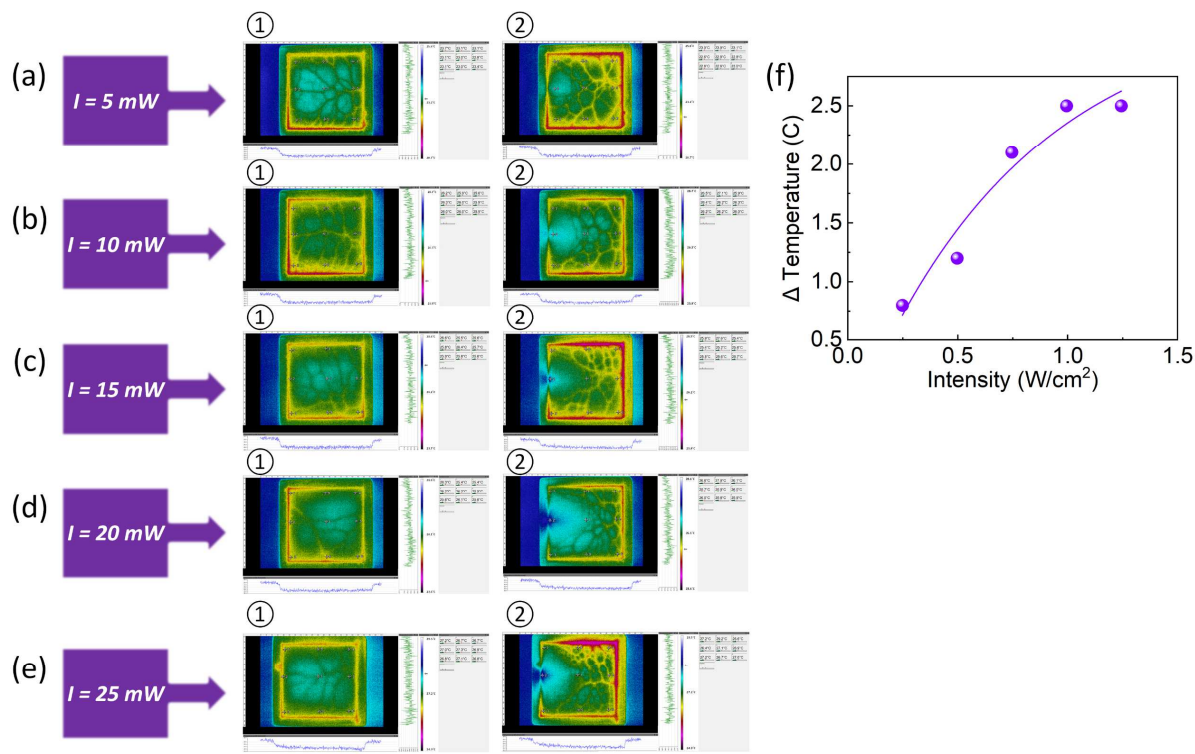

**Figure S19.** (a-c-d-e-f) Thermal camera photos obtained at different laser beam strengths ($5 - 25 \ mW$ with $5 \ mW$ increments) at a wavelength of 405 nm. Adjacent photographs illustrating results before laser application (a)①-(b)①-(c)①-(d)①-(e)①-(f)① and subsequent to laser application (a)②-(b)②-(c)②-(d)②-(e)②-(f)②. The reported temperature increase closely matches data from a thermal camera.

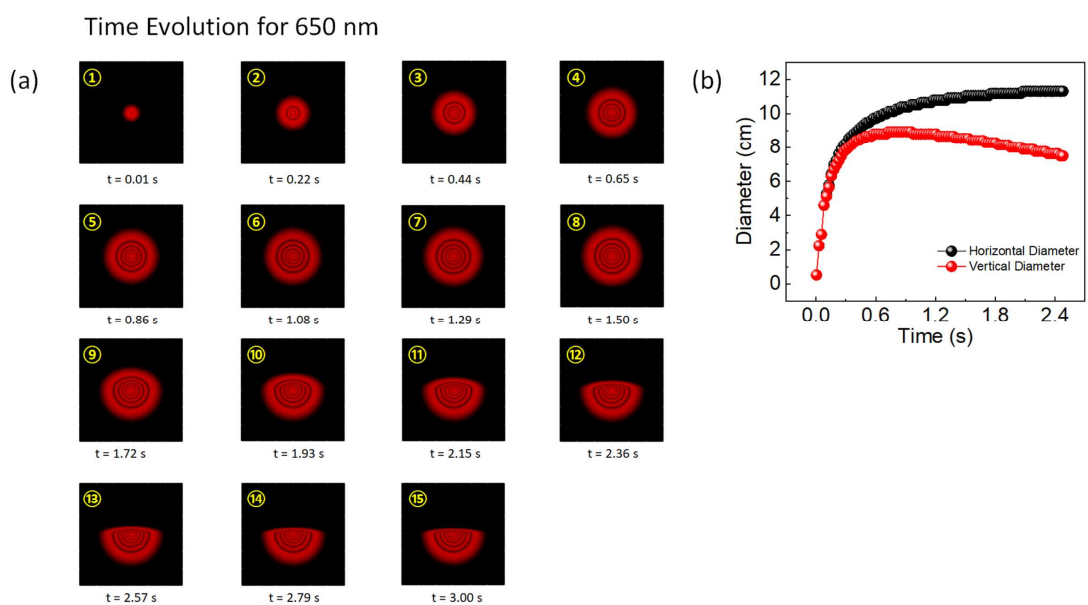

**Figure S20.** (a) Theoretical time evolution and contraction process. (b) Horizontal diameter and vertical diameter during the expansion and contraction process.

# Section: 8 Theoretical calculation of $\chi^{(3)}_{monolayer}$ of 2D BiTe

Since inversion symmetry breaking is necessary for second-order nonlinear processes like second-harmonic production, centrosymmetric 2D BiTe does not have a second-order nonlinear optical coefficient $\chi^{(2)}$. Photonic and optoelectronic applications are directly impacted by the third-order nonlinear susceptibility $\chi^{(3)}$, which is essential for comprehending optical Kerr phenomena, third-harmonic production, and two-photon absorption in materials. By solving the rank-4 $\chi^{(3)}$ A tensor with up to 81 independent components can be directly computed using density functional perturbation theory (DFPT) or time-dependent DFT (TD-DFT). This requires evaluating thousands of intricate matrix elements that are difficult to interpret experimentally. Conversely, semi-empirical Miller's rule uses the basic scaling equations of nonlinear optics to predict $\chi^{(3)}$ from linear susceptibility trends.[62-64] Despite its simplicity, it often yields extremely accurate estimates. The linear refractive index $n_0$ and $\chi^{(3)}$ are related by Miller's rule, as shown below.[64]

$$\chi^{(3)} = \alpha \left[\frac{n_0^2 - 1}{4\pi}\right]^4 \quad \ldots\ldots\ldots (37)$$

Here, α = 2.7 × 10$^{-10}$ e.s.u. represents Miller's coefficient, which provides a somewhat accurate approximation of $\chi^{(3)}$ for numerous ionic crystals, certain oxide/chalcogenide glasses, and various optical crystals.[64] According to the aforementioned formula, the estimated value of $\chi^{(3)}$ is 7.56×10$^{-9}$ e.s.u. for the wavelength 532 nm, which aligns closely with the experimental data.

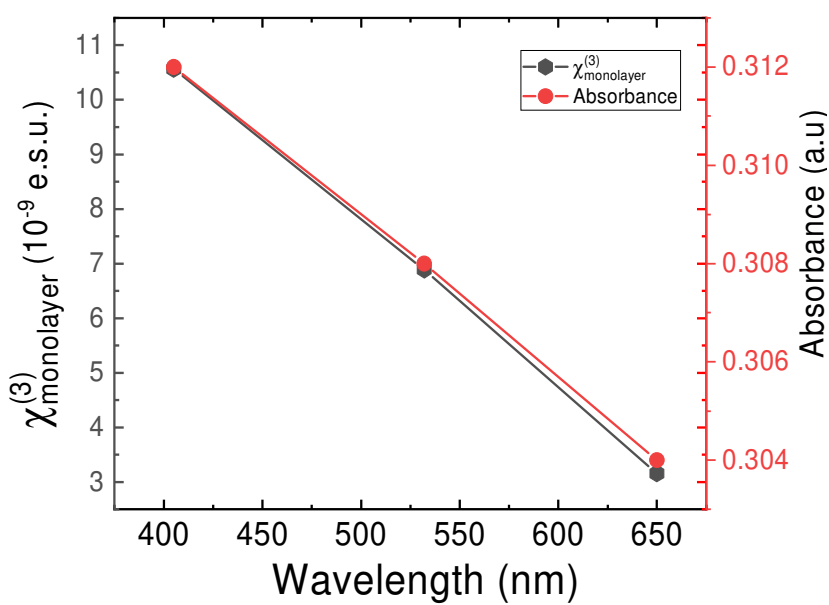

**Figure S21.** Relation between $\chi^{(3)}_{monolayer}$ and absorbance spectra of the 2D BiTe.

**Section 9: Real and Imaginary part of the z component of the dielectric function**

The dielectric function of bulk and two-dimensional BiTe structures was evaluated with the polarization vector oriented perpendicular to the plane of the structure (along the real-space c-axis), following the methodology described in this work. The imaginary part of the dielectric function for the bulk structure (Figure S22a) presents a peak located near 1.5 eV falling to zero far from this value for the photon energy. In contrast, the real part of the bulk shows a plateau at photon energies below 1.0 eV, falling rapidly to -50 at roughly 1.75 eV. Given the relationship between the imaginary part of the dielectric function and absorption, one should expect strong absorption properties near the infrared region. For the 2D slab (Figure S22 b), three distinct peaks are observed: one at roughly 1.80 eV, a large peak at 4.5 eV, and one at 9.0 eV. For other regions, the imaginary part is very close to zero. For the real part, an interesting pattern is observed for 1.80, 4.5, and 9.0 eV photon energy, showing a change in the signal of the derivative in the same regions where the imaginary part has a peak, as discussed above.

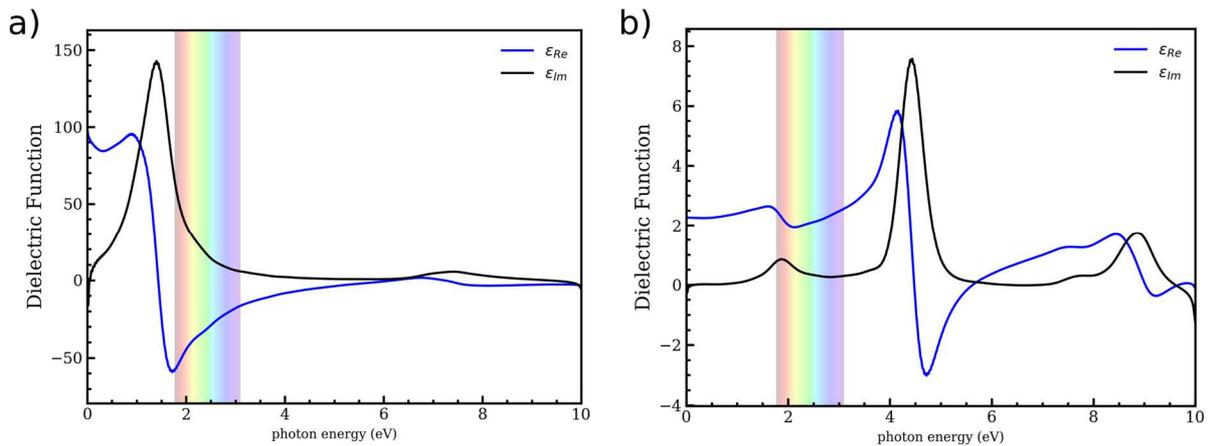

**Figure S22.** Real and Imaginary part of the z component of the dielectric function for a) Bulk and b) 2D slab of BiTe.

**Computational Details**

Spin-polarized first-principles calculations within the framework of density functional theory (DFT) were conducted using the SIESTA code. (version 5.4).[65-66] The exchange and correlation functional interactions were approximated by the revised Perdew Burke and Ernzerhoff [67] functional for solids PBESOL [68]. The Khon-Sham orbitals were expanded by a double-$\zeta$ polarized basis set, constructed by pseudoatomic orbitals of finite range with polarization orbitals. A cutoff of 300 Ry was applied for the real-space grid fineness [69]. Electron-ion interactions were modelled by pseudopotentials employing norm-conserving Vanderbilt pseudopotentials[70] for collinear spin calculations without spin-orbit coupling. For the non-collinear case, fully relativistic pseudopotentials were used. All the pseudopotentials were collected from the pseudo dojo library [71]. The following valence was considered: Bi ($5d^{10}$ $6s^2$ $6p^3$); Te ($5s^2$ $5p^4$). For tellurium, the fully relativistic valence considered was $4d^{10}$ $5s^2$ $5p^4$. Self-consistency was achieved with a 10-6 eV threshold, and forces were minimized to below 0.02 eV/Å. Brillouin zone sampling was performed under the Monkhorst and Pack scheme [72] using a (8×8×4) and (4×8×2) k-point grid for the bulk and 2D slab. For the Dielectric function calculation, the simplest approach based on dipolar transition matrix elements between different eigenfunctions on the self-consistency Hamiltonian was employed, as implemented in the SIESTA code. The energy range considered in these calculations was 0 to 10 eV. To obtain a smooth curve, a Gaussian broadening of 0.2 eV was applied to the frequency values. Integration across the Brillouin zone was performed using a 10×10×10 mesh size. We considered a polarization case where an electric field were applied in the z direction using the Optical. Vector block, defined as (0.0 0.0 1.0) as implemented in SIESTA code.

Table S5. Calculated values for the imaginary part of the dielectric function at the desired wavelengths:

| Wavelength (nm) | Energy (eV) | Bulk | 2D |
| --- | --- | --- | --- |
| 650 | 1.907 | 43.42 | 0.89 |
| 532 | 2.331 | 20.09 | 0.40 |
| 405 | 3.061 | 6.03 | 0.31 |

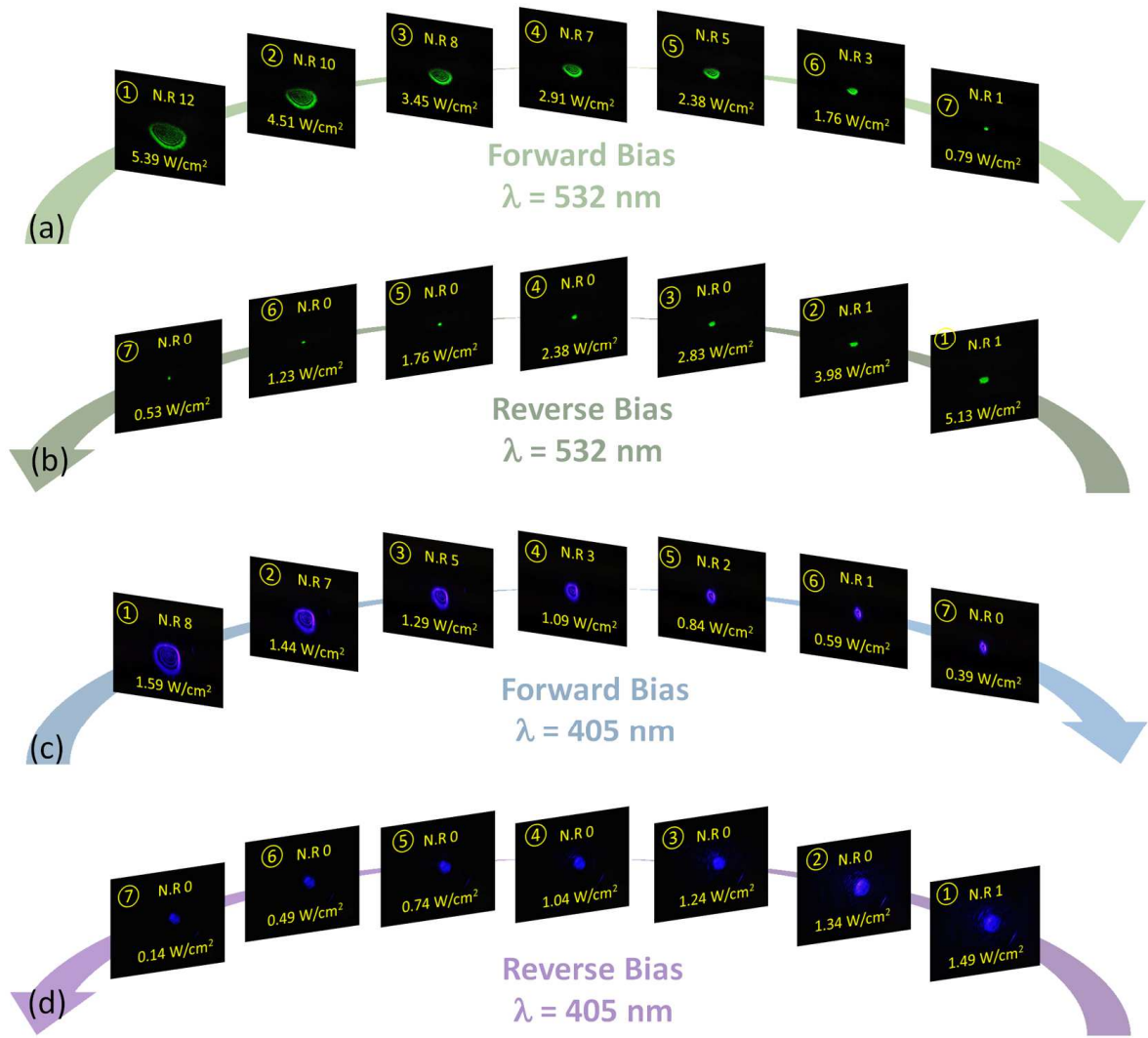

**Figure S23.** (a) and (c) Figures illustrating the diffraction rings captured for the forward bias configuration at the far screen across multiple intensities and distinct wavelengths ($\lambda$ = 532 and 405 nm). (b) and (d) Figures depicting the far-field diffraction pattern for changing intensity for reverse bias arrangement for excitation wave source ($\lambda$ = 532 and 405 nm).

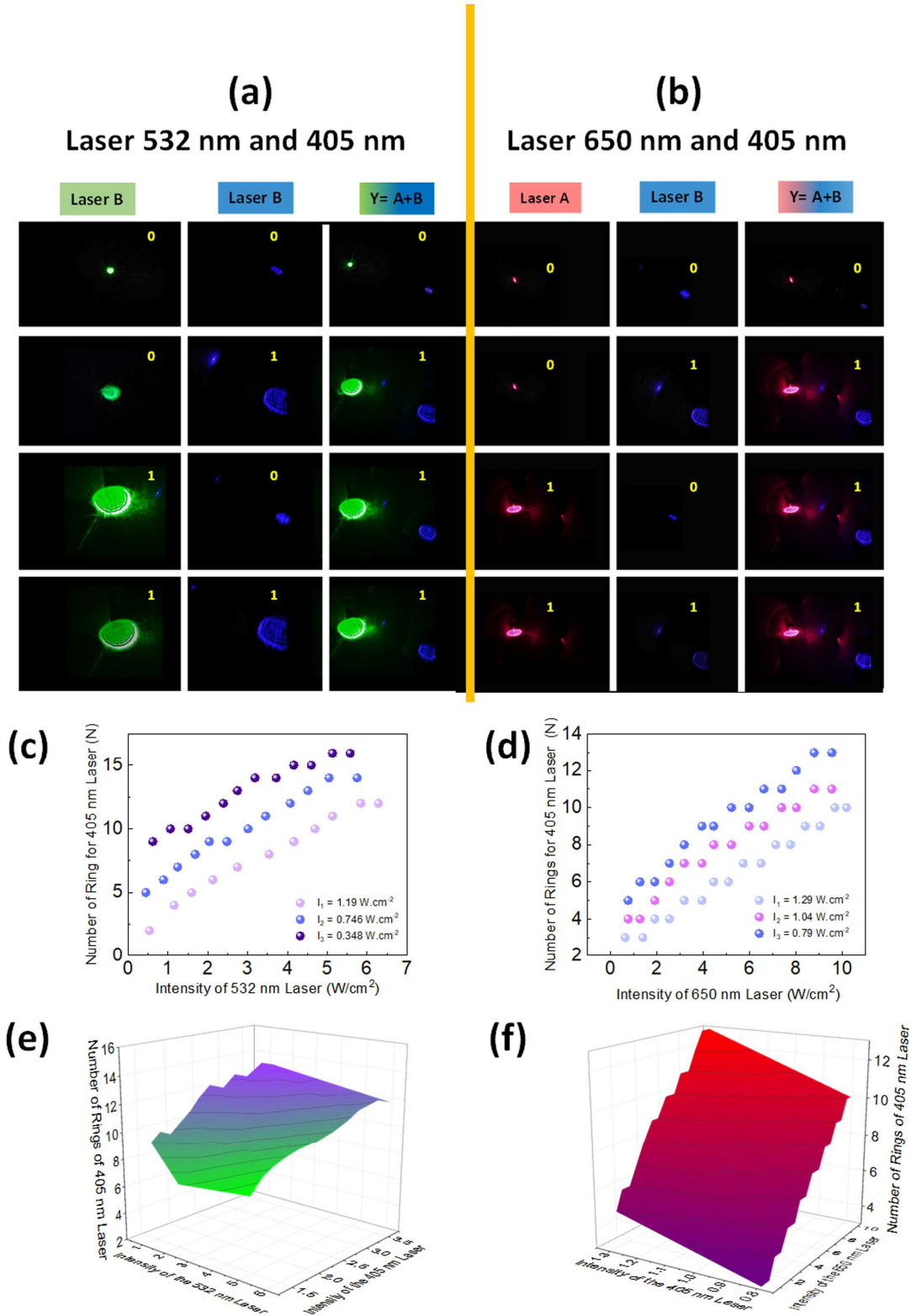

**Figure S24.** (a-b) Experimental demonstration of photonic OR gate operation using a two-color XPM scheme with 2D BiTe, tested for the laser wavelength pairs λ = 405 & 532 nm, and 405 & 650 nm. (c-d) Dependence of probe laser diffraction ring count on variation in pump laser intensity (λ = 532 and 405 nm and λ = 650 and 405 nm). (e-f) Three-dimensional representation of two-color XPM for the corresponding wavelength pairs. (λ = 532 and 405 nm and λ = 650 and 405 nm)

References

[1]     R. Wu, Y. Zhang, S. Yan, et al.," Purely coherent nonlinear optical response in solution dispersions of graphene sheets," *Nano letters* (2011), 11, 5159

[2]     Y. Wu, Q. Wu, F. Sun, et al.," Emergence of electron coherence and two-color all-optical switching in $MoS_2$ based on spatial self-phase modulation," *Proceedings of the National Academy of Sciences* (2015), 112, 11800

[3]     L. Wu, X. Jiang, J. Zhao, et al.," 2D MXene: MXene-Based Nonlinear Optical Information Converter for All-Optical Modulator and Switcher (Laser Photonics Rev. 12 (12)/2018)," *Laser & Photonics Reviews* (2018), 12, 1870055

[4]     L. Wu, W. Huang, Y. Wang, et al.," 2D tellurium based high-performance all-optical nonlinear photonic devices," *Advanced Functional Materials* (2019), 29, 1806346

[5]     L. Lu, X. Tang, R. Cao, et al.," Broadband nonlinear optical response in few-layer antimonene and antimonene quantum dots: a promising optical kerr media with enhanced stability," *Advanced Optical Materials* (2017), 5, 1700301

[6]     B. Shi, L. Miao, Q. Wang, et al.," Broadband ultrafast spatial self-phase modulation for topological insulator $Bi_2Te_3$ dispersions," *Applied Physics Letters* (2015), 107.10.1063/1.4932590

[7]     S. Xiao, Y.-l. He, Y.-l. Dong, et al.," Near-Infrared Spatial Self-Phase Modulation in Ultrathin Niobium Carbide Nanosheets," *Frontiers in Physics* (2021), Volume 9 - 2021.10.3389/fphy.2021.674820

[8]     Y. Shan, J. Tang, L. Wu, et al.," Spatial self-phase modulation and all-optical switching of graphene oxide dispersions," *Journal of Alloys and Compounds* (2019), 771, 900

[9]     L. Hu, F. Sun, H. Zhao, J. Zhao," Nonlinear optical response spatial self-phase modulation in MoTe$_2$: correlations between χ (3) and mobility or effective mass," *Optics Letters* (2019), 44, 5214

[10]    Y. Jia, Y. Liao, L. Wu, et al.," Nonlinear optical response, all optical switching, and all optical information conversion in NbSe$_2$ nanosheets based on spatial self-phase modulation," *Nanoscale* (2019), 11, 4515

[11]    X. Li, R. Liu, H. Xie, et al.," Tri-phase all-optical switching and broadband nonlinear optical response in Bi$_2$Se$_3$ nanosheets," *Optics Express* (2017), 25, 18346

[12]    J. Zhang, X. Yu, W. Han, et al.," Broadband spatial self-phase modulation of black phosphorous," *Opt. Lett.* (2016), 41, 1704.10.1364/OL.41.001704

[13]    C. Wang, S. Xiao, X. Xiao, et al.," Nonlinear Optical Response of SbSI Nanorods Dominated with Direct Band Gaps," *The Journal of Physical Chemistry C* (2021), 125, 15441.10.1021/acs.jpcc.1c04282

[14]    Y. Jia, Y. Shan, L. Wu, et al.," Broadband nonlinear optical resonance and all-optical switching of liquid phase exfoliated tungsten diselenide," *Photonics Research* (2018), 6, 1040

[15]    Y. Liao, Q. Ma, Y. Shan, et al.," All-optical applications for passive photonic devices of TaS$_2$ nanosheets with strong Kerr nonlinearity," *Journal of Alloys and Compounds* (2019), 806, 999

[16]    Y. Shan, L. Wu, Y. Liao, et al.," A promising nonlinear optical material and its applications for all-optical switching and information converters based on the spatial self-phase modulation (SSPM) effect of TaSe$_2$ nanosheets," *Journal of Materials Chemistry C* (2019), 7, 3811

[17]    Y. Jia, Z. Li, M. Saeed, et al.," Kerr Nonlinearity in germanium selenide nanoflakes measured by Z-scan and spatial self-phase modulation techniques and its applications in all-optical information conversion," *Optics Express* (2019), 27, 20857

[18]   C. Song, Y. Liao, Y. Xiang, X. Dai," Liquid phase exfoliated boron nanosheets for all-optical modulation and logic gates," *Science Bulletin* (2020), 65, 1030

[19]   L. Wu, Z. Xie, L. Lu, et al.," Few-layer tin sulfide: a promising black-phosphorus-analogue 2D material with exceptionally large nonlinear optical response, high stability, and applications in all-optical switching and wavelength conversion," *Advanced Optical Materials* (2018), 6, 1700985

[20]   Y. Shan, Z. Li, B. Ruan, et al.," Two-dimensional $Bi_2S_3$-based all-optical photonic devices with strong nonlinearity due to spatial self-phase modulation," *Nanophotonics* (2019), 8, 2225

[21]   W. Wang, Y. Wu, Q. Wu, J. Hua, J. Zhao," Coherent nonlinear optical response spatial self-phase modulation in $MoSe_2$ nano-sheets," *Scientific reports* (2016), 6, 22072

[22]   Y. Huang, H. Zhao, Z. Li, et al.," Laser-Induced Hole Coherence and Spatial Self-Phase Modulation in the Anisotropic 3D Weyl Semimetal TaAs," *Advanced Materials* (2023), 35, 2208362

[23]   S. Goswami, C. C. de Oliveira, B. Ipaves, et al.," Exceptionally High Nonlinear Optical Response in Two-dimensional Type II Dirac Semimetal Nickel Di-Telluride ($NiTe_2$)," *Laser & Photonics Reviews*, n/a, 2400999. https://doi.org/10.1002/lpor.202400999

[24]   P. Dey, N. Chakraborty, M. Samanta, B. Das, K. K. Chattopadhyay," Strong light–matter interaction and non-linear effects in organic semiconducting CuPc nanotubes: realization of all-optical diode and switching applications," *Physical Chemistry Chemical Physics* (2024), 26, 20112

[25]   N. Sen, N. Chakraborty, B. Das, K. K. Chattopadhyay," Strong non-linear optical response of $Sb_2Se_3$ nanorods in a liquid suspension based on spatial self-phase modulation and their all-optical photonic device applications," *Nanoscale* (2023), 15, 19671. 10.1039/D3NR04623K

[26] X. Xu, Z. Cui, Y. Yang, et al.," Large Optical Nonlinearity Enhancement and All-Optical Logic Gate Implementation in Silver-Modified Violet Phosphorus," *Laser & Photonics Reviews* (2025), 19, 2401521

[27] Z. Xu, H. Wang, W. Niu, et al.," Strong Nonlinear Response and All-Optical Applications in Hybrid Bismuth Halide with Spatial Self-Phase Modulation," *Laser Photonics Rev.*, n/a, 2401929.https://doi.org/10.1002/lpor.202401929

[28] D. Weng, C. Ling, Y. Gao, et al.," Spatially Asymmetric Optical Propagation and All-Optical Switching Based on Spatial Self-Phase Modulation of Semimetal MoP Microparticles," *Laser Photonics Rev.*, n/a, 2401587.https://doi.org/10.1002/lpor.202401587

[29] S. Bera, S. Kalimuddin, A. Bera, et al.," Nonlinear Optical Properties of 2D vdW Ferromagnetic Nanoflakes for Magneto-Optical Logic Applications," *Advanced Optical Materials* (2025), 13, 2402318.https://doi.org/10.1002/adom.202402318

[30] S. Goswami, B. Ipaves, J. G. Quispe, et al.," All Photonic Isolator using Atomically Thin (2D) Bismuth Telluride ($Bi_2Te_3$)," *arXiv preprint arXiv:2508.03319* (2025),

[31] J. Xu, C. Zhang, Y. Wang, et al.," All-in-one, all-optical logic gates using liquid metal plasmon nonlinearity," *Nature Communications* (2024), 15, 1726

[32] J. Li, W. Niu, X. Xu, et al.," Laser-induced hole coherence in 2D antiferromagnet MPS3 through spatial self-phase modulation," *Optics Express* (2025), 33, 1044

[33] Y. Wu, L. Zhu, Q. Wu, et al.," Electronic origin of spatial self-phase modulation: evidenced by comparing graphite with C60 and graphene," *Applied physics letters* (2016), 108

[34] X. Xu, Z. Cui, Y. Yang, et al.," Large Optical Nonlinearity Enhancement and All-Optical Logic Gate Implementation in Silver-Modified Violet Phosphorus," *Laser Photonics Rev.* (2025), 19, 2401521.https://doi.org/10.1002/lpor.202401521

[35]    W. G. Z. S. U. FA," Cheng X. Dong N. Coghlan D. Cheng Y. Zhang L. Blau WJ Wang J.,"Tunable effective nonlinear refractive index of graphene dispersions during the distortion of spatial self-phase modulation,""" *Appl. Phys. Lett* (2014), 104, 141909

[36]    L. Wu, Y. Dong, J. Zhao, et al.," Kerr nonlinearity in 2D graphdiyne for passive photonic diodes," *Adv. Mater.* (2019), 31, 1807981

[37]    Y. Liao, Y. Shan, L. Wu, Y. Xiang, X. Dai," Liquid-Exfoliated Few-Layer InSe Nanosheets for Broadband Nonlinear All-Optical Applications," *Advanced Optical Materials* (2020), 8, 1901862

[38]    K. Sk, B. Das, N. Chakraborty, et al.," Nonlinear Coherent Light–Matter Interaction in 2D MoSe$_2$ Nanoflakes for All-Optical Switching and Logic Applications," *Advanced Optical Materials* (2022), 10, 2200791

[39]    Y. Wu, Q. Wu, F. Sun, et al.," Emergence of electron coherence and two-color all-optical switching in MoS$_2$ based on spatial self-phase modulation," *Proceedings of the National Academy of Sciences* (2015), 112, 11800.doi:10.1073/pnas.1504920112

[40]    L. Cheng, Y. Liu," What limits the intrinsic mobility of electrons and holes in two dimensional metal dichalcogenides" *Journal of the American Chemical Society* (2018), 140, 17895

[41]    M. Orlita, C. Faugeras, P. Plochocka, et al.," Approaching the Dirac point in high-mobility multilayer epitaxial graphene," *Physical review letters* (2008), 101, 267601

[42]    G. Wang, S. Zhang, F. A. Umran, et al.," Tunable effective nonlinear refractive index of graphene dispersions during the distortion of spatial self-phase modulation," *Applied Physics Letters* (2014), 104.10.1063/1.4871092

[43]    G. Long, D. Maryenko, J. Shen, et al.," Achieving ultrahigh carrier mobility in two-dimensional hole gas of black phosphorus," *Nano Letters* (2016), 16, 7768

[44]    S.-L. Li, K. Tsukagoshi, E. Orgiu, P. Samorì," Charge transport and mobility engineering in two-dimensional transition metal chalcogenide semiconductors," *Chemical Society Reviews* (2016), 45, 118

[45]    S. Yu, H. D. Xiong, K. Eshun, H. Yuan, Q. Li," Phase transition, effective mass and carrier mobility of $MoS_2$ monolayer under tensile strain," *Applied Surface Science* (2015), 325, 27

[46]    A. Allain, A. Kis," Electron and hole mobilities in single-layer $WSe_2$," *ACS nano* (2014), 8, 7180

[47]    S. Kumar, U. Schwingenschlogl," Thermoelectric response of bulk and monolayer $MoSe_2$ and $WSe_2$," *Chemistry of Materials* (2015), 27, 1278

[48]    D. Ovchinnikov, A. Allain, Y.-S. Huang, D. Dumcenco, A. Kis," Electrical transport properties of single-layer $WS_2$," *ACS nano* (2014), 8, 8174

[49]    G. Wang, S. Zhang, X. Zhang, et al.," Tunable nonlinear refractive index of two-dimensional $MoS_2$, $WS_2$, and $MoSe_2$ nanosheet dispersions," *Photonics Research* (2015), 3, A51

[50]    R. Maiti, M. A. S. R. Saadi, R. Amin, et al.," Strain-induced spatially resolved charge transport in 2H-$MoTe_2$," *ACS Applied Electronic Materials* (2021), 3, 3781

[51]    N. F. Hinsche, K. S. Thygesen," Electron–phonon interaction and transport properties of metallic bulk and monolayer transition metal dichalcogenide $TaS_2$," *2D Materials* (2017), 5, 015009

[52]    Y. Xu, H. Zhang, H. Shao, et al.," First-principles study on the electronic, optical, and transport properties of monolayer GeSe," *Physical Review B* (2017), 96, 245421.10.1103/PhysRevB.96.245421

[53]    C. Xin, J. Zheng, Y. Su, et al.," Few-Layer Tin Sulfide: A New Black-Phosphorus-Analogue 2D Material with a Sizeable Band Gap, Odd–Even Quantum Confinement Effect,

and High Carrier Mobility," *The Journal of Physical Chemistry C* (2016), 120, 22663.10.1021/acs.jpcc.6b06673

[54] S. Goswami, C. C. de Oliveira, B. Ipaves, et al.," Exceptionally High Nonlinear Optical Response in Two-dimensional Type II Dirac Semimetal Nickel Di-Telluride (NiTe$_2$)," *Laser & Photonics Reviews* (2025), 2400999

[55] F. Dabby, T. Gustafson, J. Whinnery, Y. Kohanzadeh, P. Kelley," Thermally self-induced phase modulation of laser beams," *Applied Physics Letters* (1970), 16, 362

[56] Y. Liao, Y. Shan, L. Wu, Y. Xiang, X. Dai," Liquid-Exfoliated Few-Layer InSe Nanosheets for Broadband Nonlinear All-Optical Applications," *Advanced Optical Materials* (2020), 8, 1901862.https://doi.org/10.1002/adom.201901862

[57] K. Sk, B. Das, N. Chakraborty, et al.," Nonlinear Coherent Light–Matter Interaction in 2D MoSe$_2$ Nanoflakes for All-Optical Switching and Logic Applications," *Advanced Optical Materials* (2022), 10, 2200791.https://doi.org/10.1002/adom.202200791

[58] J. Wang, Y. Hernandez, M. Lotya, J. N. Coleman, W. J. Blau," Broadband nonlinear optical response of graphene dispersions," *Advanced Materials* (2009), 21, 2430

[59] L. Wu, X. Yuan, D. Ma, et al.," Recent advances of spatial self-phase modulation in 2D materials and passive photonic device applications," *Small* (2020), 16, 2002252

[60] J. Whinnery, D. Miller, F. Dabby," Thermal convention and spherical aberration distortion of laser beams in low-loss liquids," *IEEE Journal of Quantum Electronics* (1967), 3, 382

[61] R. Karimzadeh," Spatial self-phase modulation of a laser beam propagating through liquids with self-induced natural convection flow," *Journal of optics* (2012), 14, 095701

[62] C. C. Wang," Empirical relation between the linear and the third-order nonlinear optical susceptibilities," *Physical Review B* (1970), 2, 2045

[63]     M. T. Meyer, A. Schindlmayr," Derivation of Miller's rule for the nonlinear optical susceptibility of a quantum anharmonic oscillator," *Journal of Physics B: Atomic, Molecular and Optical Physics* (2024), 57, 095001

[64]     T. Wang, X. Gai, W. Wei, et al.," Systematic z-scan measurements of the third order nonlinearity of chalcogenide glasses," *Optical Materials Express* (2014), 4, 1011

[65]     J. M. Soler, E. Artacho, J. D. Gale, et al.," The SIESTA method for ab initio order-Nmaterials simulation," *Journal of physics: Condensed matter* (2002), 14, 2745

[66]     A. García, N. Papior, A. Akhtar, et al.," Siesta: Recent developments and applications," *The Journal of Chemical Physics* (2020), 152.10.1063/5.0005077

[67]     J. P. Perdew, K. Burke, M. Ernzerhof," Generalized Gradient Approximation Made Simple," *Physical Review Letters* (1996), 77, 3865.10.1103/PhysRevLett.77.3865

[68]     J. P. Perdew, A. Ruzsinszky, G. I. Csonka, et al.," Restoring the Density-Gradient Expansion for Exchange in Solids and Surfaces," *Physical Review Letters* (2008), 100, 136406.10.1103/PhysRevLett.100.136406

[69]     E. Anglada, J. M. Soler, J. Junquera, E. Artacho," Systematic generation of finite-range atomic basis sets for linear-scaling calculations," *Physical Review B* (2002), 66, 205101.10.1103/PhysRevB.66.205101

[70]     D. R. Hamann," Optimized norm-conserving Vanderbilt pseudopotentials," *Physical Review B* (2013), 88, 085117.10.1103/PhysRevB.88.085117

[71]     M. J. van Setten, M. Giantomassi, E. Bousquet, et al.," The PseudoDojo: Training and grading a 85 element optimized norm-conserving pseudopotential table," *Computer Physics Communications* (2018), 226, 39.https://doi.org/10.1016/j.cpc.2018.01.012

[72]     H. J. Monkhorst, J. D. Pack," Special points for Brillouin-zone integrations," *Physical Review B* (1976), 13, 5188.10.1103/PhysRevB.13.5188